\documentclass[11pt]{article}
\usepackage{jheppub}

\usepackage{amssymb}
\usepackage{amsmath}
\usepackage{amsfonts}
\usepackage{graphicx, subfigure, placeins, float}

\usepackage{array} 
\usepackage{longtable} 
\usepackage{colortab} 
\usepackage{colortbl}
\usepackage{arydshln}

\usepackage{dsfont,bbm}
\usepackage{mathrsfs}  
\usepackage{xfrac}
\usepackage{slashed}
\usepackage{mathabx}
\usepackage{enumitem}
\usepackage{shuffle}
\usepackage{stmaryrd}
\usepackage{pifont}

\makeatletter
\newcommand\xleftrightarrow[2][]{%
  \ext@arrow 9999{\longleftrightarrowfill@}{#1}{#2}}
\newcommand\longleftrightarrowfill@{%
  \arrowfill@\leftarrow\relbar\rightarrow}
\makeatother

\newcommand{\Rome}[1]{\uppercase\expandafter{\romannumeral#1}}

\newcommand{\itbf}[1]{\textbf{\textit{#1}}}

\newcolumntype{C}[1]{>{\centering\arraybackslash}m{#1}}

\makeatletter
\newcommand{\mathleft}{\@fleqntrue\@mathmargin0pt}
\newcommand{\mathcenter}{\@fleqnfalse}
\makeatother

\usepackage{extarrows}

\newcommand{\QQ}{{\bf q}}

\preprint{USTC-ICTS/PCFT-22-29}
\title{Double copy for tree-level form factors. Part I. Foundations}

\author[a,e]{Guanda Lin,}
\emailAdd{linguandak@pku.edu.cn}
\author[a,b,c,d]{Gang Yang}
\emailAdd{yangg@itp.ac.cn}
\affiliation[a]{CAS Key Laboratory of Theoretical Physics, Institute of Theoretical Physics, \\Chinese Academy of Sciences, Beijing 100190, China}
\affiliation[b]{School of Fundamental Physics and Mathematical Sciences, Hangzhou Institute for Advanced Study, UCAS, Hangzhou 310024, China}
\affiliation[c]{International Centre for Theoretical Physics Asia-Pacific, Beijing/Hangzhou, China}
\affiliation[d]{Peng Huanwu Center for Fundamental Theory, Hefei, Anhui 230026, China}
\affiliation[e]{Department of Physics, University of California, Berkeley, CA 94720, U.S.A.}

\abstract{
The double-copy construction for form factors was reported in our previous work, in which a novel mechanism of turning spurious poles in Yang-Mills theory into physical poles in gravity is observed. 
This paper is the first of a series of two papers providing the details as well as various generalizations on the double-copy construction of tree-level form factors. 
In this paper, we establish the generic formalism by focusing on the form factor of ${\rm tr}(\phi^2)$ in the Yang-Mills-scalar theory.
A thorough discussion is given on the emergence of the ``spurious"-type poles and various related properties. 
We also discuss two generalizations: the Higgs amplitudes in QCD, and the ${\rm tr}(\phi^2)$ form factors with multiple external scalar states.
}

\begin{document}

\maketitle

\setcounter{footnote}{0}

\section{Introduction}

Gauge and gravity theories are  known to be closely related in spite of their very different nature.
The double copy relation ``gravity = (gauge theory)$\otimes$(gauge theory)'' reveals profound connections  between certain quantities in gauge and gravity theories. 
This was originally inspired by the study of open and closed string amplitudes, which  resulted in the
Kawai, Lewellen and Tye (KLT) relations \cite{Kawai:1985xq}. 
The double copy construction was later generalized to amplitudes in gauge and gravity theories as related but different formalisms, including
 the Bern, Carrasco, and Johansson (BCJ) double copy \cite{Bern:2008qj, Bern:2010ue} and the Cachazo, He, and Yuan (CHY) formula \cite{Cachazo:2013hca, Cachazo:2014xea}. 
In particular, for amplitudes, the BCJ double copy, which starts from a color-kinematics (CK) duality representation of gauge-theory amplitudes and performs double copy thereby, turns out to be a huge success and uncovers relations between amplitudes in a wide range of gauge and gravity theories (see \cite{Bern:2019prr,Bern:2022wqg,Adamo:2022dcm} for reviews).

+In this paper, we will focus on the double copy of form factors using the BCJ and KLT formalism. Form factors \cite{Maldacena:2010kp,Brandhuber:2010ad, Bork:2010wf} can be taken as a generalization of amplitudes and are matrix elements between on-shell states and a gauge invariant local operator ${\cal O}(x)$ (see \cite{Yang:2019vag} for a recent review):
\begin{equation}
\itbf{F}_{{\cal O},n}(1,..,n) =  \int d^{D} x \, e^{-i q \cdot x}\langle 1, .. , n |{\cal O}(x)| 0\rangle = \delta^{D} (q-\sum_{i=1}^n p_i) \langle 1, .. , n |{\cal O}(0)| 0\rangle \,,
\end{equation} 
where $p_i,\  i=1,..,n,$ are on-shell momenta associated with $n$ external particles, and $q=\sum_i p_i$ is the off-shell momentum associated with the operator.

How to perform ``double-copy" for form factors is an interesting open question and deserves exploration.
In previous studies, the CK duality has been used as a powerful tool for computing high-loop form factors in gauge theories, like ${\cal N}=4$ SYM \cite{Boels:2012ew,Yang:2016ear,Lin:2021kht,Lin:2021qol, Lin:2021lqo,Lin:2020dyj} and pure YM theory \cite{Li:2022tir}; however, the double-copy side was still unclear. 
The problem is that the form factor double copy can not be derived straightforwardly from the previously obtained duality satisfying representations (where only the Jacobi-type relations were considered), and as we will see in the main text, to get a consistent double copy for form factors, it is necessary to consider some new operator-induced relations.

Another motivation to study the form factor double copy is to have a better insight into the double copy for amplitudes with ``colorless'' particles. 
Here the connection is that the insertion of  a gauge-invariant operator in a form factor can be interpreted as a color-singlet particle. Therefore, solving the form factor double copy problem can also help to understand the double copy for amplitudes involving color-singlet particles, which has not been well explored in literature. 

A concrete step towards understanding the double copy of form factors was reported in our recent paper in \cite{Lin:2021pne}.
The most intriguing observation there is that new poles (corresponding to new Feynman diagrams) arise in pursuing a diffeomorphism invariant double copy. In this and a forthcoming paper \cite{treepaper2}, we will give more elaborate and complete explanations on the novelties in the form factor double copy, and show that the construction applies to a large class of form factors.

We outline several salient features of the form factor double copy as follows. 
\begin{enumerate}[topsep=3pt,itemsep=-1ex,partopsep=1ex,parsep=1ex]
    \item  The new poles mentioned above are ``spurious" poles for gauge-theory form factors, in the sense that the gauge-theory form factors do not diverge on such poles. 
    After double copy, interestingly, these poles become real poles (as physical propagators) in the gravity theory:
\begin{equation} 
\textrm{spurious poles in gauge theory} \ 
\xlongrightarrow[\mbox{}]{\scriptstyle \textrm{double-copy}} \ 
\textrm{physical propagators in gravity} . 
\nonumber
\end{equation}
More concretely, these ``spurious"-type poles appear in the numerators in the CK-dual representation but cancel in full form factors summing up all diagrams. 
However, such poles not only survive after double copy but also bear nice factorization behavior. Hence they become real physical propagators in the gravity theory.

\item For amplitudes, one usually expresses the double copy $\mathcal{M}$ in terms of a bilinear form of gauge amplitudes $\mathcal{A}$ as $\mathcal{M}=\vec{\mathcal{A}}\cdot \mathbf{S}^{\cal A}\cdot \vec{\mathcal{A}}$, where $\mathbf{S}^{\cal A}$ is the double copy kernel (KLT kernel).
We obtain a similar bilinear representation for form factors. The form factor kernel $\mathbf{S}^{\cal F}$ contains the ``spurious"-type poles mentioned above (note that they are real poles in $\mathbf{S}^{\cal F}$). There is a nice matrix decomposition for $\mathbf{S}^{\cal F}$ as
\begin{equation}
 \textrm{Residue of} \   \mathbf{S}^{\cal F}\  \text{on}\  \textrm{spurious poles} = \mathbf{V}^{\scriptscriptstyle \rm T}\cdot (\mathbf{S}_{m}^{\cal F}\otimes \mathbf{S}_{m'}^{\cal A})\cdot \mathbf{V} \,,
\end{equation}
where $\mathbf{V}$ is a rectangular matrix as a function of Mandelstam variables, and $\mathbf{S}_{m}^{\cal F}$ and $\mathbf{S}_{m'}^{\cal A}$ are (smaller) kernels of  a lower-point form factor and amplitude.
The $\mathbf{V}$ also plays an important role in the following third point. 

\item For gauge-theory form factors, there are hidden factorization structures.
Concretely, when taking the kinematics in the limit that a ``spurious"-type pole goes to zero, a certain linear combination of form factors will factorize as the product of a lower-point form factor and an amplitude, which can be schematically given as
\begin{equation}
\label{eq:generalfactorization}
\vec{v} \cdot \vec{\mathcal{F}}_n  \big|_{\textrm{spurious pole}}= \mathcal{F}_{m}\times \mathcal{A}_{m'}\,,
\end{equation}
where $\vec{v}$ is a vector as a row of the matrix $\mathbf{V}$, $\vec{\mathcal{F}}_n$ is a list of color-ordered form factor basis, and $\mathcal{F}_{m}$/$ \mathcal{A}_{m'}$ are lower point form factors/amplitudes. 
We stress that there is no singular behavior on the LHS of \eqref{eq:generalfactorization}, and thus this equation is very different from the usual factorizations on physical propagators.

\end{enumerate}

\begin{figure}[htbp]
\centering
\includegraphics[clip,scale=0.85]{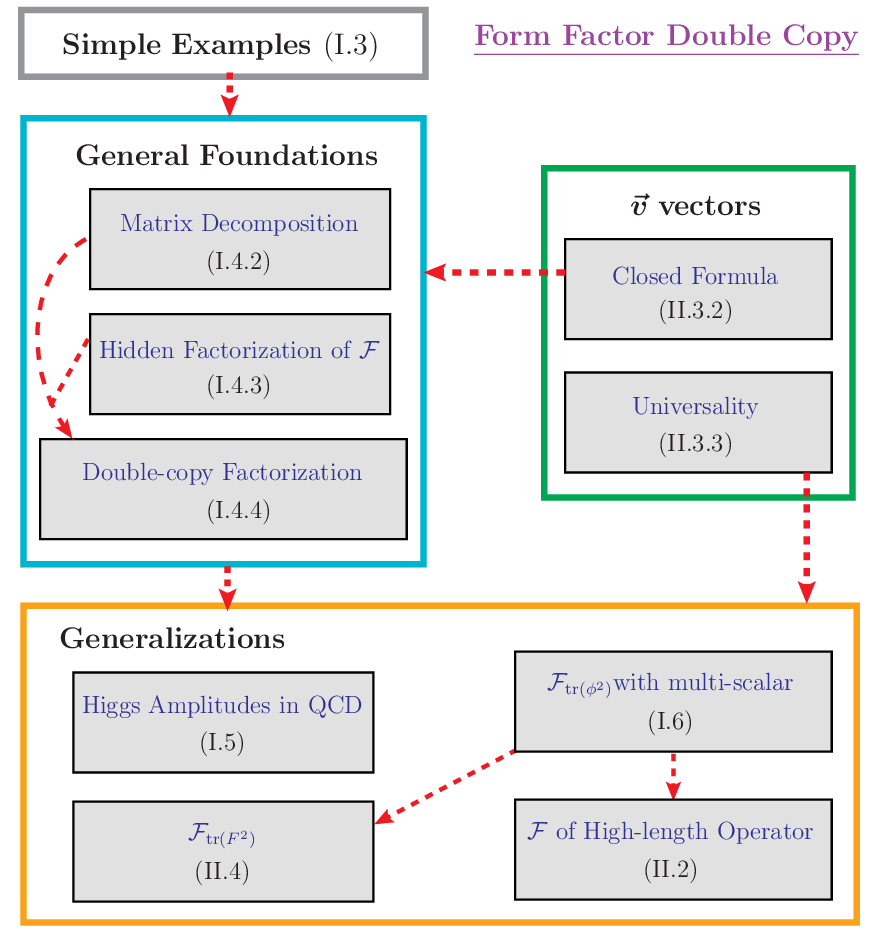}
\caption[a]{The main structure of the two papers for the form factor double copy. The topics are organized into five major parts.
For each topic, we also label the corresponding section in the papers. The roman number I/II refers to the first/second paper.
}
\label{fig:paperstructure}
\end{figure}

All the above and the related stories, as well as several generalizations, will be explained in detail in this and the forthcoming paper  \cite{treepaper2}. 
For the reader's convenience, we outline the structure of these two papers in Figure~\ref{fig:paperstructure}.
In this first paper, we will explain the basic properties of the double copy for form factors. We use the ${\rm tr}(\phi^2)$ form factors in the Yang-Mills-Scalar (YMS) theory as main examples, which provide prototypes for all the  generalizations in this and the second paper. 

We explain the structure of this paper in more detail as follows.

We first give a brief review of the well-known double copy prescription for tree-level amplitudes in Section~\ref{sec:review4gluon}. 

In Section~\ref{sec:34ptexample}, we consider the double-copy for the form factors of ${\rm tr}(\phi^2)$ in the Yang-Mills-scalar (YMS) theory.
The focus of this section is the three- and four-point form factors. With these concrete and relatively simple examples, various features in the double-copy construction can be discussed in explicit expressions. 

In Section~\ref{sec:nptscalar}, we generalize the discussion to higher points.
This section contains the {major results} of this paper, in the sense that the general foundation, which is applicable to other generalizations, as well as the three central new features (see Figure~\ref{fig:paperstructure}) will be thoroughly discussed. 

In Section~\ref{sec:generalize1}, we take the perspective of regarding form factors as amplitudes with a color-singlet particle and discuss the generalization on the double copy of a class of Higgs amplitudes.

In Section~\ref{sec:generalize2}, we discuss the generalization of the form factor of ${\rm tr}(\phi^2)$ with more than two external scalars. In this case, the propagator matrices and kernels will take new forms, but the salient features maintain.

We conclude with some discussions, outlooks, and the forecast of the topics in the second paper in Section~\ref{sec:discussion}.

We also give three appendices including explicit expressions of the ${\vec v}$ vectors in Appendix~\ref{app:vectors}, further details on the factorizations on physical poles in Appendix~\ref{ap:physical}, and more discussions on the master numerators in Appendix~\ref{ap:nums}.

\section{Review of the amplitudes double copy}
\label{sec:review4gluon}

Before studying form factors, we give a review of the double-copy of amplitudes in this section. We use the four-point amplitudes to present various basic concepts and properties, including the CK duality and double copy, the BCJ relation, the propagator matrix, as well as the KLT formula.
This section will also help to set up some notations.

\subsubsection*{CK duality: from gauge invariance to diffeomorphism invariance}

\begin{figure}[t]
\centering
\includegraphics[clip,scale=0.6]{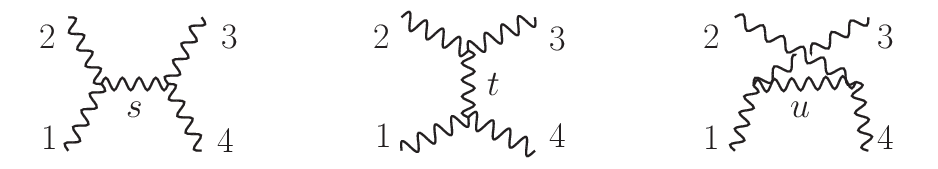}
\caption[a]{Trivalent graphs for the four-gluon tree amplitude.}
\label{fig:tree4CK}
\end{figure}

For scattering amplitudes, it is well-known that the structure of color-kinematics duality and the gauge invariance of gauge-theory amplitudes jointly lead to the diffeomorphism invariance of the gravity amplitudes from the double copy construction \cite{Bern:2019prr}.
Consider the four-gluon amplitude in the CK-dual representation
\begin{equation}\label{eq:CKrep}
		\itbf{A} _{4}(1^{g},2^{g},3^{g},4^{g})=\frac{C_{s}N_{s}}{s}+\frac{C_{t}N_{t}}{t}+\frac{C_{u}N_{u}}{u}\,,
\end{equation}
where
\begin{equation}
C_s = {f}^{a_1 a_2 \text{x}}{f}^{\text{x} a_3 a_4}\,, \qquad C_t = {f}^{a_4 a_1 \text{x}}{f}^{\text{x} a_2 a_3} \,, \qquad C_u = {f}^{a_1 a_3 \text{x}}{f}^{\text{x} a_2 a_4} \,,
\end{equation}
with $f^{abc}={\rm tr}(T^aT^bT^c)-{\rm tr}(T^aT^cT^b)$, and $N_{s,t,u}$ are the color factors and kinematics numerators of the $s, t, u$-channel cubic diagrams respectively; see Figure~\ref{fig:tree4CK}.

The CK duality imposes the condition that the kinematics numerators satisfy the same (dual) Jacobi relation as the color Jacobi relation:
\begin{equation}\label{eq:4ptck}
C_{s}=C_{t}+C_{u} \quad \Rightarrow  \quad N_s=N_t+N_u \,.
\end{equation}
An explicit solution of the numerators can be given as 
\begin{align}\label{eq:4ptcksolution}
\nonumber
& N_{s}=\left(\mathcal{E}_{12} {\bf p}_{12}^{\mu}+2\mathcal{P}_{21} \varepsilon_{2}^{\mu}-2 \mathcal{P}_{12} \varepsilon_{1}^{\mu}\right) \left(\mathcal{E}_{34} {\bf p}_{34,\mu}+2\mathcal{P}_{43} \varepsilon_{4,\mu}-2\mathcal{P}_{34} \varepsilon_{3,\mu}\right)+  s_{12}\left(\mathcal{E}_{13}\mathcal{E}_{24}-\mathcal{E}_{14}\mathcal{E}_{23}\right) ,\\
& N_{t}=\left.N_{s}\right|_{1 \leftrightarrow 3}, \qquad N_{u}=\left.N_{s}\right|_{2 \leftrightarrow 3},
\end{align}
where $\mathcal{E}_{ij}=\varepsilon_i\cdot \varepsilon_j$, $\mathcal{P}_{ij}=p_i\cdot \varepsilon_j$,
and ${\bf p}_{ij}=p_i{-}p_j$, with $p_i$ and $\varepsilon_i$ the momentum and polarization vector of the $i$-th particle. The double copy is obtained by replacing the color factors in \eqref{eq:CKrep} with the corresponding CK numerators
\begin{equation}\label{eq:4ptampGRA}
	\mathcal{M} _{4}(1^{h},2^{h},3^{h},4^{h})=\frac{N_{s}^2}{s}+\frac{N_{t}^2}{t}+\frac{N_{u}^2}{u}\,,
\end{equation}
where $\mathcal{M}_4$ is exactly the four-graviton amplitude. 

Now we explain that, in this example, the diffeomorphism invariance of the double copy is a direct consequence of the dual algebraic structure \eqref{eq:4ptck} between the color factors and the kinematic numerators. 
Starting from the gauge invariance of $\itbf{A}_{4}$ (\emph{i.e.}~the Ward identity), one gets
 \begin{equation}
 		0=\frac{C_{s}N_{s}\big|_{\varepsilon_1\shortrightarrow p_1}}{s}+\frac{C_{t}N_{t}\big|_{\varepsilon_1\shortrightarrow p_1}}{t}+\frac{C_{u}N_{u}\big|_{\varepsilon_1\shortrightarrow p_1}}{u}\,.
\end{equation}
An explicit calculation shows that 
\begin{equation}
\label{eq:tradGT4pt}
    N_{s}\big|_{\varepsilon_1\shortrightarrow p_1}=-s \mathcal{V}_{234}, \quad  N_{t}\big|_{\varepsilon_1\shortrightarrow p_1}=t\mathcal{V}_{234}, \quad  N_{u}\big|_{\varepsilon_1\shortrightarrow p_1}=u\mathcal{V}_{234}\,,
\end{equation}
with $\mathcal{V}_{234}=\left(\varepsilon_{2}\cdot {\bf p}_{34}\right)\mathcal{E}_{34}+\left(\varepsilon_{3}\cdot {\bf p}_{42}\right)\mathcal{E}_{42}+\left(\varepsilon_{4}\cdot {\bf p}_{23}\right)\mathcal{E}_{23}$, and as a result,
\begin{equation}\label{eq:4ptgaugetrans}
   \frac{C_{s}N_{s}\big|_{\varepsilon_1\shortrightarrow p_1}}{s}+\frac{C_{t}N_{t}\big|_{\varepsilon_1\shortrightarrow p_1}}{t}+\frac{C_{u}N_{u}\big|_{\varepsilon_1\shortrightarrow p_1}}{u}=(C_{s}-C_{t}-C_{u})\mathcal{V}_{234}=0 \,.
\end{equation}
The Jacobi relation satisfied by the color factors is required to realize the gauge invariance. 

Therefore, by replacing the color factors with the CK numerators satisfying the dual Jacobi relation, the double-copy result in \eqref{eq:4ptampGRA} has the following property
\begin{equation}\label{eq:4ptdiffeoinv}
\begin{aligned}
   \frac{N_s\big|_{\varepsilon_1\shortrightarrow \xi}N_{s}\big|_{\varepsilon_1\shortrightarrow p_1}}{s}&+\frac{N_t\big|_{\varepsilon_1\shortrightarrow \xi}N_{t}\big|_{\varepsilon_1\shortrightarrow p_1}}{t}+\frac{N_u\big|_{\varepsilon_1\shortrightarrow \xi}N_{u}\big|_{\varepsilon_1\shortrightarrow p_1}}{u}\\
	&=(N_s\big|_{\varepsilon_1\shortrightarrow \xi}-N_t|_{\varepsilon_1\shortrightarrow \xi}-N_u|_{\varepsilon_1\shortrightarrow \xi})\mathcal{V}_{234}=0,
\end{aligned}
\end{equation}
due to the fact that one can replace any polarization vector $\varepsilon_i$ in \eqref{eq:4ptcksolution} with a transverse vector $\xi$ satisfying $\xi\cdot p_i=0$ without sabotaging the dual relation
\begin{equation}
    N_s|_{\varepsilon_1\shortrightarrow \xi}=N_t|_{\varepsilon_1\shortrightarrow \xi}+N_u|_{\varepsilon_1\shortrightarrow \xi}\,.
\end{equation}

The reason why we need \eqref{eq:4ptdiffeoinv} is that it is exactly the equation required by the diffeomorphism invariance of the four-graviton amplitude: the gravity amplitude is invariant under the following transformation
\begin{equation}\label{eq:diffeoshift}
		\varepsilon_i^{\mu\nu}(p_i)\rightarrow \varepsilon_i^{\mu\nu}(p_i)+\alpha \ p_i^{(\mu} \xi_i^{\nu)} \,,
\end{equation}
which is from the linearized diffeomorphism of the asymptotic (weak) graviton field $h_{\mu\nu}$, $\delta h_{\mu\nu} = \partial_\mu \eta_\nu + \partial_\nu \eta_\mu$, with $\partial^{\mu}\eta_\mu=0$  to keep $h^{\mu\nu}$ trace-less.
From the double copy construction \eqref{eq:4ptampGRA}, if we shift  $\varepsilon_{1}^{\mu\nu}=\varepsilon_1^{(\mu}\varepsilon_1^{\nu)}$ according to \eqref{eq:diffeoshift}, we get precisely \eqref{eq:4ptdiffeoinv} that is actually zero. 

In this way, we have shown that double-copy result \eqref{eq:4ptampGRA} is diffeomorphism invariant, and the invariance can be derived from the gauge invariance and the CK duality structure \eqref{eq:4ptck} of the gluon amplitude.

\subsection*{The propagator matrix and BCJ relation}

We provide some alternative representations of the double copy which will be convenient for the discussions on the form factor double copy. 

First, we find the connection between CK numerators and ordered amplitudes and introduce the propagator matrix. 
The full-color amplitude can be decomposed as
\begin{equation}
\itbf{A} _{4}= C_{s} \mathcal{A}_{4}(1,2,3,4) + C_{u} \mathcal{A}_{4}(1,3,2,4)  \,,
\end{equation}
where $\mathcal{A}_{4}(i,j,k,l)$ are color-ordered amplitudes.
Comparing with \eqref{eq:CKrep}, one has 
\begin{equation}
\vec{\mathcal{A}}_4 = \Theta^{\mathcal{A}}_{4} \cdot \vec N_4 \,, 
\end{equation}
where
\begin{equation}
\vec{\mathcal{A}}_4 =
\begin{pmatrix}
\mathcal{A}_{4} (1,2,3,4) \\
\mathcal{A}_{4} (1,3,2,4)
\end{pmatrix} ,
\qquad
 \vec N_4 =
\begin{pmatrix}
N_s \\
N_u
\end{pmatrix} ,
\qquad 
\Theta^{\mathcal{A}}_{4} = 
\begin{pmatrix}
{1\over s} +  {1\over t}& -{1\over t} \\
-{1\over t} &  {1\over t} + {1\over u}
\end{pmatrix} .
\end{equation}
It is important to notice that the propagator matrix \cite{Vaman:2010ez} $\Theta^{\mathcal{A}}_{4}$ has zero determinant and rank one, and thus is not invertible. 
This implies that the above two color-ordered amplitudes are not independent, and they satisfy the so-called BCJ relation \cite{Bern:2008qj}:
\begin{equation}\label{eq:BCJ4pt}
s_{12} \mathcal{A}_{4}(1,2,3,4) + (s_{12}+s_{23}) \mathcal{A}_{4}(1,3,2,4) = 0 \,.
\end{equation}
In the general $n$-point case, we have a $(n-2)!$ by $(n-2)!$ propagator matrix with rank $(n-3)!$. The fact that a propagator matrix which is not full-ranked implies relations between color-ordered amplitudes will play an important role in the form factor discussions later.

\subsection*{The KLT double copy}

The four-graviton amplitude \eqref{eq:4ptampGRA} can be written in the KLT formula as \cite{Kawai:1985xq}
\begin{equation}\label{eq:KLT4pt}
\hskip -3pt
{\cal M}_4 = \mathbf{S}_4^{\cal A}[1,2,3,4|1,2,4,3]\mathcal{A}_{4}(1,2,3,4) \mathcal{A}_{4}(1,2,4,3) \equiv - s_{12} \mathcal{A}_{4}(1,2,3,4) \mathcal{A}_{4}(1,2,4,3).
\end{equation}
We emphasize here that because of the BCJ relations like \eqref{eq:BCJ4pt}, $n$-point amplitudes have a minimal basis with $(n-3)!$ elements. The KLT kernel is a $(n-3)!$ by $(n-3)!$ matrix and the KLT double copy is to use the kernel to ``square" the minimal basis, which can be given schematically as
\begin{equation}
{\cal M}_n = \sum_{\alpha, \beta \in S_{n-3}} \mathcal{A}_{n}[\alpha] \mathbf{S}_n^{\cal A}[\alpha|\beta] \mathcal{A}_{n}[\beta] \,.
\end{equation}

\subsubsection*{Factorization properties}

Last but not least, we point out that the factorization property is a crucial physical requirement for the double-copy quantity.
In our four-point example, beginning with the KLT double copy \eqref{eq:KLT4pt} might be the easiest way to get the factorization of the double copy. 
For gauge-theory amplitudes
\begin{equation}
\begin{aligned}
    \lim_{s_{12}\rightarrow 0} s_{12} \times \mathcal{A}_4(1^{g},2^{g},3^{g},4^{g})=\mathcal{A}_3(1^{g},2^{g},-P)\, {\scriptstyle  \circ} \,\mathcal{A}_3(P,3^{g},4^{g}), \\
    \lim_{s_{12}\rightarrow 0} s_{12} \times \mathcal{A}_4(1^{g},2^{g},4^{g},3^{g})=\mathcal{A}_3(1^{g},2^{g},-P)\, {\scriptstyle  \circ} \,\mathcal{A}_3(P,4^{g},3^{g}),
\end{aligned}
\end{equation}
where $P=p_1{+}p_2$, and correspondingly
\begin{equation}\label{eq:M4fact}
\begin{aligned}
    \lim_{s_{12}\rightarrow 0} s_{12}\times {\cal M}_4 (1^{h},2^{h},3^{h},4^{h}) & = -\lim_{s_{12}\rightarrow 0} s_{12}^2\times \mathcal{A}_{4}(1^{g},2^{g},3^{g},4^{g}) \mathcal{A}_{4}(1^{g},2^{g},4^{g},3^{g})\\
    & =\mathcal{A}_3(1^{g},2^{g},-P)^2 \, {\scriptstyle  \circ} \, \mathcal{A}_3(P,3^{g},4^{g})^2 \\
    & = \mathcal{M}_3(1^{h},2^{h},-P)\, {\scriptstyle  \circ} \,\mathcal{M}_3(P,3^{h},4^{h})\,,
\end{aligned}
\end{equation}
where we have used the three-point gravity amplitude $\mathcal{M}_3=\mathcal{A}_3^2$, and the ``$\circ$" is the helicity summation for the internal $P$-leg.\footnote{The helicity sum relations for $(\varepsilon^g)^{\mu} $ and $(\varepsilon^h)^{\mu\nu}=(\varepsilon^g)^{(\mu}(\varepsilon^g)^{\nu)}$ are:
\begin{equation}
(\varepsilon^g)_P^\mu {\scriptstyle \circ} (\varepsilon^g)_P^{* \nu} \equiv \sum_{\rm hel.} (\varepsilon^g)_P^\mu (\varepsilon^g)_P^{* \nu} = \eta^{\mu\nu} \,, \quad  (\varepsilon^h)_P^{\mu\nu} {\scriptstyle \circ} (\varepsilon^h)_P^{*\rho\sigma} \equiv  
\sum_{\rm hel.} (\varepsilon^h)_P^{\mu\nu} (\varepsilon^h)_P^{* \rho\sigma} = {1\over2} (\eta^{\mu\rho}\eta^{\nu\sigma} + \eta^{\mu\sigma}\eta^{\nu\rho}) \,.
\end{equation}
The tensor product of gluon polarization vectors also contains the antisymmetric and trace parts, which are identified with an antisymmetric tensor field and the dilaton. They do not contribute to the tree-level examples. }
The message conveyed here is that the double copy factorization requires the factorization of gauge-theory amplitudes as well as the special property of the kernel.\footnote{In this example, the kernel itself is just $s_{12}$ and goes to zero 
 in the $s_{12}{\rightarrow}0$ limit.} We will see this is also true for the form factor double copy later.

\section{Warm-up: the three- and four-point form factors}\label{sec:34ptexample}

In this section, we consider  some simple examples for the double-copy of form factors in the Yang-Mills-Scalar (YMS) theory with the Lagrangian given as
\begin{equation}\label{eq:LagrangianYMS}
{\cal L}^{{\rm YMS}} = -{1\over4}{\rm tr}(F_{\mu\nu}F^{\mu\nu}) +  {1\over2}{\rm tr}(D^\mu \phi D_\mu \phi) \,. 
\end{equation}
The gauge field $A_\mu = A_\mu^a T^a$ and the scalar $\phi = \phi^a T^a$ are both in the adjoint representation, where $T^a$ are the generators of gauge group satisfying $[T^a, T^b] = i \sqrt{2} f^{abc} T^c$.
The covariant derivative acts as
$D_\mu \  \cdot = \partial_\mu \cdot + i g_{\rm \scriptscriptstyle YM} [A_\mu, \cdot \ ]$, and $[D_\mu, D_\nu] \ \cdot = i g_{\rm \scriptscriptstyle YM} [F_{\mu\nu}, \cdot \ ]$.

We define the $n$-point form factor of the operator ${\rm tr}(\phi^2)$, which will be our target in Section~\ref{sec:34ptexample} and Section~\ref{sec:nptscalar}:
\begin{equation}\label{eq:phi2ffdef}
\itbf{F}_n(1^{\phi},2^{\phi},3^{g}, \ldots, n^g) =  \int d^{D} x \, e^{-i q \cdot x}\langle \phi(p_1) \, \phi(p_2) \, g(p_3) \ldots g(p_n) |{\rm tr}(\phi^2)(x)| 0\rangle \,,
\end{equation} 
where we have two external scalars, with momenta always labeled as $p_{1,2}$. As an example, the two-point minimal form factor is proportional to a delta function in the color space:
\begin{equation}
\itbf{F}_2(1^{\phi},2^{\phi}) = (2\pi)^4 \delta^{(4)}(q - p_1 - p_2) \delta^{a_1 a_2}  \,,
\end{equation}  
which has a trivial kinematic part equal to one, and one can make a double copy directly.\footnote{In the following, we will omit the momentum delta function for simplicity. We will also ignore the gauge and gravitational couplings in this paper.} 
Thus, the first interesting case is the three-point case, with two scalars and one gluon, and we will also discuss the four-point double copy. In these two simple examples, most of the interesting features of the form factor double copy will be exposed.

\subsection{The three-point example}\label{ssec:3pttrphi2}

\begin{figure}[t]
    \centering
 \includegraphics[height=0.13\linewidth]{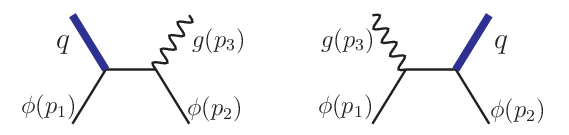}
    \caption{Feynman diagrams for the three-point form factor in gauge theory. The thick blue line represents operator insertion. The black straight line and wiggle line are (light) scalars and gluons respectively.  }
    \label{fig:F3tree}
\end{figure}

We first consider the three-point form factor.
At the tree level, there are two cubic Feynman diagrams $\Gamma_{1,2}$, as shown in Figure \ref{fig:F3tree}. The form factor can be expressed as 
\begin{equation}\label{eq:3pttreefullcolor}
	\itbf{F} _{3}(1^{\phi},2^{\phi},3^{g})=\frac{C_1 N_1}{s_{23}}+\frac{C_2 N_2}{s_{13}}\,,
\end{equation}
where the color factors are
\begin{equation}
\label{eq:3ptcolor}
C_1 = C_2 = f^{a_1 a_2 a_3}  \,,
\end{equation}
and kinematic numerators from Feynman diagrams are 
\begin{equation}
N_1^{\rm \scriptscriptstyle Feyn} = - 2\varepsilon_3 \cdot p_2 \,, \qquad N_2^{\rm \scriptscriptstyle Feyn} =  2\varepsilon_3 \cdot p_1 \,.
\end{equation}
One can also obtain the color-ordered form factor (associated with the color factor ${\rm tr}(T^{a_1} T^{a_3} T^{a_2})$):
\begin{equation}
\label{eq:F3colorordered}
{\cal F}_3(1^{\phi},3^{g},2^{\phi}) = -\frac{2\varepsilon_3 \cdot p_2}{s_{23}}+\frac{2\varepsilon_3 \cdot p_1}{s_{13}} \  
\xLongrightarrow[\mbox{{\tiny $\varepsilon_3^+$}}]{\mbox{{\tiny 4-dim}}} \
{\langle 2 1 \rangle \over \langle 13 \rangle \langle 32 \rangle} = {\cal F}_3(1^{\phi},3^+,2^{\phi}) \,,
\end{equation} 
where in the second equation we express the result in the spinor helicity form in four dimensions.

Let us consider the double copy. We note first that a well-defined quantity in gravity should preserve  the diffeomorphism invariance, in other words, it should be invariant under a transformation of graviton polarization tensor: $\varepsilon_{3}^{\mu\nu} \rightarrow \varepsilon_{3}^{\mu\nu} + p_3^{(\mu} \xi^{\nu)}$ as in \eqref{eq:diffeoshift}. 
However, a naive double-copy of \eqref{eq:3pttreefullcolor} gives
\begin{equation}\label{eq:naiveG3}
{\cal G}_3^{\rm naive} = \frac{(\varepsilon_3 \cdot p_2)^2}{s_{23}}+\frac{(\varepsilon_3 \cdot p_1)^2}{s_{13}} 
= \frac{\varepsilon_3^{\mu\nu}  p_{2\mu} p_{2\nu}}{s_{23}}+\frac{\varepsilon_3^{\mu\nu}  p_{1\mu} p_{1\nu}}{s_{13}} \,,
\end{equation}
which obviously breaks the diffeomorphism invariance. 
To understand this point in more detail, we can see that a gauge transformation in \eqref{eq:3pttreefullcolor} gives
\begin{equation}
\itbf{F} _{3}|_{\varepsilon_3 \rightarrow p_3} =     C_1 \frac{-2 p_3\cdot p_2 }{s_{23}} + C_2 \frac{2 p_3\cdot p_1 }{s_{13}} = -(C_1-C_2)=0\,,
\end{equation}
which is consequence of the color relation \eqref{eq:3ptcolor}, similar to the situation in \eqref{eq:4ptgaugetrans}. However, the numerators from Feynman diagrams do not follow a similar dual relation. 

To solve the problem of diffeomorphism invariance, one can follow the previous four-point amplitude example and impose the color-kinematics duality condition for the form factor.
In particular, with the color relation \eqref{eq:3ptcolor}, we try to require that
\begin{equation}\label{eq:CK3ptexample}
C_1 = C_2  \quad \Rightarrow \quad N_1^{\rm \scriptscriptstyle CK} = N_2^{\rm \scriptscriptstyle CK}  \,,
\end{equation}
where we add a superscript `CK' in the numerators to distinguish the numerators of Feynman diagrams.
Given this requirement, the form factor can be written as
\begin{equation}\label{eq:3pttreefullcolor-CKansatz}
	\itbf{F} _{3}(1^{\phi},2^{\phi},3^{g})=\Big(\frac{1}{s_{23}}+\frac{1}{s_{13}} \Big)C_1 N_1^{\rm \scriptscriptstyle CK} = C_1 \mathcal{F} _{3}(1^{\phi},3^{g},2^{\phi}) \,,
\end{equation}
where in the second equation we apply the color decomposition and  $\mathcal{F} _{3}$ is the color-ordered three-point form factor given in \eqref{eq:F3colorordered}.
Thus one finds the CK-dual numerator solution as: 
\begin{equation}\label{eq:3ptnumsols}
	N_{1}^{\rm \scriptscriptstyle CK}=N_{2}^{\rm \scriptscriptstyle CK}=\frac{s_{13}s_{23}}{s_{13}+s_{23}}\mathcal{F} _{3}(1^{\phi},3^{g},2^{\phi})=-\frac{2 f_3^{\mu\nu}p_{1,\mu}p_{2,\nu}}{s_{12}-q^2}\,,
\end{equation}
where $f^{\mu\nu}_{i}{=} \varepsilon_i^{\mu}p_i^{\nu}{-}\varepsilon_i^{\nu}p_i^{\mu}$. 
We stress that the numerators are \emph{uniquely} determined by the requirement of CK duality and are also manifestly gauge invariant.

When applying \eqref{eq:3ptnumsols} to the double copy for \eqref{eq:3pttreefullcolor}, we obtain
\begin{equation}\label{eq:G3fromCK}
	\mathcal{G}_3= \frac{(N_{1}^{\rm \scriptscriptstyle CK})^2}{s_{23}}+\frac{(N_{2}^{\rm \scriptscriptstyle CK})^2}{s_{13}}  =\frac{s_{13}s_{23}}{s_{13}+s_{23}} \Big(\mathcal{F} _{3}(1^{\phi},3^{g},2^{\phi}) \Big)^2 \,.
\end{equation}
The gauge invariance of the numerators immediately implies the diffeomorphism invariance of $\mathcal{G}_3$ under the transformation $\varepsilon_{3}^{\mu\nu} \rightarrow \varepsilon_{3}^{\mu\nu} + p_3^{(\mu} \xi^{\nu)}$.

The alert reader may find it disturbing that the numerator solutions \eqref{eq:3ptnumsols} have a pole $s_{13}+s_{23}$: in the gauge-theory form factor \eqref{eq:3pttreefullcolor}, this pole is not an actual pole but must cancel in the summation; however, after double copy of the numerators, this ``spurious" pole will not disappear in ${\cal G}_3$ in \eqref{eq:G3fromCK}. 
It turns out that this is a crucial point in the double-copy story of the form factors.

One can first note that the $s_{13}{+}s_{23}=  {-}( s_{12} {-} q^2)$ is a simple pole, and it looks like a (massive) Feynman propagator with mass square $q^2$. Moreover, the residue on this pole can be evaluated as 
\begin{equation}\label{eq:3ptspuriousfactorize0}
	\mathrm{Res}_{s_{12}=q^2}\left[\mathcal{G}_3 \right]=(2\varepsilon_3\cdot q)^2 \,, 
\end{equation}
which can be nicely rewritten as
\begin{equation}\label{eq:3ptspuriousfactorize}
	\mathrm{Res}_{s_{12}=q^2}\left[\mathcal{G}_3 \right]= \big(\mathcal{F} _{2}(1^{\phi},2^{\phi})\big)^2\times (\mathcal{A} _{3}(\QQ_2^{S}, 3^g, -q^{S}))^2 \,.
\end{equation}
Here $\mathcal{F} _{2}(1^{\phi},2^{\phi})$ is just $1$ as the minimal form factor of $\operatorname{tr}(\phi^2)$, and 
\begin{equation}\label{eq:A3exp}
\mathcal{A} _{3}(\QQ_2^{S}, 3^g, -q^{S})=2 \varepsilon_3\cdot q \,, \quad \text{with }\QQ_2=p_1+p_2 \,,
\end{equation}
is the three-point planar amplitude of a gluon and one pair of massive scalar particles $S$ with mass $m^2=q^2=\QQ_2^2$, see \emph{e.g.}~\cite{Badger:2005zh}.
In this way, \eqref{eq:3ptspuriousfactorize} can be interpreted as a factorization formula
\begin{equation}\label{eq:3pttreeGRA2}
	\mathrm{Res}_{s_{12}=q^2}\left[\mathcal{G}_3 \right]=\mathcal{G}_2 (1^\phi,2^\phi) \  \mathcal{M}_{3}(\QQ_2^{S}, -q^{S}, 3^h) \,,
\end{equation} 
where $\mathcal{G}_2 =\big(\mathcal{F} _{2}\big)^2$ is the double copy of the minimal form factor, and $\mathcal{M}_{3}=(\mathcal{A} _{3})^2$  is the double copy of the three-point amplitude.  

\begin{figure}[t]
    \centering
 \includegraphics[height=0.15\linewidth]{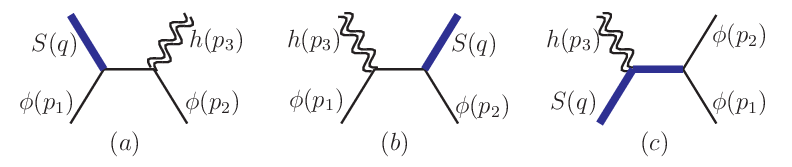}
    \caption{Feynman diagrams for the double copy (in gravity theory) of the three-point form factor. The blue double line in this case is the massive scalar with mass $m^2=q^2$. The black straight line is still the (light) scalar while  we doubled the spring line to represent gravitons.}
    \label{fig:F3treeGRA}
\end{figure}

Clearly, \eqref{eq:3pttreeGRA2} represents the factorization of a new Feynman diagram $ \Gamma_c$ (the third one) in Figure \ref{fig:F3treeGRA}. Furthermore, one can check that \eqref{eq:3ptspuriousfactorize} also gives a consistent factorization on the $s_{13}$ and $s_{23}$ poles (which also appear in the gauge form factor):
\begin{align}\label{eq:3ptfactorizationphysical}
	\mathrm{Res}_{s_{23}=0}\left[\mathcal{G}_3 \right]=\mathcal{G}_2 (1^\phi,(p_2{+}p_3)^\phi) \  \mathcal{M}_{3}(-(p_2{+}p_3)^\phi, 2^\phi, 3^h) \,, \\
	\mathrm{Res}_{s_{13}=0}\left[\mathcal{G}_3 \right]=\mathcal{G}_2 (2^\phi,(p_1{+}p_3)^\phi) \  \mathcal{M}_{3}(-(p_1{+}p_3)^\phi, 1^\phi, 3^h) \,,
\end{align} 
and they correspond to $\Gamma_a$ and $\Gamma_b$ respectively. Thus we conclude that the double copy \eqref{eq:G3fromCK} is physically meaningful in the sense that it satisfies the diffeomorphism invariance and the unitarity requirement|having the desired factorizations on all poles. 

For later purposes, we can rewrite \eqref{eq:G3fromCK} in an alternative form as:
\begin{equation}\label{eq:G3asKLT}
\mathcal{G}_3 = \mathcal{F} _{3}(1^{\phi},3^{g},2^{\phi})\, {\bf S}^{{\cal F}}_{3} \,  \mathcal{F} _{3}(1^{\phi},3^{g},2^{\phi}) \,, \quad \text{ with } \ \  {\bf S}^{{\cal F}}_{3} = \frac{s_{13}s_{23}}{s_{13}+s_{23}} \,.
\end{equation}
This is the KLT form as ``squared" colored-ordered form factors to get the gravitational double copy. 
Another nice property is that by taking the ``square root'' of the double copy factorization \eqref{eq:3ptspuriousfactorize},  one can get a relation for the gauge-theory form factor:
\begin{equation}
s_{13}\mathcal{F}_3(1^{\phi},3^{g},2^{\phi})  \big|_{s_{12}=q^2}= \mathcal{F}_2(1^{\phi},2^{\phi}) \  \mathcal{A} _{3}(\QQ_2^{S}, 3^g, -q^{S}) \,.
\end{equation}
Because of the factorized result on the RHS, such a new relation, which is different from the ordinary physical factorizations obtained by taking residues on propagators, will be referred to as ``hidden factorization relation'' in the story of form factor double copy.
In this three-point example, one may wonder that these properties could be accidental, given the particular simplicity of the three-point results.
Remarkably, as we will see shortly, these nice properties apply to more non-trivial higher-point cases as well.

Finally, it is worth mentioning that $\mathcal{G}_3$ actually represents a four-point gravitational amplitude. 
By checking that $\mathcal{G}_3 $ exactly matches the expression from the Feynman diagrams in Figure \ref{fig:F3treeGRA}, we see that $\mathcal{G}_3$ can be understood as a four-point tree-level amplitude
$$\mathcal{G}_3 = {\cal M}_4(1^\phi, 2^\phi, 3^h,q^S)$$
in the gravitational theory involving massless scalars $\phi$ and a new massive scalar $S$, and the operator in $\itbf{F}_3$ is interpreted as a three-scalar vertex $S\phi^2$ in $\mathcal{G}_3$. 
This fact holds for higher points, and we will come back to this later in Section~\ref{sec:nptscalar} and give the Lagrangian in Section~\ref{sec:generalize1}.

\subsection{The four-point example}\label{ssec:4pteg}

We consider next the four-point form factor, which can capture most of the characteristics and help to clarify the generalization. 

\begin{figure}[t]
    \centering
 \includegraphics[width=0.7\linewidth]{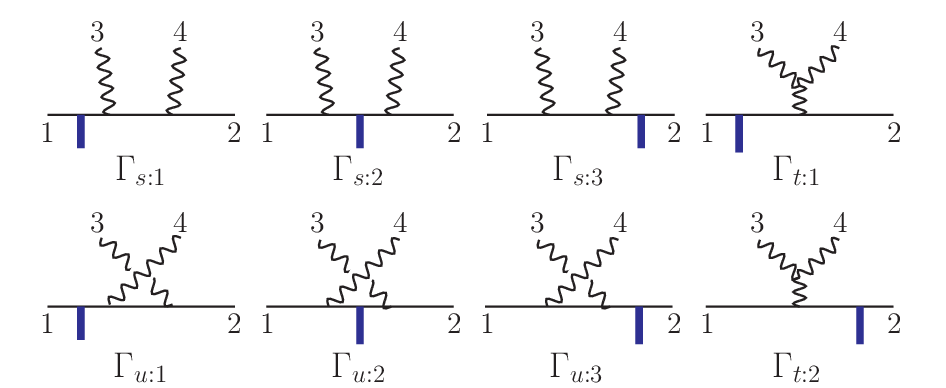}
    \caption{Cubic diagrams for the double copy of the four-point form factor of ${\rm tr}(\phi^2)$.}
    \label{fig:F4tree}
\end{figure}

\subsubsection*{The color-kinematics duality for the four-point form factor}

The four-point full-color form factor of $\operatorname{tr}(\phi^2)$ contains eight cubic diagrams, which are given in Figure~\ref{fig:F4tree}. 
The form factor can be expanded as 
\begin{equation}\label{eq:4pttreefullcolor}
\itbf{F} _{4}(1^{\phi},2^{\phi},3^{g},4^{g})= \sum_{\Gamma_i} \frac{C_{i}N_{i}}{D(\Gamma_i)}\,,
\end{equation}
where $C_i$ and $N_i$ are color and kinematic numerators, respectively, and $D(\Gamma_i)$ denote the propagators. For example, for the first two diagrams 
\begin{equation}
D(\Gamma_{s:1}) = s_{24} s_{234} \,, \qquad D(\Gamma_{s:2}) = s_{13} s_{24} \,.
\end{equation}
These diagrams can be combined into three groups, each of which includes graphs with the same color factor. For instance, the first three graphs $\Gamma_{s:1,2,3}$ share the same color factor which is the same with a $s$-channel four-point amplitude:
\begin{equation}\label{eq:simple4ptcolor}
C_{s:1}=C_{s:2}=C_{s:3} = f^{a_1a_3\text{x}}f^{\text{x} a_4a_2} \,,
\end{equation}
because the operator insertion (the blue arrow) is a $\delta$ function in color space.
We impose the same relation for the three numerators as
\begin{equation}\label{eq:simple4ptnum}
    N_{s:1}=N_{s:2}=N_{s:3} \,.
\end{equation}
Similar relations are also imposed for the other $t$- and $u$-channel diagrams. Because of these numerator relations, it is meaningful to define a color factor and a numerator for a group, such as  $C_{\{s\}}=C_{s:i}$ and $N_{\{s\}}=N_{s:i}$. 
Note that the three groups $s,t,u$ actually correspond to three $4$-point cubic diagrams without the operator insertion. 
We stress that for higher-point form factors of $\operatorname{tr}(\phi^2)$, a parallel argument shows that we can also group the diagrams according to their color factors and every group corresponds to an $n$-point cubic diagram with no operator insertion. 

The numerators should also satisfy dual Jacobi relations to give a consistent double copy. Here we have 
\begin{equation}\label{eq:simple4ptJacobi}
C_{\{s\}}=C_{\{t\}}+C_{\{u\}} \quad \Rightarrow \quad N_{\{s\}}=N_{\{t\}}+N_{\{u\}} \,,
\end{equation}
and $N_{\{s\}},N_{\{u\}}$ can be selected as ``master numerators''.
The important point here is that by imposing the  two types of numerator relations, the operator-induced relation like \eqref{eq:simple4ptnum} and the Jacobi relation \eqref{eq:simple4ptJacobi}, we can express the numerators of the eight diagrams in terms of the two master numerators.

Then the form factor can be written as
\begin{equation}
\label{eq:4pttreefullcolor-CK}
\itbf{F} _{4}= {C_{\{s\}} N_{\{s\}} \over P_{\{s\}}} +{C_{\{t\}} N_{\{t\}} \over P_{\{t\}}}+{C_{\{u\}} N_{\{u\}} \over P_{\{u\}}} \,,
\end{equation}
with $P_{\{s\}}^{-1}\equiv \sum_{i=1}^{3}(D(\Gamma_{s:i}))^{-1}$, and $P_{{\{t\}},{\{u\}}}$ are defined likewise. 
$C_{\{s\}}$ and $C_{\{u\}}$ form the four-point Del Duca-Dixon-Maltoni (DDM) color basis \cite{DelDuca:1999rs}, which fixes the position of $1^{\phi},2^{\phi}$ and permute $3^g,4^g$ as indicated in Figure \ref{fig:F4tree}, and the form factor can be also expanded as
\begin{equation}
\label{eq:4pttreefullcolor-DDM}
\itbf{F} _{4}= C_{\{s\}} \mathcal{F}_{4} (1, 3,4,2) + C_{\{u\}} \mathcal{F}_{4} (1,4,3,2) \,,
\end{equation}
where $\mathcal{F}_{4}$ are  color-ordered form factors and $\{\mathcal{F}_{4} (1, 3,4,2),\mathcal{F}_{4} (1, 4,3,2)\}$ also form a basis for color-ordered form factors. 

By comparing \eqref{eq:4pttreefullcolor-CK} and \eqref{eq:4pttreefullcolor-DDM}, we relate the master numerators with the (basis) color-ordered form factors as
\begin{equation}\label{eq:4ptFNeq0}
\vec{\mathcal{F}}_4 = \Theta^{\mathcal{F}}_{4} \cdot \vec N_4 \,, 
\end{equation}
where
\begin{equation}
\vec{\mathcal{F}}_4 =
\begin{pmatrix}
\mathcal{F}_{4} (1, 3,4, 2) \\
\mathcal{F}_{4} (1, 4,3, 2)
\end{pmatrix} ,
\qquad
 \vec N_4 =
\begin{pmatrix}\label{eq:vFandvN}
N_{\{s\}} \\
N_{\{u\}}
\end{pmatrix} ,
\end{equation}
and in particular, $\Theta^{\mathcal{F}}_{4}$ is a matrix of propagators as
\begin{align}\label{eq:4ptThetaMatrix}
& \Theta^{\mathcal{F}}_{4} = 
\begin{pmatrix}
{1\over P_{\{s\}}} +  {1\over P_{\{t\}}}& - {1\over P_{\{t\}}} \\
-{1\over P_{\{t\}}} &  {1\over P_{\{t\}}} + {1\over P_{\{u\}}}
\end{pmatrix}\\
&= \begin{pmatrix}
{1\over s_{24}s_{234}} + {1\over s_{24}s_{13}} + {1\over s_{13}s_{134}} +  \frac{1}{s_{34}}\Big(\frac{1}{s_{134}}+\frac{1}{s_{234}}\Big)& - \frac{1}{s_{34}}\Big(\frac{1}{s_{134}}+\frac{1}{s_{234}}\Big)\\
- \frac{1}{s_{34}}\Big(\frac{1}{s_{134}}+\frac{1}{s_{234}}\Big) &  \frac{1}{s_{34}}\Big(\frac{1}{s_{134}}+\frac{1}{s_{234}}\Big) + {1\over s_{23}s_{234}} + {1\over s_{23}s_{14}} + {1\over s_{14}s_{134}}
\end{pmatrix} .\nonumber
\end{align}
It is convenient to rewrite \eqref{eq:4ptFNeq0} in a component form as
\begin{equation}\label{eq:4ptFNeqelement}
{\mathcal{F}}_4[\alpha] = \sum_{\beta\in S_2}  \Theta^{\mathcal{F}}_{4}[\alpha|\beta] N_4[\beta] \,, 
\end{equation}
where $\alpha, \beta$ label the vector/matrix components in \eqref{eq:vFandvN}-\eqref{eq:4ptThetaMatrix} and are related to the $S_2$ permutation of gluon legs.

The propagator matrix $\Theta^{\mathcal{F}}_{4}$ has many intriguing properties, as will be discussed in more detail later. Here we just mention that $\Theta^{\mathcal{F}}_{4}$ has full rank. This is different from the amplitude case. For $n$-point amplitudes,  a similar $(n-2)!\times(n-2)!$ matrix $\Theta_n$ can be defined, which has rank $(n-3)!$. This implies that there are $(n-3)(n-3)!$ BCJ relations among the planar amplitudes. But for form factors, the propagator matrix has full rank so that the numerators can be solved as 
\begin{equation}\label{eq:4ptnumsols}
	N_4[\alpha]=\sum_{\beta\in S_2} 
	{\bf S}^{{\cal F}}_{4}[\alpha|\beta] \, \mathcal{F}_{4} [\beta] , \ \text{ with }\ {\bf S}^{{\cal F}}_{4}\equiv\left(\Theta^{\mathcal{F}}_{4}\right)^{-1} .
\end{equation}

\subsubsection*{BCJ and KLT Double-copy}

From \eqref{eq:4pttreefullcolor-CK},  the double copy (the BCJ form) is
\begin{equation}\label{eq:4pttreeGRA1}
	\mathcal{G} _{4}={N_{\{s\}}^2 \over P_{s}}+{N_{\{t\}}^2 \over P_{t}}+{N_{\{u\}}^2 \over P_{u}} = 
	\sum_{\alpha,\beta\in S_2}  N_4[\alpha] \Theta^{\mathcal{F}}_{4}[\alpha|\beta] N_4[\beta] \,.
\end{equation}
Using \eqref{eq:4ptnumsols}, the above result can be rewritten as
\begin{equation}\label{eq:4pttreeGRA2}
	\mathcal{G} _{4}= \sum_{\alpha,\beta\in S_2}\hskip -4pt \mathcal{F}_{4} [\alpha] {\bf S}^{\mathcal{F}}_{4}[\alpha|\beta] \mathcal{F}_{4} [\beta] \,,
\end{equation}
which is similar to the KLT form for amplitudes, 
and ${\bf S}^{{\cal F}}_{4}$ serves as a (four-point) KLT kernel.
It is also manifestly diffeomorphism invariant.

\subsubsection*{Comment on the propagator matrix and the KLT kernel}

Now we discuss the property of the propagator matrix $\Theta^{\mathcal{F}}_{4}$.
From \eqref{eq:4ptThetaMatrix}, we obtain the determinant in a simple factorized form as
\begin{equation}
\det (\Theta^{\mathcal{F}}_{4}) = {(q^2 - s_{12})(q^2 - s_{123})(q^2 - s_{124}) \over s_{13} s_{14} s_{23} s_{24} s_{34} s_{134} s_{234} }\,,
\end{equation}
which is non-zero and thus $\Theta^{\mathcal{F}}_{4}$ is invertible. 
Besides, the numerators of $\det (\Theta^{\mathcal{F}}_{4})$ is a product of the ``spurious'' poles $s_{12}{-}q^2$,  $s_{123}{-}q^2$ and  $s_{124}{-}q^2$, while the denominator is a product of all the ``physical'' poles, which are the propagators of $\itbf{F} _{4}$. 

Next, we consider the KLT kernel $\mathbf{S}_4^{\cal F}$, which is the inverse of the propagator matrix. 
The zeros of $\det (\Theta^{\mathcal{F}}_{4})$ provide information about poles of ${\bf S}^{\mathcal{F}}_{4}$.
Explicitly, from \eqref{eq:4ptThetaMatrix} we have
\begin{align}\label{eq:KLTkernel4pt}
{\bf S}_{4}^{\cal F} = 
\begin{pmatrix}
\frac{s_{13}s_{24}s_{34}(s_{134}s_{234}+s_{14}s_{134}+s_{23}s_{234})}{(s_{12}-q^2)(s_{123}-q^2)(s_{124}-q^2)}- \Delta_{t} & \Delta_{t}\\
 \Delta_{t}  &  \frac{s_{14}s_{23}s_{34}(s_{134}s_{234}+s_{24}s_{234}+s_{13}s_{134})}{(s_{12}-q^2)(s_{123}-q^2)(s_{124}-q^2)}- \Delta_{t}
\end{pmatrix}\,,
\end{align}
where 
\begin{equation}
\Delta_{t}=  \frac{s_{13}s_{14}s_{23}s_{24}}{s_{12}-q^2}\left({1\over   s_{123}-q^2} {+} {1\over s_{124}-q^2}\right) \,.
\end{equation}
It is interesting to note that ${\bf S}^{\mathcal{F}}_{4}$ has only simple poles $s_{12}{-}q^2$,  $s_{123}{-}q^2$ and  $s_{124}{-}q^2$; in particular, none of the propagators like  $s_{13}$ appear in the denominators of elements of  ${\bf S}^{\mathcal{F}}_{4}$.\footnote{This is understandable because ${\bf S}^{\mathcal{F}}_{4}\sim \det(\Theta^{\mathcal{F}}_{4})^{-1}\times$ (Cofactors) and $\det({\Theta}^{\mathcal{F}}_{4})^{-1}$ contain enough power of propagators such as $s_{13}$ to cancel the propagators appearing in the denominators of the cofactors.} 

As a direct consequence of the structure of $\mathbf{S}^{\cal F}_4$,  the numerators $N_4[\alpha]$ given in \eqref{eq:4ptnumsols} have only simple poles $s_{12}{-}q^2$,  $s_{123}{-}q^2$ and  $s_{124}{-}q^2$. 
For example, $N_{\{s\}}$ in \eqref{eq:vFandvN} is
\begin{equation}
	N_{\{s\}}=-\frac{2\left(f_{3}^{\mu\nu}f_{4,\nu\rho}p_{1,\mu}p_{2}^{\rho}\right)}{(s_{12}-q^2)}+ 
	\frac{4\left(f_{3}^{\mu\nu}p_{1,\mu}p_{2,\nu}\right)\left(f_{4}^{\mu\nu}p_{2,\mu}q_{\nu}\right)}{(s_{12}-q^2)(s_{123}-q^2)}
	+\frac{4\left(f_{4}^{\mu\nu}p_{1,\mu}p_{2,\nu}\right)\left(f_{3}^{\mu\nu}p_{1,\mu}q_{\nu}\right)}{(s_{12}-q^2)(s_{124}-q^2)}\,.
	\label{eq:4ptNsSolution}
\end{equation}
We will see in Section~\ref{ssec:FFDCnpt} that the above properties are also true for more general $n$-point cases: there are also no ``physical'' poles (such as $s_{1i_1\cdots i_m}$) appearing in numerators $N_n[\alpha]$, and the  poles in the numerators (given by the zeros of $\det(\Theta^{\mathcal{F}}_{n})$) are always simple poles like $(s_{(12\cdots)}{-}q^2)$, where ``${\dots}$" represents gluon momenta.

\subsubsection*{Factorization}

\begin{figure}[t]
    \centering
 \includegraphics[width=.7\linewidth]{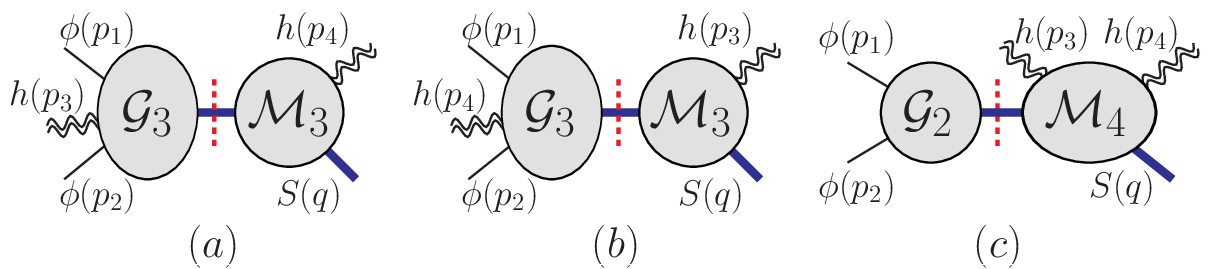}
    \caption{Factorization of $\mathcal{G}_4$ on the three new poles.} 
    \label{fig:4ptfactorizationGRA}
\end{figure}

\begin{figure}[t]
    \centering
 \includegraphics[width=.9\linewidth]{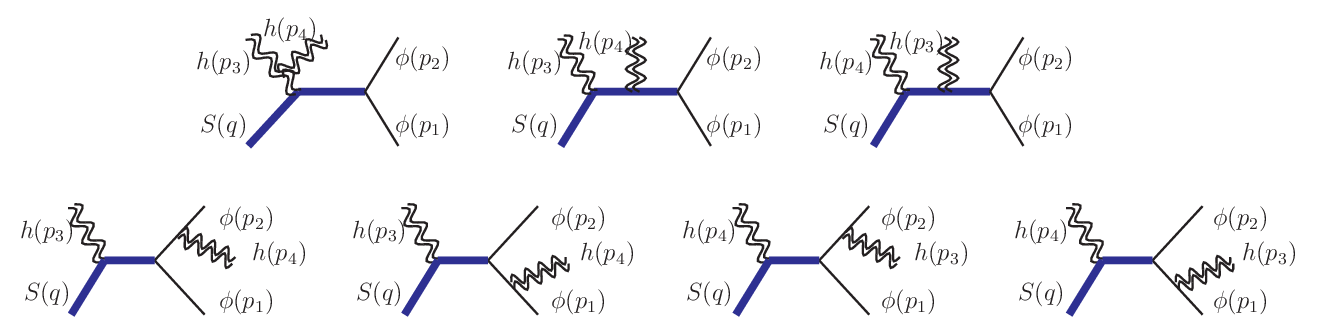}
    \caption{New Feynman diagrams  in $\mathcal{G}_4$ after double copy.} 
    \label{fig:F4treeGRA}
\end{figure}

After performing the double copy, the hidden ``spurious'' poles, looking like massive Feynman propagators, become real (simple) poles of $\mathcal{G}_{4} $. 
The factorizations can be explicitly checked as 
\begin{align}
	\mathrm{Res}_{q^2=s_{12}}\left[\mathcal{G}_4 \right] & = \mathcal{M}_{4}(\QQ_2^{S}, -q^{S}, 3^h,4^h) \times \mathcal{G}_2 (1,2), \nonumber\\
	\mathrm{Res}_{q^2=s_{123}}\left[\mathcal{G}_4 \right] & = \mathcal{M}_{3}(\QQ_3^{S}, -q^{S},4^h) \times \mathcal{G}_3 (1,2,3), 
	\label{eq:4ptfactorizationGRA}
\end{align}
where $\QQ_2=p_1+p_2$, $\QQ_3=p_1+p_2+p_3$ and the $s_{124}-q^2$ factorization is similar; they are shown in Figure~\ref{fig:4ptfactorizationGRA}. These factorization properties mean that new Feynman diagrams, as those listed in Figure~\ref{fig:F4treeGRA}, also contribute to $\mathcal{G}_4 $.

\subsubsection*{Hidden factorization relations}

As in the three-point case, the factorization relations on the spurious poles \eqref{eq:4ptfactorizationGRA} imply hidden relations for the gauge form factor:
\begin{equation}
\label{eq:4ptbcjB}
(\vec{v}_4 \cdot \vec{\mathcal{F}}_{4})\big|_{s_{123}=q^2}
 =\mathcal{F}_{3}(1^{\phi},3^{g},2^{\phi}) \ \mathcal{A}_{3}(\QQ_{3}^{S},4^{g},-q^{S}) \,,
\end{equation}
where the (row) vector $\vec{v}_4=\{s_{42}, s_{42} + s_{43}\}$ and (column) vector $\vec{\mathcal{F}}_4 $ is defined in \eqref{eq:vFandvN}.
Explicitly, this gives
\begin{equation}\label{eq:4ptbcjexpand}
\big[ s_{42} \mathcal{F}_{4}(1,3,4,2) + (s_{42}+s_{43}) \mathcal{F}_{4}(1,4,3,2) \big] \big|_{s_{123}=q^2}
 =\mathcal{F}_{3}(1^{\phi},3^{g},2^{\phi}) \ \mathcal{A}_{3}(\QQ_{3}^{S},4^{g},-q^{S}) \,,
\end{equation}
which is reminiscent of the BCJ relation for four-point amplitudes in \eqref{eq:BCJ4pt}:
\begin{equation}
s_{42} \mathcal{A}_{4}(1,3,4,2) + (s_{42}+s_{43}) \mathcal{A}_{4}(1,4,3,2) = 0 \,.
\end{equation}
Here we emphasize that the RHS of the BCJ-like equation \eqref{eq:4ptbcjexpand} is not zero. Instead, \eqref{eq:4ptbcjB} is a relation connecting form factors with different numbers of external states, which will be a generic feature for the generalized BCJ relations for form factors. Later we will also call relations similar to \eqref{eq:4ptbcjexpand} as hidden factorization relations.

Let us see how the relation \eqref{eq:4ptbcjB} is related to the double copy, especially the factorization property \eqref{eq:4ptfactorizationGRA}. By taking the square of the RHS of  \eqref{eq:4ptbcjB} and dressing a factor $s_{13}s_{23}/(s_{13}+s_{23})$, we reproduce the RHS of \eqref{eq:4ptfactorizationGRA}. As a result, we have
\begin{equation}
\begin{aligned}
	\mathrm{Res}_{s_{123}=q^2}\left[\mathcal{G}_4 \right]&={s_{13}s_{23}\over s_{13}+s_{23}} \big[ s_{42} \mathcal{F}_{4}(1,3,4,2) + (s_{42}+s_{43}) \mathcal{F}_{4}(1,4,3,2) \big]^2 \big|_{s_{123}=q^2}\\
    &=\mathbf{S}^{\cal F}_3 \, \mathbf{S}^{\cal A}_3 \, (\vec{v}_4 \cdot \vec{\mathcal{F}}_{4})^2 \big|_{s_{123}=q^2}\,,
\end{aligned}
\end{equation}
where ${\bf S}^{\mathcal{F}}_{3} = s_{13}s_{23}/(s_{13}{+}s_{23})$ in \eqref{eq:G3asKLT}, and ${\bf S}^{\mathcal{A}}_{3} = 1$ is the KLT kernel for three-point amplitudes.
On the other hand, from \eqref{eq:4pttreeGRA2}, we also have 
\begin{equation}
	\mathrm{Res}_{s_{123}=q^2}\left[\mathcal{G}_4 \right]=
  \vec{\cal F}_4^{\scriptscriptstyle \rm T}\cdot {\rm Res}_{s_{123}=q^2}[{\bf S}^{\mathcal{F}}_{4}] \cdot \vec{\cal F}_{4} \big|_{s_{123}=q^2} \,.
\end{equation}
By comparing the above two equations, we actually see that the following relation should holds
\begin{equation}\label{eq:4ptSfdecomp}
	\mathrm{Res}_{s_{123}=q^2}\left[{\bf S}^{\mathcal{F}}_{4}\right]=\vec{v}^{\rm \scriptscriptstyle T} \cdot ({\bf S}^{\mathcal{F}}_{3}\otimes {\bf S}^{\mathcal{A}}_{3}) \cdot \vec{v}\big|_{s_{123}=q^2}\,,
\end{equation}
where ${\bf S}^{\mathcal{F}}_{3}$ and ${\bf S}^{\mathcal{A}}_{3} $ can be regarded both are $1\times 1$ matrices and the $\vec{v}$ should also be viewed as an $1\times 2$ matrix. Thus, \eqref{eq:4ptSfdecomp} is a matrix decomposition relating the large $2\times 2 $ $\mathbf{S}^{\cal F}_4$ with the small ${\bf S}^{\mathcal{F}}_{3}\otimes{\bf S}^{\mathcal{A}}_{3} $ with a rectangular matrix $\vec{v}$. Such a decomposition will also be a central topic in the next section. 

Synthesizing everything above, the factorization \eqref{eq:4ptfactorizationGRA} is equivalent to the combination of \eqref{eq:4pttreeGRA2}, \eqref{eq:4ptbcjB} and \eqref{eq:4ptSfdecomp} as:
\begin{equation}
\begin{aligned}
\label{eq:4ptFactorization}
\mathrm{Res}_{\QQ_3^2=q^2}\left[\mathcal{G}_{4} \right] 
& = \mathcal{F} _{4} \cdot \mathrm{Res}_{\QQ_3^2=q^2}\left[\mathbf{S}^{\mathcal{F}}_{4} \right] \cdot \mathcal{F} _{4}  \big|_{\QQ_3^2=q^2} =(\vec{v}_4\cdot \mathcal{F}_4) (\mathbf{S}^{\mathcal{F}}_{3} \otimes \mathbf{S}^{\mathcal{A}}_{3}) (\vec{v}_4\cdot \mathcal{F}_4) \big|_{\QQ_3^2=q^2} \\
& =  (\mathcal{F}_3 {\cal A}_{3})  (\mathbf{S}^{\mathcal{F}}_{3} \otimes \mathbf{S}^{\mathcal{A}}_{3})  (\mathcal{F}_3 {\cal A}_{3}) = (\mathcal{F} _{3} \cdot \mathbf{S}^{\mathcal{F}}_{3} \cdot \mathcal{F} _{3}) (\mathcal{A} _{3} \cdot \mathbf{S}^{\mathcal{A}}_{3} \cdot \mathcal{A} _{3}) \\
&= \mathcal{G}_{3} \times  \mathcal{M}_{3}. 
\end{aligned}
\end{equation}
To summarize, when the ``spurious''-type propagators are on-shell, a generalized version of BCJ relations for form factors and a decomposition of ${\bf S}^{\mathcal{F}}$ are uncovered, which lead to the factorization property of the double copy $\mathcal{G}$.

\section{General $n$-point cases in YMS Theory}\label{sec:nptscalar}

In this section, we describe the $n$-point generalization of the double copy of the ${\rm tr}(\phi^2)$ form factors in \eqref{eq:phi2ffdef}. 
We first give the general procedure for constructing the form factor double copy in Section~\ref{ssec:FFDCnpt}. 
A key feature is that there are new ``spurious"-type poles that become physical poles after double copy. 
Therefore, in the remaining subsections, we focus on various intriguing properties associated with these poles. 
In particular, the decomposition properties of the KLT kernels and the propagator matrices are discussed in Section~\ref{ssec:bilinearDecomp}.
Moreover, the hidden relations satisfied by the color-ordered form factors are studied in Section~\ref{ssec:Hiddenfac}. 
Finally, we explain the factorizations of the gravitational form factors on the new ``spurious"-type poles as well as the ``physical"-type poles in Section~\ref{ssec:FFDCnptfac}.

\subsection{The $n$-point form factor double copy}\label{ssec:FFDCnpt}

We generalize the previous constructions for the three- and four-point ${\rm tr}(\phi^2)$ form factors to the $n$-point case. 
We also try to make the presentation in a general form so that it can be applied to form factors of more general operators or external states.

\vspace{3pt}

\textbf{1.} First, one writes down the cubic diagram expansion of the full-color form factor as
\begin{equation}\label{eq:FCFncubic}
    \itbf{F}_n=\sum_{\Gamma_i} \frac{C(\Gamma_i)N(\Gamma_i)}{D(\Gamma_i)} \,,
\end{equation}
where  $\Gamma_i$s are diagrams with cubic interaction vertices plus one operator vertex, such as Figure~\ref{fig:F3tree} and \ref{fig:F4tree}. 
The propagators $(D(\Gamma_i))^{-1}$ and color factors $C(\Gamma_i)$ can be read out from $\Gamma_i$. (Note that the color factors depend on the color factor of the operator in general.)

\vspace{3pt}

\textbf{2.} 
Next, for the kinematic numerators $N(\Gamma_i)$, we impose the color-kinematics duality that every color relation among $C(\Gamma_i)$ has a dual relation satisfied by the corresponding $N(\Gamma_i)$. 
In particular, besides the usual dual Jacobi relations (as in amplitudes), we have the new operator-induced dual relations for form factors like \eqref{eq:CK3ptexample} and \eqref{eq:simple4ptnum} in the previous examples, which are due to that the color factor does not change when moving around the operator $q$-leg.

With these relations, one can pick up a subset of cubic diagrams, denoted by $\Delta$ (called master graphs), whose color factors or kinematic numerators form a basis. 
For example, for the form factors of  ${\rm tr}(\phi^2)$, it is convenient to choose the master graphs as the DDM basis \cite{DelDuca:1999rs} represented by the half-ladder diagrams as
$$
    \begin{aligned}
        \includegraphics[width=0.36\linewidth]{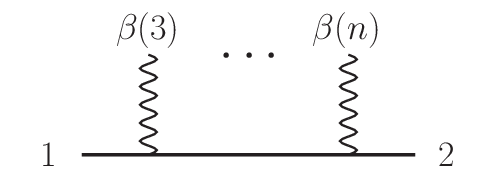}
    \end{aligned}
$$
with $\beta\in S_{n{-}2}$ permuting $\{3^{g},\ldots,n^{g}\}$. As noted above, it does not matter where we insert the operator $q$-leg since the color factor is not changed by the insertion.

\vspace{3pt}
 
\textbf{3.} Furthermore, we write down an alternative representation of the full-color form factor in terms of color-ordered form factors. Since the color-ordered form factors satisfy KK-relations, one can select a set of linearly independent ordered form factors ${\cal F}_{n}[\alpha]$ where $\alpha$ is an ordering of $\{1,\ldots,n\}$ and belongs to a set of ordering ``basis", denoted by $\Pi$. As a result, we have
\begin{equation}\label{eq:FCFnordered}
        \itbf{F}_n=\sum_{\alpha \in \Pi} \widetilde{C}_{n}[\alpha]{\cal F}_{n}[\alpha] \,,
    \end{equation}
where $\widetilde{C}_{n}[\alpha]$ are certain color factors, as the coefficient of ${\cal F}_{n}[\alpha]$ in $\itbf{F}_n$. 
Note that the number of elements in the ordering set $\Pi$ is the same as in the set $\Delta$ of cubic diagrams mentioned above. 

\vspace{3pt}

\textbf{4.}  By matching \eqref{eq:FCFncubic} with \eqref{eq:FCFnordered}, we obtain the relationship between master numerators $N(\Gamma_i)$($i\in\Delta$) and color-ordered form factors ${\cal F}_{n}[\alpha]$($\alpha\in\Pi$).

For the ${\rm tr}(\phi^2)$ form factor, we can pick $\alpha$ to be the ordering $\{1,\beta,2\}$ with $\beta\in S_{n{-}2}$ permuting $\{3^{g},\ldots,n^{g}\}$. 
    This choice makes $\widetilde{C}_{n}[\alpha]$ exactly the same as the color basis $C(\Gamma_{i})$($i\in\Delta$) above, reading 
    \begin{equation}\label{eq:ddmC}
       \widetilde{C}_n[1,\beta,2]= C\bigg( \hskip -3pt \begin{aligned}
        \includegraphics[width=0.28\linewidth]{figure/2scalars.eps}
        \end{aligned}\hskip -6pt\bigg)={f}^{1 \beta(3) \text{x}_1}{f}^{\text{x}_1 \beta(4) \text{x}_2}\cdots {f}^{ \text{x}_{n-3} \beta(n) 2}\,.
    \end{equation}
    And, because of the color-kinematics duality, we define accordingly\footnote{The $N[\beta]$ defined here are equivalent to the definition of ``pre-numerators" in \cite{Brandhuber:2021bsf,Chen:2022nei}.}
    \begin{equation}\label{eq:ddmN}
        N_{n}[1,\beta,2]\equiv N\bigg( \hskip -3pt \begin{aligned}
        \includegraphics[width=0.28\linewidth]{figure/2scalars.eps}
        \end{aligned}\hskip -6pt\bigg)\,.
    \end{equation}
We emphasize that \eqref{eq:ddmC} and \eqref{eq:ddmN} are due to the speciality of the ${\rm tr}(\phi^2)$ operator and the choice of DDM basis; for more general form factors, things can be more complicated, as will be covered in \cite{treepaper2}.\footnote{When dealing with general high-length operators, usually the ``new" color basis $\widetilde{C}_{n}[\alpha],\ \alpha\in \Pi$ and the ``old" color basis $C(\Gamma_j),\ j\in \Delta$ are not the same. Similarly, $N[\alpha]$, the numerator defined for an ordering $\alpha$, and $N(\Gamma)$, the numerator defined for a cubic diagram $\Gamma$ are not supposed to coincide.}

    Expanding all $C(\Gamma_i)$ and $N(\Gamma_i)$ in \eqref{eq:FCFncubic} in terms of $\widetilde{C}_{n}[\beta]$ and $N_{n}[\beta]$ and comparing with \eqref{eq:FCFnordered}, we get 
    \begin{equation}\label{eq:FthetaNn}
        {\cal F}_{n}(1,\beta,2)=\sum_{\gamma\in S_{n-2}}\Theta_n^{\cal F}[1,\beta,2|1,\gamma,2] N_{n}[1,\gamma,2]
    \end{equation}
    with $\Theta_n^{\cal F}$ is the propagator matrix for form factors, whose matrix elements are sum of $(D(\Gamma_i))^{-1}$ for some $\Gamma_i$, such as \eqref{eq:4ptThetaMatrix} in the previous four-point example. 
    We will discuss the propagator matrix in more detail in Section~\ref{ssec:bilinearDecomp}.
    
    \vspace{3pt}

  \textbf{5.} Finally, the double copy of the form factor $\itbf{F}_n$ can be defined by replacing $C(\Gamma_i)$ with $N(\Gamma_i)$ in \eqref{eq:FCFncubic} as 
  \begin{equation}
      \mathcal{G}_n\equiv \sum_{\Gamma_i} \frac{N(\Gamma_i)N(\Gamma_i)}{D(\Gamma_i)} \,.
  \end{equation}
One can check that this definition is equivalent to the following form of $\mathcal{G}_{n}$ by substituting $\widetilde{C}_{n}[\alpha]$ with $N_{n}[\alpha]$ in \eqref{eq:FCFnordered}:
  \begin{align}
      {\cal G}_{n}=\sum_{\alpha \in \Pi} N_{n}[\alpha] {\cal F}_{n}[\alpha] &= \sum_{\alpha_{1,2}\in \Pi}N _{n}[\alpha_1] \Theta_{n}^{\mathcal{F}}[\alpha_1|\alpha_2] N_{n}[\alpha_2] \label{eq:genericBCJKLT}\\
      &= \sum_{\alpha_{1,2}\in \Pi}\mathcal{F} _{n}[\alpha_1] \mathbf{S}_{n}^{\mathcal{F}}[\alpha_1|\alpha_2] \mathcal{F} _{n}[\alpha_2]\,,\label{eq:genericKLT}
    \end{align}
   which is because the relations between $N(\Gamma_i)$ and $N_n[\alpha]$ and between $C(\Gamma_i)$ and $\widetilde{C}_n[\alpha]$ are exactly the same set of relations. 
   Here the KLT kernel $\mathbf{S}_{n}^{\mathcal{F}} = (\Theta_{n}^{\mathcal{F}})^{-1}$ is the inverse of the propagator matrix $\Theta_{n}^{\mathcal{F}}$.

\subsubsection*{Outline of the properties related to the form factor double copy}

Below we present some important properties related to the double-copy quantities $\mathcal{G}_n$. 

\begin{itemize}[itemsep=3pt]
    \item The bilinear form  \eqref{eq:genericKLT} in terms of gauge-invariant form factors $\mathcal{F} _{n}[\alpha]$ makes it clear that $\mathcal{G}_{n} $ is  diffeomorphism invariant. 
    \item The master numerators $N[\alpha]$ contain the  poles $s_{12 i_{1}\cdots i_r}{-}q^2$, where $i_1,\ldots,i_r$ are some gluons. We have the following closed formula for the master numerators for the form factor of ${\rm tr}(\phi^2)$ \cite{Chen:2022nei}
\begin{equation}\label{eq:BCJnumF}
    {N}[1^{\phi},3\ldots n,2^{\phi}]=\sum_{k=1}^{n-2}\sum_{\tau\in \mathbf{P}_{\mathbf{g}}^{(k)}}(-2)^k{\prod\limits_{i=1}^k \Big(p_{1\Xi_L(i)}\cdot f_{(\tau_i)}\cdot p_{2 {\Xi}_R(i)}\Big)\over  (p_{12}^2{-}q^2) (p_{12\tau_1}^2{-}q^2)\cdots (p_{12\tau_1\cdots \tau_{k-1}}^2{-}q^2)},
\end{equation}
where $\mathbf{P}_{\mathbf{g}}^{(r)}$ denotes all the ordered partitions of  the gluon set $\mathbf{g}{=}\{3,\ldots, n\}$ into $r$ subsets, and $\Xi_{L,R}(i)$ denote two special subsets of $\Xi(i){\equiv}\{\tau_1,\tau_2,\ldots, \tau_{i-1}\}$: $\Xi_{L}(i)$ contain the elements in $\Xi(i)$ which are smaller than the first element in $\tau_{i}$;  $\Xi_{R}(i)$ contain the elements in $\Xi(i)$ which are bigger than the last element in $\tau_{i}$.
See the explicit four-point case given in \eqref{eq:4ptNsSolution}.
    
    \item The poles $s_{12 i_{1}\cdots i_r}{-}q^2$ in \eqref{eq:BCJnumF} are particularly interesting because they look like massive Feynman propagators. Although they are not poles in the gauge-theory form factor $\itbf{F}_n$, they become real poles in $\mathcal{G}_n$ after double copy.
    Furthermore, from the $\mathcal{G}_n=\sum_{\alpha} N[\alpha] {\cal F}_n[\alpha]$ form, we can see that $s_{12i_1\cdots i_r} {-} q^2$ appear only as simple poles in ${\cal G}_n$. Besides, the massless poles $s_{1 i_1\cdots i_s}$ or $s_{2 i_1\cdots i_s}$ which are already included in $\itbf{F}_n$ also appear in ${\cal G}_n$ as simple poles. These are the only two types of poles for ${\cal G}_n$. 
    Below we will refer to the $s_{12i_1\cdots i_r} - q^2$ poles as the ``spurious"-type poles and the massless poles $s_{1 i_1\cdots i_s}$ ($s_{2 i_1\cdots i_s}$) as the ``physical"-type poles.

    \item Importantly, ${\cal G}_n$ has nice factorization properties on both the ``spurious"-type and the ``physical"-type poles:
    \begin{equation}
    \begin{aligned}
        &\mathcal{G}_{n}\sim \frac{1}{\delta}\mathcal{G}_{m} \times \mathcal{M}_{n-m+2}
        \quad \textbf{when}\   s_{123\cdots m}{-}q^2=\delta \rightarrow 0 \,, \\
        &\mathcal{G}_{n}\sim \frac{1}{\delta}\mathcal{G}_{n-m+2} \times \mathcal{M}^{\prime}_{m} \quad \textbf{when}\  s_{13\cdots m}=\delta \rightarrow 0 \,,
    \end{aligned} 
    \end{equation}
    where $\mathcal{M}$ and $\mathcal{M}^{\prime}$ are (two different types of) amplitudes with two scalars coupling to multiple gravitons. A detailed discussion of these factorization properties will be given in Section~\ref{ssec:FFDCnptfac}.
    We comment that the fact that $\mathcal{G}_{n} $ has factorization properties on the new ``spurious"-type poles indicates contributions of new Feynman diagrams, such as in Figure~\ref{fig:F3treeGRA} and \ref{fig:F4treeGRA}.
\end{itemize}

In summary, $\mathcal{G}_{n}$ has the properties of locality, unitarity (factorization), and diffeomorphism invariance so that it is indeed the physically meaningful double copy of the form factor $\itbf{F}_n$.
Among all these aspects, the most interesting one is the factorization relations, as well as other intriguing observations leading us to them, which will be the central topic of the remaining part of this section.

\subsection{Matrices and decomposition properties}\label{ssec:bilinearDecomp}

The property of the propagator matrix $\Theta^{\cal F}$ and the KLT kernel $\mathbf{S}^{\cal F}$ involved in the double copy are crucial in understanding the factorization properties mentioned above. 
In this subsection, we will discover that (1) these matrices involved in the form factor double copy are in general invertible; and (2) they take nice factorization structures when considering special kinematics with the ``spurious"- or ``physical"-type poles on-shell. 
In particular, a systematic bilinear decomposition will be raised, which involves some special vectors $\vec{v}$ as generalizations of BCJ vectors. 
In Section~\ref{ssec:FFDCnptfac}, we will come to how to apply these properties; and the thorough study of the $\vec{v}$ vectors will be presented in the second paper  \cite{treepaper2}. 

\subsubsection{Propagator matrix and KLT kernels }\label{subsec:PmatrixFF}

Before discussing the decomposition properties, we spend some time learning more about what are the important matrices $\Theta^{\cal F}$ and $\mathbf{S}^{\cal F}$.

\subsubsection*{Propagator matrix $\Theta^{\cal F}$ as scalar form factors}

The matrix elements of the propagator matrix $\Theta^{\cal F}$ are sums of certain propagators of Feynman diagrams involved in $\itbf{F}_n$. This is very similar to the propagator matrix for amplitudes which can be interpreted as amplitudes in the bi-adjoint $\phi^3$ theory. 
For $\Theta^{\cal F}$, the matrix elements can be understood as form factors in the bi-adjoint $\phi^3$ theory. In particular, we need two different kinds of scalar fields $\{ \phi, \Phi \}$, and the Lagrangian of the theory is
\begin{equation}\label{eq:ymsL}
\begin{aligned}
    \mathcal{L}^{\phi^3}=&\frac{1}{2} \operatorname{tr}_{\mathrm{C}}\left(D_{\mu} \phi^{I} D^{\mu} \phi^{I}\right)
    +\frac{1}{2} \operatorname{tr}_{\mathrm{C}}\left(D_{\mu} \Phi^{I} D^{\mu} \Phi^{I}\right)
    -\frac{\lambda_3}{3 !} f^{I J K} f^{abc} \phi^{I, a} \phi^{J, b} \Phi^{K, c} \\
    &-\frac{\lambda_1}{3 !} f^{I J K} f^{abc} \phi^{I, a} \phi^{J, b} \phi^{K, c}-\frac{\lambda_2}{3 !} f^{I J K} f^{abc} \Phi^{I, a} \Phi^{J, b} \Phi^{K, c}
    \,, 
\end{aligned}
\end{equation}
where we use $\{I, J, K\}$ to denote the flavor (FL) index and $\{a, b, c\}$ to denote the color (C) index. 
We define the gauge-invariant operator in the way that it couples only to $\phi$ with $\mathcal{O}_{\phi}=\sum_{I}{\rm tr}_{\rm C}(\phi^I\phi^I)$.

The matrix elements of $\Theta^{\cal F}$ can be given as form factors defined as (which is similar to the propagator matrix $\Theta^{\cal A}$ for amplitudes in the bi-adjoint scalar theory \cite{Du:2011js, Bjerrum-Bohr:2012kaa, Cachazo:2013iea}) 
\begin{align}\label{eq:deftheta}
    \Theta_n^{\cal F}[\alpha &|\beta]
    =\int d^{D}x\  e^{\mathrm{i}q\cdot x} \langle 1^\phi, 2^\phi, 3^\Phi \ldots n^\Phi | \mathcal{O}_{\phi}(x) | 0 \rangle \big|_{\operatorname{tr}_{\rm C}(\alpha) \operatorname{tr}_{\rm FL}(\beta)}\,,
\end{align}
where $\alpha$ refers to an ordering of color indices and $\beta$ to an ordering of flavor indices.
Note that for the purpose of this section, we can take the external state $\{1,2\}$ to be $\phi$ and the rest as $\Phi$. 
As an example, the four-point matrix can be found in \eqref{eq:4ptThetaMatrix}. One of the features in these matrix elements is that $s_{12}$ or  $s_{12\cdots }$ never appear in the poles of the matrix elements, due to the special structure of the Feynman diagram that the scalar line $1,2$ are always on the different sides of the $q$-leg.

Since $\Theta^{\cal F}[\alpha|\beta]$ are actually scalar form factors, which means they are special Feynman diagrams with the ``physical"-poles and trivial numerators, they satisfy factorization properties when the propagators become on-shell, which will be of use later.

\subsubsection*{From the propagator matrix $\Theta^{\cal F}$ to the KLT kernel $\mathbf{S}^{\cal F}$}

An important observation is that the propagator matrix $\Theta_{n}^{\mathcal{F}}$ is a \emph{full-ranked} $(n-2)!\times (n-2)!$ matrix, so that directly taking the inverse $\mathbf{S}^{\mathcal{F}}=(\Theta^{\mathcal{F}})^{-1}$ is legal. 
We stress that this is very different from usual amplitude cases in which the propagator matrices $\Theta^{\mathcal{A}}$ have zero determinant.

It turns out that the determinant of the propagator matrix   $\Theta_{n}^{\mathcal{F}}$ takes a very simple form.
Here we directly give the  compact expression for both the numerator and the denominator of $\det(\Theta^{\cal F}_{n})$:
\begin{equation}\label{eq:det}
\begin{aligned}
	\text{numerator of }   \det(\Theta^{\cal F}_{n}) &={\prod_{r=0}^{n-3}\prod_{\substack{\{i_1,\ldots,i_r\}\\
	\ \ \subset \{3,\ldots,n\}}}}\left(s_{12i_1\cdots i_r}-q^2\right)^{r!\times (n-3-r)! }\,,\\
	\text{denominator of }   \det(\Theta^{\cal F}_{n}) &=\left({\prod_{r=1}^{n-2}\prod_{\substack{\{i_1,\ldots,i_r\}\\
	\ \ \subset \{3,\ldots,n\}}}}(s_{1 i_1 \cdots i_r})^{(r-1)! \times (n-r-2)!}\right)\times (p_1\leftrightarrow p_2)\,.\\
\end{aligned}
\end{equation}
We have explicitly checked them up to the $n=8$ case, which corresponds to a $6!\times 6!$ propagator matrix.

With the non-zero determinant, we can take the inverse and get $\mathbf{S}^{\cal F}_{n}$ as
\begin{equation}
    \mathbf{S}^{\cal F}_n=(\Theta_{n}^{\cal F})^{-1}=\frac{\text{denominator of }\det(\Theta^{\cal F}_{n})}{\text{numerator of }\det(\Theta^{\cal F}_{n})}\times \text{(the cofactor matrix)}\,,
\end{equation}
where we have used the cofactor formula of the matrix inverse.

We have the  following observations from the above equations: 

1) The denominator of $\det(\Theta^{\cal F}_{n})$ is composed of high powers of all ``physical"-type poles, and they will cancel all  poles in the cofactor matrix, such that $\mathbf{S}^{\cal F}_n$ has no ``physical"-type poles. 

2) The zeros of the numerator of $\det(\Theta^{\cal F}_{n})$ give the poles of $\mathbf{S}_{n}^{\mathcal{F}}$. Since the zeros take solely the $s_{12\cdots}{-}q^2$ form, all the poles involved in $\mathbf{S}_{n}^{\mathcal{F}}$ are ``spurious"-type simple poles, which look like massive Feynman propagators with the mass square $M^2=q^2$.

These features can be seen from some simple examples. For the four-point case,  we quote the ${\bf S}_{4}^{\cal F}$  in \eqref{eq:KLTkernel4pt} as a  2 by 2 matrix, and one can see that the above observations are indeed valid. 
As another more complicated example of a five-point matrix element, one has
\begin{align}
  \mathbf{S}^{\cal F}_{5}[1,3, 4,&5,2 |1,5,3,4,2] = \frac{\tau_{31}\tau_{51}\tau_{42}\tau_{52}}{s_{12}-q^2}\times \\
 & \bigg\{ \frac{\tau_{41}\tau_{3(2+4)}}{s_{124}-q^2}\Big(\frac{1}{s_{1245}-q^2}{+}\frac{1}{s_{1234}-q^2}\Big) {+}  \frac{\tau_{32}\tau_{4(1+3)}}{s_{123}-q^2}\Big(\frac{1}{s_{1235}-q^2}{+}\frac{1}{s_{1234}-q^2}\Big)\nonumber \\
 &{+}\frac{\tau_{4(2+5)}\tau_{3(1+5)}}{s_{12}-q^2}\Big(\frac{1}{s_{1245}-q^2}{+}\frac{1}{s_{1234}-q^2}\Big) {+} \frac{\tau_{34}}{s_{1234}-q^2}{+} \frac{\tau_{4(2+5)}}{s_{1235}-q^2}{+}\frac{\tau_{3(2+4)}}{s_{1245}-q^2}\bigg\} , \nonumber 
 \end{align}
which has only the desired $(s_{12\ldots}-q^2)$ pole, and no double pole ever appears.

\subsubsection{Matrix decomposition of the KLT kernel}\label{sssec:KLTkerneldecomp}

To understand the factorizations of the double-copy form factor, it is important to understand the matrix decomposition of the KLT kernel $\mathbf{S}_{n}^{\mathcal{F}}$  on the ``spurious"-type poles $s_{12 i_1\cdots i_r}{-}q^2$. As we will see, the residue of the kernel matrix takes a nice decomposition form as the tensor product of two lower-point matrices.

We can first have a simple look at the rank of the matrix. Upon taking the residue, the $(n-2)!$ by $(n-2)!$ kernel matrix has rank smaller than $(n-2)!$: 
\begin{equation}
\begin{aligned}
	&\text{rank}\left(\text{Res}_{s_{12 i_1 .. i_r}=q^2}\left[\mathbf{S}_{n}^{\mathcal{F}}\right]\right)=r!\times (n-r-3)!\,.
\end{aligned}
\end{equation}
Furthermore, the ranks of the matrices on the poles satisfy
\begin{equation}
\begin{aligned}
		\text{rank}\left(\text{Res}_{s_{12 i_1.. i_r}=q^2}\left[\mathbf{S}_{n}^{\mathcal{F}}\right]\right)&=\text{rank}\left(\mathbf{S}_{{r+2}}^{\mathcal{F}}\right)\times \text{rank}\left(\mathbf{S}_{n-r}^{\mathcal{A}}\right)=\text{rank}\left(\mathbf{S}_{{r+2}}^{\mathcal{F}}\otimes \mathbf{S}_{n-r}^{\mathcal{A}}\right)\,.
\end{aligned}
\end{equation}
These rank conditions, which can be inspected by explicit calculations, strongly indicate the connection between $\text{Res}_{s_{12 i_1.. i_r}=q^2}\left[\mathbf{S}_{n}^{\mathcal{F}}\right]$ and the small factorized kernels $\mathbf{S}_{{r+2}}^{\mathcal{F}}\otimes \mathbf{S}_{n-r}^{\mathcal{A}}$. 

We now present the decomposition of the kernel matrix:
\begin{equation}\label{eq:SFDecomp}
	\text{Res}_{s_{12 i_1.. i_r}=q^2}\left[\mathbf{S}_{n}^{\mathcal{F}}\right]=\mathbf{V}^{\scriptscriptstyle \rm T} \cdot \left(\mathbf{S}_{{r+2}}^{\mathcal{F}}\otimes \mathbf{S}_{n-r}^{\mathcal{A}}\right) \cdot \mathbf{V} \,,
\end{equation}
where $\mathbf{V}$ is a rectangular $[r!\ (n-r-3)!]$ by $(n-2)!$ matrix with matrix elements rational functions of Lorentz products, and $\mathbf{V}^{\scriptscriptstyle \rm T}$ is its transpose. 
To make this abstract expression less obscure, we give some examples below.

We will often use short notation $\QQ_m = \sum_{i=1}^m p_i$, and $\tau_{ij} = 2 p_i \cdot p_j$.
\subsubsection*{1. Four-point case}
For the four-point kernel  $\mathbf{S}_{4}^{\mathcal{F}}$ given in \eqref{eq:KLTkernel4pt}, the allowed ``spurious"-type poles are $s_{12}{-}q^2$, $s_{123}{-}q^2$ and $s_{124}{-}q^2$. Due to the symmetry of permuting $\{3,4\}$, we only need to consider the first two cases. 
We inspect one by one for these two cases. 

For the pole $\QQ_2^2{-}q^2=s_{12}{-}q^2$, we are looking for the relation between the $2\times 2$ matrix $\mathrm{Res}_{\QQ_2^2=q^2}\left[{\bf S}^{\mathcal{F}}_{4}\right]$ and the $1\times 1 $ matrix $\left({\bf S}^{\mathcal{F}}_{2}\otimes{\bf S}^{\mathcal{A}}_{4}\right) $. 
Here we have
\begin{equation}
\mathbf{S}^{\cal F}_{2}=1 \,, \qquad \mathbf{S}^{\cal A}_{4}=\mathbf{S}^{\cal A}_{4}[\QQ_2,3,4,-q|\QQ_2,3,4,-q]= { \tau_{3 ,(1+2)} \tau_{34} \over \tau_{4,(1+2)} }\,,
\end{equation} 
in which the latter is the inverse of the bi-adjoint scalar amplitude $\Theta_4^{\cal A}[\QQ_2,3,4,-q|\QQ_2,3,4,-q]={\tau_{3,(1+2)}^{-1}}+{\tau_{34}^{-1}}$.\footnote{We would like to point out that here for the amplitude KLT kernel $\mathbf{S}^{\cal A}$, we will use a less-commonly used representation which is different from the conventional  ``polynomial" form $\mathbf{S}^{\cal A}[1,\ldots,(n{-}1),n|1,\ldots,n,(n{-}1)]$ \cite{Bern:1998sv}.}

Now given the $\vec{v}$ vector, which can also be regarded as an $1\times 2$ matrix $\mathbf{V}_4^{A}$, as 
\begin{equation}
    \vec{v}^{A}_4 =\Big\{ \frac{\tau_{31}\tau_{42}}{\tau_{3\QQ_2}}, \frac{ \tau_{32}\tau_{41}}{\tau_{3\QQ_2}} \Big\}, \qquad \mathbf{V}_4^{A}=\left(\vec{v}^{A}_4 \right),
\end{equation}
one can directly verify that the $2\times 2$ matrix $\mathrm{Res}_{s_{12}=q^2}\left[{\bf S}^{\mathcal{F}}_{4}\right]$ satisfies
\begin{equation}\label{eq:4ptmatdecpA}
\begin{aligned}
    \mathrm{Res}_{\QQ_2^2=q^2}\left[{\bf S}^{\mathcal{F}}_{4}\right]&= \begin{pmatrix}
        \frac{\tau_{31}\tau_{42}}{\tau_{3\QQ_2}} \\   \frac{ \tau_{32}\tau_{41}}{\tau_{3\QQ_2}}
    \end{pmatrix}
\cdot \left(1\times { \tau_{3 ,(1+2)} \tau_{34} \over \tau_{4,(1+2)} }\right)\cdot 
    \left( \frac{\tau_{31}\tau_{42}}{\tau_{3\QQ_2}} \ \   \frac{ \tau_{32}\tau_{41}}{\tau_{3\QQ_2}} \right)
    \\
    &=({\bf V}^{A}_4)^{\rm \scriptscriptstyle T}\cdot \left({\bf S}^{\mathcal{F}}_{2}\otimes{\bf S}^{\mathcal{A}}_{4}\right) \cdot  {\bf V}^{A}_4\,.
\end{aligned}
\end{equation}

Similarly, for the pole $\QQ_3^2-q^2=0$, one has the decomposition relation as 
\begin{equation}\label{eq:4ptmatdecpB}
\begin{aligned}
     \mathrm{Res}_{\QQ_3^2=q^2}\left[{\bf S}^{\mathcal{F}}_{4}\right]&=
     \begin{pmatrix}
        \tau_{42} \\ \tau_{4,(2+3)}
    \end{pmatrix}
\cdot \left({\tau_{13}\tau_{23} \over s_{12}{-}q^2} \times 1\right)\cdot     \left(\tau_{42} \ \ \tau_{4,(2+3)} \right)\\
    &=({\bf V}^{B}_4)^{\rm \scriptscriptstyle T}\cdot \left({\bf S}^{\mathcal{F}}_{3}\otimes{\bf S}^{\mathcal{A}}_{3}\right) \cdot  {\bf V}^{B}_4\,,
\end{aligned}
\end{equation}
where
\begin{equation}
    \vec{v}^{B}_4 =\left\{\tau_{42}, \tau_{4,(2+3)}\right\}, \quad \mathbf{V}_4^{B}=\left(\vec{v}^{B}_4 \right)\,.
\end{equation}

\subsubsection*{2. Five-point case} 

We give more non-trivial examples at five points. Once again, from the permutation symmetry of the gluons $\{3,4,5\}$, it suffices to consider $s_{12}{-}q^2$, $s_{123}{-}q^2$ and $s_{1234}{-}q^2$. 

Beginning with the $\QQ_2^2{-}q^2=s_{12}{-}q^2$ pole, we have a decomposition connecting the large 6 by 6 matrix $\text{Res}_{\QQ_2^2=q^2}[\mathbf{S}^{\cal F}_{5}]$ with the small 2 by 2 matrix $\mathbf{S}^{\cal F}_{2}\otimes \mathbf{S}^{\cal A}_{5}$ via a rectangular 2 by 6 matrix $\mathbf{V}^{A}_5$. We give directly the matrix $\mathbf{V}^{A}_5$ first (and we will explain the subscripts for $\vec{v}$ shortly)
\begin{equation}
\label{eq:bfVmatrix5pt}
{\bf V}^{A}_5=\begin{pmatrix}
{\vec v}^{\scriptscriptstyle A}_{5,(\mathbf{1},\mathbf{1})} \\
{\vec v}^{\scriptscriptstyle A}_{5,(\mathbf{1},\sigma_2)}
\end{pmatrix} ,
\end{equation}
with 
{\small 
\begin{align}\label{eq:v5A}
\hskip -0.1cm
& {\vec v}^{ A}_{5,(\mathbf{1},\mathbf{1})} = \\
\hskip -0.1cm
&\bigg\{ 
\frac{\tau_{31}\tau_{52}}{\tau_{5 q}}\Big(\frac{\tau_{4,(1+3)}}{\tau_{3 \QQ_2}}+1\Big), \frac{\tau_{31}\tau_{42}\tau_{5,(1+3)}}{\tau_{3 \QQ_2}\tau_{5 q}} , -\frac{\tau_{32}\tau_{41}\tau_{52}}{\tau_{3 \QQ_2}\tau_{5 q}}, -\frac{\tau_{32}\tau_{41}\tau_{5,(2+3)}}{\tau_{3 \QQ_2}\tau_{5 q}} , \frac{\tau_{31}\tau_{42}\tau_{51}}{\tau_{3 \QQ_2}\tau_{5 q}} , -\frac{\tau_{32}\tau_{51}}{\tau_{5 q}}\Big(\frac{\tau_{4,(2+3)}}{\tau_{3 \QQ_2}}+1\Big) 
\bigg\} , \nonumber \\
\hskip -0.5cm
&{\vec v}^{ A}_{5,(\mathbf{1},\sigma_2)} = \nonumber \\
\hskip -0.5cm
&\bigg\{ 
\frac{\tau_{31}\tau_{52}\tau_{4,(1+3)}}{\tau_{3 \QQ_2}\tau_{4 q}} , \frac{\tau_{31}\tau_{42}}{\tau_{4 q}}\Big(\frac{\tau_{5,(1+3)}}{\tau_{3 \QQ_2}}+1\Big) , \frac{\tau_{31}\tau_{41}\tau_{52}}{\tau_{3 \QQ_2}\tau_{4 q}} ,  -\frac{\tau_{32}\tau_{41}}{\tau_{4 q}}\Big(\frac{\tau_{5,(2+3)}}{\tau_{3 \QQ_2}}+1\Big) , -\frac{\tau_{32}\tau_{42}\tau_{51}}{\tau_{3 \QQ_2}\tau_{4 q}} ,  -\frac{\tau_{32}\tau_{51}\tau_{4,(2+3)}}{\tau_{3 \QQ_2}\tau_{4 q}}
\bigg\} \nonumber  .
\end{align}
}

Then we have the following matrix decomposition
\begin{equation}\label{eq:5ptmatdecpA}
	 \mathrm{Res}_{\QQ_2=q^2}\left[{\bf S}^{\mathcal{F}}_{5}\right]=({\bf V}^{A}_5)^{\rm \scriptscriptstyle T} \cdot \left({\bf S}^{\mathcal{F}}_{2}\otimes{\bf S}^{\mathcal{A}}_{5}\right) \cdot{\bf V}^{A}_5\,.
\end{equation}
In this example, we need to pay attention to the ordering of matrix elements. For example, when it comes to the 2 by 2 matrix $\mathbf{S}^{\cal A}_5$, we need to specify what each matrix element exactly means. They can be explicitly given as follows: \begin{equation}
	{\bf S}^{\mathcal{A}}_{5}=\begin{pmatrix}
{\bf S}^{\mathcal{A}}_5[\mathbf{q}_2,3,4,5,-q|\mathbf{q}_2,3,4,5,-q] & \ {\bf S}^{\mathcal{A}}_5[\mathbf{q}_2,3,4,5,-q|\mathbf{q}_2,3,5,4,-q]  \\
{\bf S}^{\mathcal{A}}_5[\mathbf{q}_2,3,5,4,-q|\mathbf{q}_2,3,4,5,-q]  &\ {\bf S}^{\mathcal{A}}_5[\mathbf{q}_2,3,5,4,-q|\mathbf{q}_2,3,5,4,-q]
\end{pmatrix} ,
\end{equation}
and the representation that we are using  for this KLT kernel is 
\begin{equation}
	{\bf S}^{\mathcal{A}}_{5}=\begin{pmatrix}
\Theta^{\mathcal{A}}_5[\mathbf{q}_2,3,4,5,-q|\mathbf{q}_2,3,4,5,-q] & \ \Theta^{\mathcal{A}}_5[\mathbf{q}_2,3,4,5,-q|\mathbf{q}_2,3,5,4,-q]  \\
\Theta^{\mathcal{A}}_5[\mathbf{q}_2,3,5,4,-q|\mathbf{q}_2,3,4,5,-q]  &\ \Theta^{\mathcal{A}}_5[\mathbf{q}_2,3,5,4,-q|\mathbf{q}_2,3,5,4,-q]
\end{pmatrix}^{-1} ,
\end{equation}
where the $\Theta^{\mathcal{A}}[\alpha|\beta]$ is nothing but the (double-color-ordered) bi-adjoint scalar amplitudes.\footnote{The only thing to notice here is that $\QQ_2$ and $-q$ are not light-like momenta, but this fact does not matter to the definition of bi-adjoint scalar amplitudes.} 
This is also another example of our using the symmetric representation for the amplitude KLT kernel. As a concrete example of the matrix elements, we have 
\begin{equation}
    \Theta^{\mathcal{A}}_5[\mathbf{q}_2,3,4,5,-q|\mathbf{q}_2,3,5,4,-q]=-\frac{1}{\tau_{3,(1+2)}s_{45}}-\frac{1}{s_{34}s_{345}} \,.
\end{equation}

We now introduce some notations that are convenient to use later. To abbreviate the kind of lengthy notations, we notice that the two orderings $\{\mathbf{q}_2,3,4,5,-q\}$ and $\{\mathbf{q}_2,3,5,4,-q\}$ involved above can be related by a permutation $\sigma_2$ permuting $4,5$ (for an amplitude KLT kernel, we need to fix the positions of three particles in the ordering). 
In other words, these two orderings are obtained by acting the permutation group $S_2$ on the ``standard ordering'' $\{\mathbf{q}_2,3,4,5,-q\}$. Thus, we label the two orderings $\{\mathbf{q}_2,3,4,5,-q\}$ and $\{\mathbf{q}_2,3,5,4,-q\}$ in terms of the permutations  $\{\mathbf{1},\sigma_2\}\in S_2$, so that we can rewrite ${\bf S}^{\mathcal{A}}_{5}$ as 
\begin{equation}
	{\bf S}^{\mathcal{A}}_{5}=\begin{pmatrix}
{\bf S}^{\mathcal{A}}_{5}[\mathbf{1}|\mathbf{1}] &{\bf S}^{\mathcal{A}}_{5}[\mathbf{1}|\sigma_2]  \\
{\bf S}^{\mathcal{A}}_{5}[\sigma_2|\mathbf{1}]  & {\bf S}^{\mathcal{A}}_{5}[\sigma_2|\sigma_2]
\end{pmatrix}\,,
\end{equation}
where the rows and columns of the matrix $\mathbf{S}$ are labeled by permutations. 
The other instances are the form factor propagator matrix $\Theta_n^{\cal F}$ and KLT kernel $\mathbf{S}_{n}^{\cal F}$: they are also defined on two orderings $\alpha_1, \alpha_2$ which can be obtained by acting $\beta\in S_{n-2}$ on the standard ordering $\{1,3,\ldots,n,2\}$. Thus, we can drop the $1,2$ (in the $1,*,2$) in  \eqref{eq:FthetaNn}. 

To make the notations consistent, we are also labeling rows and columns of the matrix $\mathbf{V}_{n}$ in terms of permutations, which is exactly the reason why we put the subscript $(\mathbf{1},\mathbf{1})$ and $(\mathbf{1},\sigma_{2})$ to label the $\vec{v}$ vectors in \eqref{eq:v5A}.

Adopting these notations, we have 
\begin{align}\label{eq:5ptSFdecompositionA}
	&{\rm Res}_{\QQ_{2}^2=q^2}\left[\mathbf{S}^{\mathcal{F}}_{5}[\alpha_1|\alpha_2]\right]=\sum_{\bar{\rho}_{1,2}\in S_{2} } { v}^{A}_{5,(\mathbf{1},\bar\rho_1)}[\alpha_1]\mathbf{S}^{\mathcal{F}}_{2}[\mathbf{1}|\mathbf{1}]  \mathbf{S}^{\mathcal{A}}_{5}[\bar{\rho}_1|\bar{\rho}_2] { v}^{A}_{5,(\mathbf{1},\bar\rho_2)}[\alpha_2],
\end{align}
with $\alpha_{1,2}\in S_{n-2}$ 
and the matrix elements $v_{(*,*)}[*]$ are the vector elements in \eqref{eq:v5A}, \emph{e.g.} 
\begin{equation}
\begin{aligned}
    &v_{5,(\mathbf{1},\mathbf{1})}^{\scriptscriptstyle A}[\mathbf{1}]=\frac{\tau_{31}\tau_{52}}{\tau_{5 q}}\Big(\frac{\tau_{4,(1+3)}}{\tau_{3 \QQ_2}}+1\Big),\  v_{5,(\mathbf{1},\mathbf{1})}^{\scriptscriptstyle A}[(4,5)]=\frac{\tau_{31}\tau_{42}\tau_{5,(1+3)}}{\tau_{3 \QQ_2}\tau_{5 q}},\\
    &v_{5,(\mathbf{1},\mathbf{1})}^{\scriptscriptstyle A}[(3,4,5)]= -\frac{\tau_{32}\tau_{41}\tau_{5,(2+3)}}{\tau_{3 \QQ_2}\tau_{5 q}},\   v_{5,(\mathbf{1},\sigma_2)}^{\scriptscriptstyle A}[(3,4,5)]= -\frac{\tau_{32}\tau_{41}}{\tau_{4 q}}\Big(\frac{\tau_{5,(2+3)}}{\tau_{3 \QQ_2}}+1\Big)\,.
\end{aligned}
\end{equation}

Next, we move on to the $\QQ_3^2{-}q^2=s_{123}{-}q^2$  pole.
In this case, we give directly the $\mathbf{V}^{B}_5$ matrix as $\mathbf{V}^{B}_5=(\vec{v}^{\scriptscriptstyle \, B}_{5,(\mathbf{1},\mathbf{1})})$ with 
\begin{equation}
   \vec{v}^{\scriptscriptstyle \, B}_{5,(\mathbf{1},\mathbf{1})} = \left\{\frac{\tau_{4,(1+3)}\tau_{52}}{\tau_{4\QQ_3}},\frac{\tau_{5,(1+3)}\tau_{42}}{\tau_{4\QQ_3}},\frac{\tau_{41}\tau_{52}}{\tau_{4\QQ_3}},\frac{\tau_{41}\tau_{5,(2+3)}}{\tau_{4\QQ_3}},\frac{\tau_{42}\tau_{51}}{\tau_{4\QQ_3}},\frac{\tau_{4,(2+3)}\tau_{51}}{\tau_{4\QQ_3}}\right\},
\end{equation}
so that 
\begin{equation}
\begin{aligned}
    {\rm Res}_{\QQ_3^2=q^2}\left[\mathbf{S}^{\mathcal{F}}_{5}\right] &= ({\bf V}^{ B}_{5})^{\scriptscriptstyle {\rm T}} \cdot (\mathbf{S}^{\mathcal{F}}_{3}\otimes \mathbf{S}^{\mathcal{A}}_{4}) \cdot {\bf V}^{ B}_{5} \,, \\
    {\rm Res}_{\QQ_3^2=q^2}\left[\mathbf{S}^{\mathcal{F}}_{5}[\alpha_1|\alpha_2]\right]& = { v}^{\scriptscriptstyle \, B}_{5,(\mathbf{1},\mathbf{1})}[\alpha_1]\mathbf{S}^{\mathcal{F}}_{3}[\mathbf{1}|\mathbf{1}]  \mathbf{S}^{\mathcal{A}}_{4}[\mathbf{1}|\mathbf{1}] { v}^{\scriptscriptstyle \, B}_{5,(\mathbf{1},\mathbf{1})}[\alpha_2] \,.
\end{aligned}
\end{equation}
In this formula, we are connecting the 6 by 6 kernel $\text{Res}_{\QQ_3^2=q^2}[{\bf S}^{\cal F}]$ with the 1 by 1 matrix $\mathbf{S}^{\mathcal{F}}_{3} \otimes \mathbf{S}^{\mathcal{A}}_{4}$ with the 1 by 6 matrix ${\bf V}_5^{B}$. Both the $\mathbf{S}^{\mathcal{F}}_{3}$ and $\mathbf{S}^{\mathcal{A}}_{4}$ are 1 by 1 matrices so that only one matrix element ${\bf S}[\mathbf{1}|\mathbf{1}]$ appears.

We end up with the last case on the $\QQ_4{-}q^2=s_{1234}{-}q^2$ pole:
\begin{equation}
\begin{aligned}
    \mathrm{Res}_{\QQ_4=q^2}\left[{\bf S}^{\mathcal{F}}_{5}\right]&= {\bf V}^{ C}_{5}\cdot ({\bf S}^{\mathcal{F}}_{4} \otimes  {\bf S}^{\mathcal{A}}_{3}) \cdot  {\bf V}^{ C}_{5}\\
    \mathrm{Res}_{\QQ_4=q^2}\left[{\bf S}^{\mathcal{F}}_{5}[\alpha_1|\alpha_2]\right]&=\sum_{\bar{\kappa}_{1,2}\in S_{2} } { v}^{\scriptscriptstyle \, C}_{5,(\bar\kappa_1,\mathbf{1})}[\alpha_1] {\bf S}^{\mathcal{F}}_{4}[\bar\kappa_1|\bar\kappa_2]{\bf S}^{\mathcal{A}}_{3}[\mathbf{1}|\mathbf{1}] { v}^{\scriptscriptstyle \, C}_{5,(\bar\kappa_2,\mathbf{1})}[\alpha_2]\,,
\end{aligned}
\end{equation}
where 
\begin{equation}
\begin{aligned}
    {\bf V}^{C}_{5}=\begin{pmatrix}
        {\vec v}^{\scriptscriptstyle \, C}_{5,(\mathbf{1},\mathbf{1})}\\
        {\vec v}^{\scriptscriptstyle \, C}_{5,(\sigma_2,\mathbf{1})}
    \end{pmatrix}
\end{aligned},\ \text{ and }\  
\begin{aligned}
{\vec v}^{\scriptscriptstyle \, C}_{5,(\mathbf{1},\mathbf{1})} & = 
\{ \tau_{52}, \tau_{5,(2+4)}, 0, 0, \tau_{5,(2+3+4)}, 0 \} ,  \\
{\vec v}^{\scriptscriptstyle \, C}_{5,(\sigma_2,\mathbf{1},)} & = 
\{ 0, 0, \tau_{52}, \tau_{5,(2+3)}, 0, \tau_{5,(2+3+4)} \} .
\end{aligned}
\end{equation}

\subsubsection*{3. General $n$-point cases}

For the general $n$-point case, without loss of generality, we consider the pole $\QQ_m^2=q^2$ with $\sum_{i=1}^{m}{p_i}=\QQ_m$ and try to connect the residue of ${\bf S}_{n}^{\cal F}$  with ${\bf S}_{m}^{\cal F}\otimes {\bf S}_{n-m+2}^{\cal A}$. 

Let us first clarify the orderings appearing in the matrices:
\begin{itemize}[itemsep=3pt]
\item 
The $n$-point form factor kernels ${\bf S}_{n}^{\cal F}$ is a $(n-2)!$ by $(n-2)!$ matrix. 
To label its rows and columns, we pick the standard ordering to be $\{1,3,\ldots,n,2\}$ and use $\{1,\alpha(3,\ldots,n),2\}$ where $\alpha\in S_{n-2}$ to label its rows and columns. Each element of ${\bf S}_{n}^{\cal F}$ will be denoted as ${\bf S}_{n}^{\cal F}[\alpha_1| \alpha_2]$.
\item
A similar labeling applies to the sub-kernel ${\bf S}_{m}^{\cal F}$: we denote the permutation as $\bar{\kappa}$ with $\bar{\kappa}\in S_{m-2}$,\footnote{The bar on permutation indicate that it is only permuting part of the particles.} and the matrix element is $\mathbf{S}^{\mathcal{F}}_{m}[\bar{\kappa}_1|\bar{\kappa}_2]$. 
\item
For the kernel $\mathbf{S}_{n-m+2}^{\cal A}$ of sub-amplitude, the dimension is $(m'-3)!$, where $m^{\prime}=n{-}m{+}2$. We choose the orderings $\{\QQ_m,m{+}1, \bar{\rho}(m{+}2,\ldots,n),-q\}$, where $\bar{\rho} \in S_{m^{\prime}-3}$  is used to label the rows and columns of $\mathbf{S}^{\cal A}_{m^{\prime}}$. The matrix element is $\mathbf{S}^{\mathcal{A}}_{m^{\prime}}[\bar{\rho}_1|\bar{\rho}_2]$. 
\item
The tensor product ${\bf S}_{m}^{\cal F}\otimes \mathbf{S}^{\cal A}_{m^{\prime}}$ is a matrix of dimension $(m-2)! \times (m'-3)!$. To label a matrix element in the tensor product ${\bf S}_{m}^{\cal F}\otimes \mathbf{S}^{\cal A}_{m^{\prime}}$, one can use a pair of permutations $(\bar{\kappa},\bar{\rho})$. 
\end{itemize}
Since we are looking for a decomposition like $\text{Res}[{\bf S}^{\cal F}_n]=\mathbf{V}^{\scriptscriptstyle \rm T}\cdot ({\bf S}_m^{\cal F}\otimes \mathbf{S}_{m'}^{\cal A}) \cdot \mathbf{V}$, the number of rows of $\mathbf{V}$ is equal to the dimension of the tensor product ${\bf S}_m^{\cal F}\otimes \mathbf{S}_{m'}^{\cal A}$ and can be labelled by $(\bar{\kappa},\bar{\rho})$, while the number of columns of $\mathbf{V}$ is the same as the dimension of $\text{Res}[{\bf S}^{\cal F}_n]$. 
Therefore, we can write $\mathbf{V}$ as
\begin{equation}
{\bf V}=\begin{pmatrix}
{\vec v}_{(\bar{\kappa},\bar{\rho})} \\
\cdots \\
{\vec v}_{(\bar{\kappa}',\bar{\rho}')}
\end{pmatrix} ,
\end{equation}
where each  $\vec{v}_{(\bar{\kappa},\bar{\rho})}$ represents a row vector with length equal $(n-2)!$. The elements in each $\vec{v}_{(\bar{\kappa},\bar{\rho})}$ can be explicitly denoted as ${v}_{(\bar{\kappa},\bar{\rho})}[\alpha]$ with $\alpha \in S_{n-2}$.

Having the above preparation, now we can present the following decomposition formula:
\begin{align}\label{eq:nptSFdecomposition}
	&{\rm Res}_{\QQ_{m}^2=q^2}\left[\mathbf{S}^{\mathcal{F}}_{n}[\alpha_1|\alpha_2]\right]=\sum_{\substack{\bar{\kappa}_{1,2}\in S_{m-2} \\ \bar{\rho}_{1,2}\in S_{m^{\prime}-3} } } { v}_{(\bar\kappa_1,\bar\rho_1)}[\alpha_1]\left(\mathbf{S}^{\mathcal{F}}_{m}[\bar{\kappa}_1|\bar{\kappa}_2] \mathbf{S}^{\mathcal{A}}_{m^{\prime}}[\bar{\rho}_1|\bar{\rho}_2]\right) { v}_{(\bar\kappa_2,\bar\rho_2)}[\alpha_2]\,, 
\end{align}
with $\alpha_{1,2}\in S_{n-2}$. 
The decomposition is saying that on the pole of $\mathbf{S}^{\cal F}$, the kernel factorizes into ${\bf S}^{\cal F}\otimes \mathbf{S}^{\cal A}$, and since the sizes of matrices before and after the ``factorization" do not match, we need a rectangular $\mathbf{V}$ matrix. 

We would like to point out that in the discussions of this section, we regard the vectors $\vec{v}$ as predetermined, and utilize these ``known" expressions to perform the matrix decomposition. 
We will not discuss the derivation of $\vec{v}$ in this paper (explicit expresses up to six points can be found in Appendix~\ref{app:vectors}) but just mention that the $\vec{v}$ vectors have many interesting properties, \emph{e.g.}~the matrix element has the mass dimension two as $(p\cdot q)$, there are connections between matrix elements for $\vec{v}_{(\bar{\kappa},\bar{\rho})}$ with different $\bar{\kappa}$ and $\bar{\rho}$, and they are null vectors of the propagator matrix $\Theta_{n}$ when evaluated at the special kinematics $\QQ_m^2=q^2$. All these properties, as well as a closed formula for $\vec{v}$ will be presented  in \cite{treepaper2}.

\subsubsection{The decomposition for the propagator matrix }\label{sssec:thetadecomp}

Apart from the decomposition of the KLT kernel ${\bf S}^{\cal F}_n$, we can do the same thing for the propagator matrix $\Theta^{\cal F}_n$. 
In this case, we have only the ``physical''-type poles $s_{1i_1\cdots i_r}$ (or $s_{2i_1\cdots i_r}$) with $i_1,\ldots,i_r$ gluons, which are poles in the gauge form factors. 
This is expected since $\Theta^{\cal F}$ can be understood as the double-ordered form factors in the bi-adjoint scalar theory as discussed in Section~\ref{subsec:PmatrixFF}. 

Taking the residue gives the following factorization (here $I$ is the intermediate particle with $p_{I}=p_{i_1}{+}\cdots {+}p_{i_r}$)
\begin{equation}\label{eq:nptmFDecomp}
	\text{Res}_{s_{1i_1 \cdots  i_r}=0}[\Theta_n^{\mathcal{F}}[1,\alpha,2|1,\beta,2]]=\Theta_{r+2}^{\mathcal{A}}[1,\bar{\alpha}_1,I|1,\bar{\alpha}_2,I] \times   \Theta_{n-r}^{\mathcal{F}} [{-}I,\bar{\beta}_1,2|{-}I,\bar{\beta}_2,2] \,,
\end{equation}
as long as the orderings $\alpha$ and $\beta$ permits the $s_{1 i_1\cdots i_r}$ pole, that is 
\begin{equation}
    \alpha=\{\alpha_1,\beta_1\}\,, \quad \beta=\{\alpha_2,\beta_2\} \,,
\end{equation}
where $\alpha_{1,2}$ are certain permutations in $S_{r}$ acting on the gluons $\{i_1,\ldots,i_r\}$ and $\beta_{1,2}$ are permutations in $S_{n-r-2}$ on the rest of gluons. 

The above relation can be understood as a matrix decomposition formula
\begin{equation}\label{eq:nptmFDecomp1}
	\text{Res}_{s_{1i_1 \cdots  i_r}=0}[\Theta_n^{\mathcal{F}}]\xrightarrow{\text{ non-zero element}}\Theta_{r+2}^{\mathcal{A}}\otimes \Theta_{n-r}^{\mathcal{F}}\,,
\end{equation}
in which $\Theta_{r+2}^{\cal A}$ is a $r!$ by $r!$ propagator matrix and $\Theta_{n-r}^{\cal F}$ is a $(n{-}r{-}2)!$ by $(n{-}r{-}2)!$ propagator matrix. 
Here, $\Theta_{n-r}^{\cal F}$ is full-ranked as a form-factor propagator matrix.
On the other hand, $\Theta_{r+2}^{\cal A}$ has only rank $(r{-}1)!$ that is smaller than its size $r!$, because its matrix elements are $(r{+2})$-point bi-adjoint scalar amplitudes and the minimal basis for $(r{+2})$-point amplitudes has $(r{-}1)!$ elements. 
These arguments lead us to the following rank condition
\begin{align}
 \text{rank}\left(\text{Res}_{s_{1i_1 \cdots  i_r}=0}\left[\Theta^{\cal F}_{n}\right]\right) & =(r-1)!\times (n-r-2)!  \nonumber\\
& = \text{rank}\left(\Theta^{\mathcal{A}}_{r+2}\right)\times \text{rank}\left(\Theta^{\mathcal{F}}_{n-r}\right)=\text{rank}\left(\Theta^{\mathcal{A}}_{r+2} \otimes \Theta^{\mathcal{F}}_{n-r}\right)\,.
\end{align}
We see that, like $\mathbf{S}^{\cal F}$, the full-ranked matrix $\Theta^{\cal F}$ can be factorized as a tensor product of lower-rank matrices on the poles. 

One can go one step further. The decomposition for $\mathbf{S}^{\cal F}$ as in \eqref{eq:SFDecomp} indicates that it is possible to ``compress" an  ${\bf a} \times {\bf a}$ matrix with rank ${\bf b}$ (${\bf b}<{\bf a}$) to a ${\bf b}\times {\bf b}$ matrix, by introducing some ${\bf a} \times {\bf b}$ rectangular matrix like $\mathbf{V}$. We can do the same thing for the $\Theta^{\cal A}$ part in \eqref{eq:nptmFDecomp1} ($\Theta^{\cal F}$ is already full-ranked).  

As a concrete example, we will consider the five-point propagator matrix $\Theta^{\cal A}_5$, which may appear in the decomposition like $\text{Res}_{s_{1345}=0}[\Theta^{\cal F}_{7}]\sim \Theta^{\cal A}_{5}\otimes \Theta^{\cal F}_{4}$ as in \eqref{eq:nptmFDecomp1}. 
One can perform the decomposition in this case and rederive an interesting connection to the BCJ relations.  
$\Theta^{\mathcal{A}}_5$ looks like
\begin{align}
&\begin{pmatrix}
 \Theta[1,3,4,5,I|1,3,4,5,I] & \ \Theta[1,3,5,4,I|1,3,4,5,I] & \ \cdots\\
 \Theta[1,3,4,5,I|1,3,5,4,I] & \ \Theta[1,3,5,4,I|1,3,5,4,I] & \ \cdots \\
 \vdots & \vdots & \ \ddots
\end{pmatrix}=\\
&\begin{pmatrix}
\frac{1}{s_{13}s_{45}} + \frac{1}{s_{I1}s_{34}} +  \frac{1}{s_{34}s_{5I}} +\frac{1}{s_{45}s_{I1}} + \frac{1}{s_{5I}s_{13}} &  - \frac{1}{s_{I1}s_{34}} -  \frac{1}{s_{34}s_{5I}} & \cdots &   \\
- \frac{1}{s_{I1}s_{34}} -  \frac{1}{s_{34}s_{5I}} &  \frac{1}{s_{14}s_{35}} + \frac{1}{s_{I1}s_{34}} +  \frac{1}{s_{34}s_{5I}} +\frac{1}{s_{35}s_{I1}} + \frac{1}{s_{5I}s_{14}}  & \ \cdots \\
 \vdots & \vdots & \ddots &
\end{pmatrix}, \nonumber 
\end{align}
where we only show its upper-left $2\times 2$ minor and  $p_I{=}p_1{+}p_3{+}p_4{+}p_5$ with $s_{1345}=p_I^2=0$. 
And of course, $\text{rank}(\Theta_5^{\cal A})=2$. 

We denote the upper-left $2\times 2$ minor shown above as $\overline{\Theta}_5^{\cal A}$.
Since the five-point amplitudes has a length $(5-3)!=2$ minimal basis, this $2$ by $2$ $\,$ $\overline{\Theta}_5^{\cal A}$ is enough for performing double copy. 
Then the full $6\times 6$ matrix $\Theta_5^{\cal A}$ can be related to $\overline{\Theta}_5^{\cal A}$ in the following decomposition form: 
\begin{equation}\label{eq:theta5decomp}
	\Theta_5^{\cal A}= \mathbf{Y}_5^{\scriptscriptstyle \rm T}\cdot \overline{\Theta}_5^{\cal A} \cdot  \mathbf{Y}_5\,, \text{ with }  \mathbf{Y}_5=
\begin{pmatrix}
1 & 0 & -\frac{s_{13} s_{45}}{s_{15} s_{4I}} & -\frac{s_{13}( s_{45}+s_{15})}{
 s_{15} s_{3I}} & \frac{s_{13}}{s_{3I}}  & \frac{s_{14}+s_{34}}{s_{4I}}\\
 0 & 1 & -\frac{s_{14} (s_{15}+s_{35})}{s_{15} s_{4I}} & -\frac{s_{14}s_{35}}{
 s_{15} s_{3I}} & \frac{s_{13}+s_{34}}{s_{3I}}  & \frac{s_{14}}{s_{4I}}
\end{pmatrix}\,,
\end{equation}
where the six orderings corresponding to the rows of $\mathbf{Y}$ are 
\begin{equation}
    \{1,3,4,5,I\},\ \{1,3,5,4,I\},\ \{1,4,3,5,I\},\ \{1,4,5,3,I\},\ \{1,5,3,4,I\},\ \{1,5,4,3,I\}\,.
\end{equation}
The $\mathbf{Y}$ matrix now plays a similar role as the $\mathbf{V}$ matrices before in \eqref{eq:SFDecomp}. 
We can see that the null space of $\mathbf{Y}_5$ is 
\begin{equation}\label{eq:nullspaceY5}
\mathbf{Y}_5\cdot 
\begin{pmatrix}
\tau_{5(1+3+4)} & \tau_{4(1+3)} & 0 & 0\\
\tau_{5(1+3)} & \tau_{4(1+3+5)} & 0 & 0\\
0 & \tau_{41} & \tau_{5(1+3+4)} & 0 \\
0 & 0 & \tau_{5(1+4)} & \tau_{41} \\
\tau_{51} & 0 & 0 & \tau_{4(1+3+5)}\\
0 & 0 & \tau_{51} & \tau_{4(1+5)}
\end{pmatrix}=0\,.
\end{equation}

Interestingly, these vectors in \eqref{eq:nullspaceY5} are the BCJ vectors for amplitudes. For example,
\begin{equation}
    \tau_{5(1+3+4)} \mathcal{A}_5(1,3,4,5,I)+ \tau_{5(1+3)} \mathcal{A}_5(1,3,5,4,I)+ \tau_{51} \mathcal{A}_5(1,5,3,4,I) = 0\,. 
\end{equation}
Since we know $\Theta_5^{\cal A}[\alpha|\beta]$ can be interpreted as bi-adjoint scalar amplitudes, they also satisfy
\begin{equation}
    \tau_{5(1+3+4)} \Theta_5^{\cal A}[\alpha|1,3,4,5,I] + \tau_{5(1+3)}  \Theta_5^{\cal A}[\alpha|1,3,5,4,I]+ \tau_{51}  \Theta_5^{\cal A}[\alpha|1,5,3,4,I] = 0\,. 
\end{equation}
In other words, the null vectors of $\mathbf{Y}_5$ are also the null vectors of the propagator matrix $\Theta^{\cal A}_5$. This is expected from the decomposition \eqref{eq:nullspaceY5}, because 
\begin{equation}
    \overline{\Theta}_{5}^{\cal A}\cdot \vec{x}=0 \quad \Leftrightarrow \quad  \mathbf{Y}_5^{\scriptscriptstyle \rm T}\cdot \overline{\Theta}_5^{\cal A} \cdot  \mathbf{Y}_5\cdot \vec{x}=0 \quad \Leftrightarrow \quad  \mathbf{Y}_5\cdot \vec{x}=0\,,
\end{equation}
where the last identity is based on the fact that $\overline{\Theta}_{5}^{\cal A}$ is full ranked.

In general, for $n$-point form factors, we have the following decomposition for the propagator matrix 
\begin{equation}\label{eq:nptmFDecomp2}
	\text{Res}_{s_{1i_1.. i_r}=0}[\Theta_n^{\mathcal{F}}]\sim (\mathbf{Y}_{r+2}^{\scriptscriptstyle \rm T}\cdot \overline{\Theta}_{r+2}^{\mathcal{A}}\cdot \mathbf{Y}_{r+2})\otimes \Theta_{n-r}^{\mathcal{F}}=( \mathbb{I}_{n-r} \otimes \mathbf{Y}_{r+2}^{\scriptscriptstyle \rm T})\cdot(\overline{\Theta}_{r+2}^{\mathcal{A}}\otimes \Theta_{n-r}^{\mathcal{F}})\cdot (\mathbf{Y}_{r+2} \otimes  \mathbb{I}_{n-r}),
\end{equation}
where the ``$\sim$" means to omit the zero elements when taking the residue and $\mathbb{I}_{n-r}$ is a $(n{-}r{-}2)!$ by $(n{-}r{-}2)!$  identity matrix. 
Moreover, (a) $\overline{\Theta}_{r+2}^{\mathcal{A}}$ is the reduced amplitudes propagator matrix, which is a $(r-1)!$ by $(r-1)!$ minor of the $r!$ by $r!$ matrix ${\Theta}_{r+2}^{\mathcal{A}}$, and contains essentially all the propagator information needed for double copy; (b) $\mathbf{Y}_{r+2}$ is a $(r{-}1)!$ by  $r!$ matrix generalizing the $\mathbf{Y}_5$ above. We give more details of the decomposition in Appendix~\ref{ap:physical}. 

We also comment that the same factorization property \eqref{eq:nptmFDecomp1} also holds for the reduced propagator matrix for amplitudes. For $n$-point amplitudes, the $(n-3)!$ by $(n-3)!$ reduced propagator matrix $\overline{\Theta}_n^{\cal A}$ factorize as\footnote{Technically, the identity matrix $\overline{\mathbb{I}}_{n-r}$ here is a little bit different from the $\mathbb{I}_{n-r}$ above. We will clarify the difference in Appendix~\ref{ap:physical}. }
\begin{equation}\label{eq:nptmADecomp}
	\text{Res}_{s_{1i_1.. i_r}=0}[\overline{\Theta}_n^{\mathcal{A}}]\sim (\mathbf{Y}_{r+2}^{\scriptscriptstyle \rm T}\cdot \overline{\Theta}_{r+2}^{\mathcal{A}}\cdot \mathbf{Y}_{r+2})\otimes \overline{\Theta}_{n-r}^{\mathcal{A}}=
 (\overline{\mathbb{I}}_{n-r} \otimes \mathbf{Y}_{r+2}^{\scriptscriptstyle \rm T})\cdot(\overline{\Theta}_{r+2}^{\mathcal{A}}\otimes \overline{\Theta}_{n-r}^{\mathcal{A}})\cdot (\mathbf{Y}_{r+2} \otimes  \overline{\mathbb{I}}_{n-r}).
\end{equation}

Finally, although we will not dig into detail about the matrix elements of $\mathbf{Y}$, we comment that they are rational functions of Mandelstam variables with no overall mass dimension. \eqref{eq:nptmFDecomp2} and \eqref{eq:nptmADecomp} are also saying that we can use a unified matrix decomposition picture to deal with all appearing kernels/matrices in the form factor double copy.\footnote{Practially, the $\mathbf{Y}$ matrix, at least at this moment, seems to be nothing but alternative bookkeeping of BCJ relations.
Later on, we will show that these vectors can also be used to "factorize" master numerators, parallel to the hidden factorization relations which will be our next focus.}

\vspace{3pt}

As a summary of this subsection, we show that there exists a factorization form of the kernel matrices $\mathbf{S}_n^{\cal F}$ as well as  $\Theta^{\cal F}_n$ involved in form factor double copy. Such a decomposition will play an important role in the later discussions on factorizations.

\subsection{Hidden factorization relations of ${\cal F}$}\label{ssec:Hiddenfac} 

A remarkable fact is that the $\vec{v}$ vectors not only induce the matrix decompositions above but also lead to intriguing factorization relations for gauge-theory form factors. 
Concretely, when taking the inner product of $\vec{v}\cdot \vec{\cal F}$ and evaluating it on the special kinematics $\QQ_m^2=q^2$, one finds that it satisfies the following relation:
\begin{equation}\label{eq:nptgeneralizedBCJ}
    \vec{v}_{(\bar{\kappa},\bar{\rho})}\cdot \vec{\mathcal{F}}_n\big|_{\QQ_m^2=q^2}
    = \sum_{\alpha \in S_{n-2}} {v}_{(\bar{\kappa},\bar{\rho})}[\alpha] {\cal F}_n[\alpha]\big|_{\QQ_m^2=q^2}=\mathcal{F} _{m}[\bar{\kappa}]\times  \mathcal{A} _{m^{\prime}}[\bar{\rho}]\,,
\end{equation}
where  $\QQ_{m}=\sum_{i=1}^{m}p_i$,  $m^{\prime}=n{-}m{+}2$, and
\begin{equation}
\begin{aligned}
	&\mathcal{F} _{m}[\bar{\kappa}]= \mathcal{F} _{m}(1, \bar{\kappa}(3,\ldots,m),2),\qquad \bar{\kappa} \in S_{m-2};\\
	& \mathcal{A} _{m^{\prime}}[\bar{\rho}]= \mathcal{A} _{m^{\prime}}(\QQ_m,m{+}1, \bar{\rho}(m{+}2,\ldots,n),-q), \qquad \bar{\rho} \in S_{m'-3}\,.
\end{aligned}
\end{equation}
Note that the sub-amplitude $\mathcal{A} _{m^{\prime}}$ contains two external adjoint massive scalars with momenta $\{\QQ_m, -q\}$ and $(m'-2)$ external gluons; see \emph{e.g.}~\eqref{eq:A3exp} for the three-point case.
The notation of permutations in \eqref{eq:nptgeneralizedBCJ} is similar to that used in \eqref{eq:nptSFdecomposition} where more explanations can be found.
We stress that on the LHS of \eqref{eq:nptgeneralizedBCJ}, ${\QQ_m^2=q^2}$ is not a pole and there is no residue taken.
We refer to \eqref{eq:nptgeneralizedBCJ} as the hidden factorization relations. As we will see below, they can be also regarded as \textsl{generalized BCJ relations} for form factors.

We will start with explicit lower-point examples. Then we discuss the general $n$-point cases and also provide a proof for the MHV form factors.

\subsubsection{Explicit examples}

To explain the notations in \eqref{eq:nptgeneralizedBCJ}, we provide explicit formulas for the form factors up to five points.

\subsubsection*{1. Three-point case}

There is only one relation for the three-point case, reading 
\begin{equation}
    v^{\scriptscriptstyle A}_{3,(\mathbf{1},\mathbf{1})}[\mathbf{1}]\mathcal{F}_{3}[\mathbf{1}]\big|_{\QQ_2^2=q^2}=\tau_{31} \mathcal{F}_3(1,3,2)\big|_{\QQ_2^2=q^2}=2\epsilon_3\cdot (p_{1}{+}p_2)=\mathcal{F}_2(1,2) \times \mathcal{A}_3(\QQ_2,3,-q)\,.
\end{equation}
For the purpose of later discussion, we give further the four-dimensional MHV case, with 
\begin{equation}\label{eq:3ptinitialcond}
    \tau_{31}\mathcal{F}_{3}(1^{\phi},3^{+},2^{\phi})\big|_{\tau_{13}+\tau_{23}=0}=\langle 13 \rangle [31] \frac{\langle 12 \rangle^2 }{\langle 13 \rangle \langle 32 \rangle \langle 21 \rangle}\Big|_{\tau_{13}+\tau_{23}=0} = \frac{ \langle 2 |q|3]}{\langle 23 \rangle} ,
\end{equation}
which can be checked by a straightforward calculation. 

\subsubsection*{2. Four-point case}
At four points, there are two kinds of relations corresponding to ${\QQ_2=q^2}$ and ${\QQ_3=q^2}$, respectively. They take the following forms:
\begin{align}\label{eq:4ptinitialcond}
    & \left(v^{\scriptscriptstyle A}_{4,(\mathbf{1},\mathbf{1})}[\mathbf{1}]\mathcal{F}_{4}[\mathbf{1}] + v^{\scriptscriptstyle A}_{4,(\mathbf{1},\mathbf{1})}[(3,4)]\mathcal{F}_{4}[(3,4)]\right)_{\QQ_2=q^2} \\
    & \hskip 1.5cm =\frac{\tau_{31}\tau_{42}}{\tau_{3 \QQ_2}}\mathcal{F}_4(1,3,4,2)+\frac{\tau_{32}\tau_{41}}{\tau_{3 \QQ_2}}\mathcal{F}_4(1,4,3,2)\big|_{\QQ_2=q^2}=\mathcal{F}_2(1,2)\times \mathcal{A}_4(\QQ_2,3,4,-q)\,,   \nonumber  
\end{align} 
\begin{align}\label{eq:4ptBCJA}
    & \left(v^{\scriptscriptstyle \, B}_{4,(\mathbf{1},\mathbf{1})}[\mathbf{1}]\mathcal{F}_{4}[\mathbf{1}] + v^{\scriptscriptstyle \, B}_{4,(\mathbf{1},\mathbf{1})}[(3,4)]\mathcal{F}_{4}[(3,4)]\right)_{\QQ_3=q^2} \\
    & \hskip 1.5cm =\tau_{42}\mathcal{F}_4(1,3,4,2)+\tau_{4,(2+3)}\mathcal{F}_4(1,4,3,2)\big|_{\QQ_3=q^2}=\mathcal{F}_3(1,3,2)\times \mathcal{A}_3(\QQ_3,4,-q)\,.\nonumber 
\end{align}
Again, we can easily verify these relations for MHV form factors. In the first one, there is a four-point amplitude involved whose expression can be found in \cite{Badger:2005zh}, reading 
\begin{equation}
    \mathcal{A}_4(\QQ_2,3,4,-q)=\frac{q^2 [34]}{ \tau_{3\QQ_2}\langle 34 \rangle}\,.
\end{equation}
We also mention that as in \eqref{eq:3ptinitialcond}, the MHV version of \eqref{eq:4ptinitialcond} serves as the initial condition for the $n$-point proof shown later. 

\subsubsection*{3. Five-point case}
Then we move on to five points. Note that in the previous discussion of the matrix decomposition, we first met the case with two $\vec{v}$ vectors in the $\mathbf{V}$ matrix. The first one satisfies 
\begin{align}\label{eq:5pthiddenA}
    &\sum_{\beta \in S_{3}}  v^{\scriptscriptstyle A}_{5,(\mathbf{1},\mathbf{1})}[\beta]\mathcal{F}_{5}[\beta]\big|_{\QQ_2^2=q^2}=\frac{\tau_{31}\tau_{52}}{\tau_{5 q}}\Big(\frac{\tau_{4,(1+3)}}{\tau_{3 \QQ_2}}+1\Big) \mathcal{F}_{5}(1,3,4,5,2)+ \\
    &\quad \frac{\tau_{31}\tau_{42}\tau_{5,(1+3)}}{\tau_{3 \QQ_2}\tau_{5 q}}\mathcal{F}_{5}(1,3,5,4,2) -\frac{\tau_{32}\tau_{41}\tau_{52}}{\tau_{3 \QQ_2}\tau_{5 q}} \mathcal{F}_{5}(1,4,3,5,2) -\frac{\tau_{32}\tau_{41}\tau_{5,(2+3)}}{\tau_{3 \QQ_2}\tau_{5 q}} \mathcal{F}_{5}(1,4,5,3,2)\nonumber\\
    & \quad +\frac{\tau_{31}\tau_{42}\tau_{51}}{\tau_{3 \QQ_2}\tau_{5 q}} \mathcal{F}_{5}(1,5,3,4,2)-\frac{\tau_{32}\tau_{51}}{\tau_{5 q}}\Big(\frac{\tau_{4,(2+3)}}{\tau_{3 \QQ_2}}+1\Big) \mathcal{F}_{5}(1,5,4,3,2)\Big|_{\QQ_2^2=q^2}\nonumber \\
    &\quad = \mathcal{F}_2(1,2) \times \mathcal{A}_{5}(\QQ_2,3,4,5,-q) \,. \nonumber 
\end{align}
As for the second one, we have the $\vec{v}^{\scriptscriptstyle A}_{5,(\mathbf{1},\sigma_2)}$ giving (note that $\sigma_2$ is acting on $(4,5)$ because we fix $\{\QQ_2,3,-q\}$ in the $\mathcal{A}_5$ below)
\begin{align}
    &\sum_{\beta \in S_{3}}  v^{\scriptscriptstyle A}_{5,(\mathbf{1},\sigma_2)}[\beta]\mathcal{F}_{5}[\beta]\big|_{\QQ_2^2=q^2}=\frac{\tau_{31}\tau_{52}\tau_{4,(1+3)}}{\tau_{3 \QQ_2}\tau_{4 q}} \mathcal{F}_{5}(1,3,4,5,2)+\\
    &\quad  \frac{\tau_{31}\tau_{42}}{\tau_{4 q}}\Big(\frac{\tau_{5,(1+3)}}{\tau_{3 \QQ_2}}+1\Big)\mathcal{F}_{5}(1,3,5,4,2) +\frac{\tau_{31}\tau_{41}\tau_{52}}{\tau_{3 \QQ_2}\tau_{5 q}} \mathcal{F}_{5}(1,4,3,5,2) -\frac{\tau_{32}\tau_{41}}{\tau_{4 q}}\Big(\frac{\tau_{5,(2+3)}}{\tau_{3 \QQ_2}}+1\Big) \mathcal{F}_{5}(1,4,5,3,2)\nonumber\\
    & \quad -\frac{\tau_{32}\tau_{42}\tau_{51}}{\tau_{3 \QQ_2}\tau_{4 q}} \mathcal{F}_{5}(1,5,3,4,2) -\frac{\tau_{32}\tau_{51}\tau_{4,(2+3)}}{\tau_{3 \QQ_2}\tau_{4 q}} \mathcal{F}_{5}(1,5,4,3,2) \Big|_{\QQ_2^2=q^2}\nonumber \\
    &\quad = \mathcal{F}_2(1,2) \times \mathcal{A}_{5}(\QQ_2,3,5,4,-q) \,. \nonumber 
\end{align}

We comment that the above two $\vec{v}$ vectors are related to each other. 
By comparing the vector elements explicitly listed in the above two equations, one can find that, \emph{e.g.}
\begin{equation}\label{eq:5ptBCJ}
    v_{(\mathbf{1},\sigma_2)}[(4,5)]=v_{(\mathbf{1},\mathbf{1})}[\mathbf{1}]\big|_{4\leftrightarrow 5},\quad  v_{(\mathbf{1},\sigma_2)}[(3,4,5)]=v_{(\mathbf{1},\mathbf{1})}[(3,4)]\big|_{4\leftrightarrow 5}\,. 
\end{equation}
These are the prototype of the permutation covariance of the $\vec{v}$ vectors, which will be explained in detail in \cite{treepaper2}. 

Moreover, we have more hidden factorization relations 
\begin{align}\label{eq:5pthiddenB}
    &\sum_{\beta \in S_{3}}  v^{\scriptscriptstyle \, B}_{5,(\mathbf{1},\mathbf{1})}[\beta]\mathcal{F}_{5}[\beta]\big|_{\QQ_3^2=q^2}=\\
    &\quad \frac{\tau_{4,(1+3)}\tau_{52}}{\tau_{4\QQ_3}} \mathcal{F}_{5}(1,3,4,5,2)+ \frac{\tau_{5,(1+3)}\tau_{42}}{\tau_{4\QQ_3}} \mathcal{F}_{5}(1,3,5,4,2) + \frac{\tau_{41}\tau_{52}}{\tau_{4\QQ_3}} \mathcal{F}_{5}(1,4,3,5,2) \nonumber\\
    & \quad + \frac{\tau_{41}\tau_{5,(2+3)}}{\tau_{4\QQ_3}} \mathcal{F}_{5}(1,4,5,3,2)+\frac{\tau_{42}\tau_{51}}{\tau_{4\QQ_3}} \mathcal{F}_{5}(1,5,3,4,2) + \frac{\tau_{4,(2+3)}\tau_{51}}{\tau_{4\QQ_3}} \mathcal{F}_{5}(1,5,4,3,2)\Big|_{\QQ_3^2=q^2}\nonumber \\
    &\quad = \mathcal{F}_3(1,3,2) \times \mathcal{A}_{4}(\QQ_3,4,5,-q) \,, \nonumber 
\end{align}
and 
\begin{align}\label{eq:5pthiddenC}
    &\sum_{\beta \in S_{3}}  v^{\scriptscriptstyle \, C}_{5,(\mathbf{1},\mathbf{1})}[\beta]\mathcal{F}_{5}[\beta]\big|_{\QQ_4^2=q^2} \\
    &\quad =\tau_{5 2} \mathcal{F}_{5}(1,3,4,5,2)+ \tau_{5,(2+4)}\mathcal{F}_{5}(1,3,5,4,2) + \tau_{5,(2+3+4)} \mathcal{F}_{5}(1,5,3,4,2) \big|_{\QQ_4^2=q^2}\nonumber\\
    &\quad  = \mathcal{F}_4(1,3,4,2) \times \mathcal{A}_{3}(\QQ_4,5,-q) \nonumber \,.
\end{align}

\subsubsection{General $n$-point relations and a proof for the MHV form factors}

The $n$-point generalization is given in \eqref{eq:nptgeneralizedBCJ}, which we reproduce here:
\begin{equation}\label{eq:npthidden}
\sum_{\alpha \in S_{n-2}} v_{n,(\bar{\kappa},\bar{\rho})}[\alpha] \mathcal{F}_{n}[\alpha] \big|_{\QQ_m^2=q^2}=\mathcal{F}_{m}[\bar{\kappa}]\times \mathcal{A}_{m'}[\bar{\rho}]\,.
\end{equation}
As we have seen in previous examples, the complexity of $\vec{v}$ vectors is related to the number of gluons in the factorized amplitude $\mathcal{A}(\QQ_m,\text{gluons},-q)$.

The simplest case is  when $m=n{-}1$, and the hidden factorization relation takes the following form:
\begin{align}\label{eq:nptbcjN}
	&\mathcal{F}_{n-1}(1,3,\ldots,n-1,2)\times \mathcal{A}_3(\QQ_{n-1},n,-q)\\
	&\quad = \bigg[ \tau_{n 2}\mathcal{F}_{n}(1,3,\ldots,n,2)+\sum_{i=3}^{n-1}\tau_{n,(2+i + \cdots + (n-1))}\mathcal{F}_{n}{\small (1,3\ldots,i-1, n,i\ldots,n-1,2)} \bigg] \bigg|_{\QQ_{n-1}^2=q^2}\nonumber ,
\end{align}
where $\QQ_{n-1}=\sum_{i=1}^{n-1}p_i$. This is the $n$-point generalization of \eqref{eq:4ptBCJA} and \eqref{eq:5ptBCJ}. One may notice that the linear combination of form factors on the RHS takes the same form as the BCJ relations for amplitudes \cite{Bern:2008qj}. Here, a special kinematic condition $\QQ_{n-1}^2{=}q^2$ is imposed, which plays a similar role as the momentum conservation $\sum {p_i}=0$ in amplitudes. Unlike the amplitude cases, we have a non-trivial contribution on the LHS, which takes a factorized form as the product of a lower-point form factor and a sub-amplitude.
The $\vec{v}$ vectors corresponding to \eqref{eq:nptbcjN} can be given as follows:
\begin{equation}
\begin{aligned}
    &v[\alpha]=0 \quad \text{\bf  unless } \beta \text{ contains the subordering  } 3,4,\ldots, (n{-}1);\\
    &v[\alpha]=\tau_{n,\Xi_{R}(n;\alpha)}\equiv\tau_{n,(2+i + \cdots + (n-1))} \text{ with } \alpha=\{ 1,3\ldots,i{-}1, n,i\ldots,n{-}1,2\}\,,
\end{aligned}
\end{equation}
where we have used the notation $\Xi_{R}(n;\alpha)$ to represent the sum over momenta on the right side of $n$ in the ordering $\alpha$.

Below we will give a proof of \eqref{eq:npthidden} by considering the MHV form factors of $\operatorname{tr}(\phi^2)$:
\begin{equation}
	{\cal F}_{n}(1^{\phi},\sigma(3^{+},\ldots,n^{+}),2^{\phi})=\frac{\langle 12\rangle^2}{\langle 1 \sigma(3)\rangle \cdots \langle \sigma(n) 2\rangle \langle 21 \rangle}\,.
\end{equation}
We will discuss the two cases in detail, which are the relations for the $\QQ_{n-1}^2{=}q^2$ and $\QQ_{n-2}^2{=}q^2$ channels.

As an outline of the strategy of the proof, we will employ the BCFW recursive method and give a proof based on induction. We first perform a standard BCFW shift. 
Then we show that after the shift, both sides have the same poles and the same residues on the $z$ plane. 

\subsubsection*{The $\QQ_{n-1}^2{=}q^2$ case}

We first proof \eqref{eq:nptbcjN}. We perform the $\langle 2 1]$-shift: 
\begin{equation}
|\hat{2}\rangle=|2\rangle-z |1\rangle \,, \qquad |\hat{1}]=|1]+z |2] \,.
\end{equation}

On the LHS, it is clear that only $\mathcal{F}_{n-1}(1,3,\ldots,n{-}1,2)$ is affected by the shift. We define   
\begin{equation}
	E_{L}(z)=\frac{1}{z}{{\cal F}}_{n-1}({\hat 1},3,\ldots,n{-}1,{\hat 2})\mathcal{A}_{3}(\QQ_{n-1},n,-q)\,,
\end{equation}
so that ${\rm Res}_{z=0}[E_{L}(z)]$ gives the LHS of \eqref{eq:nptbcjN}. Apart from $z=0$, $E_{L}$ have only one other pole on the complex plane $z_{P}=\langle 2 (n{-}1) \rangle/\langle 1 (n{-}1) \rangle$, on which the MHV form factors factorize as (here $P=p_2{+}p_{n-1}$)
\begin{equation}\label{eq:FLzp}
	{\rm Res}_{z=z_{P}}[E_{L}(z)]=-{\cal F}_{n-2}({\hat 1},3,\ldots,\hat{P})\frac{1}{s_{2(n-1)}}{\cal A}_3({\hat 2},n{-}1,-\hat{P}) \mathcal{A}_{3}(\QQ_{n-1},n,-q) \Big|_{z=z_{P}}\,.
\end{equation}
Note that there is no pole at infinity on the LHS. 

For the RHS of \eqref{eq:nptbcjN}, we define $E_{R}$ similarly as  
\begin{equation}\label{eq:nptbcjN-BB}
	E_{R}(z)=\frac{1}{z}\Big[\tau_{n {\hat 2}}\mathcal{F}_{n}({\hat 1},3,\ldots,n, {\hat 2})+\sum_{i=3}^{n{-}1}\tau_{n,({\hat 2}+i+\cdots+(n{-}1))}\mathcal{F}_{n}{\small ({\hat 1},3,\ldots,i{-}1,n,i\ldots,n{-}1, {\hat 2})}\Big] \Big|_{\QQ_{n-1}^2=q^2}\,.
\end{equation}
We examine the possible poles of $E_{R}(z)$ as follows.
\begin{itemize}[itemsep=3pt]
\item 
First, the condition $\QQ_{n-1}^2{=}q^2$ is not spoiled by the shift since $\hat{p}_1{+}\hat{p}_2=p_1{+}p_2$. 
\item
Next, for the first form factor $\mathcal{F}_{n}({\hat 1},3,\ldots,n, {\hat 2})$, one may expect a pole $\langle 2 n\rangle /\langle 1 n\rangle$; however, it is canceled by the $\tau_{n\hat{2}} = \langle 2 n \rangle [n 2] - z \langle 1 n \rangle [n 2]  $ factor. 
For the other form factors in the sum, only $z_{P}$ pole appears.
\item 
We also need to be careful about the pole at infinity. Although the MHV form factors themselves do not contribute to the pole at infinity, the $\tau$ factors do. 
One can compute the corresponding residue  for each term in $E_{R}(z)$ as 
\begin{equation}\label{eq:npt1ginfinitypole}
	{\rm Res}_{z=\infty}\big[z^{-1}\tau_{n,(\hat{2}+\cdots)}\mathcal{F}_n({\hat 1},\sigma\{3,\ldots,n\},{\hat 2})\big]=\frac{\langle 1 n \rangle [ n 2 ] \langle 2 1 \rangle}{\langle 1 \sigma(3) \rangle \cdots \langle\sigma(n) 1 \rangle }\,,
\end{equation}
where $\sigma\in S_{n-2}$ can be an arbitrary permutation. Nicely, it turns out that the sum of the residues actually vanishes: 
\begin{equation}
{\rm Res}_{z=\infty}[E_{R}(z)]= \sum_{i=3}^{n} \frac{\langle 1 n \rangle [ n 2 ] \langle 2 1 \rangle}{\langle 1 3 \rangle \cdots \langle (i{-}1)  n \rangle \langle n i \rangle \cdots \langle (n{-}1) 1 \rangle }=0\,,
\end{equation}
which is equivalent to an $(n{-}1)$-point U(1) decoupling relation (as a special case of the KK relation\cite{Kleiss:1988ne}).

\item
Therefore, we conclude that  $E_{R}$ have also only the $z_{P}$ pole (apart from the $z=0$ one). The residue is 
\begin{equation}\label{eq:FRzp}
\begin{aligned}
&	{\rm Res}_{z=z_{P}}[E_{R}(z)]\\
	&=-\sum_{i=3}^{n-1} { \tau_{n,(\hat{P}+i+\cdots+(n-2))} \over s_{2(n-1) } }
	{\mathcal{F}}_{n-1}({\hat1},3, \ldots,i{-}1,n,i\ldots,{\hat P})
	{\cal A}_3({\hat 2},n{-}1,-\hat{P})\Big|_{\QQ_{n-1}^2=q^2}^{z=z_{P}}\,.
\end{aligned}
\end{equation}
\end{itemize}

Comparing \eqref{eq:FLzp} and \eqref{eq:FRzp}, one can see that ${\rm Res}_{z=z_{P}}[E_{L}(z)]={\rm Res}_{z=z_{P}}[E_{R}(z)]$ 
by using a $(n-1)$-point relation \eqref{eq:nptbcjN}, and the residue theorem guarantees that ${\rm Res}_{z=0}[E_{L}(z)]={\rm Res}_{z=0}[E_{R}(z)]$, so that  \eqref{eq:nptbcjN} is valid for the $n$-point case.\footnote{The reader may wonder that where are we using the condition $\QQ_{n-1}^2=q^2$ condition. The answer is that it is used when confirming \eqref{eq:nptbcjN} is true for the minimal $n{=3}$ case, which is the initial condition of the inductive proof and is given in \eqref{eq:3ptinitialcond}.}

\subsubsection*{The $\QQ_{n-2}^2{=}q^2$ case}

Next we consider the $\QQ_{n-2}^2=q^2$ case. 
In this case, we will see richer structures that are not encountered in usual amplitude cases. 

The hidden factorization relation can be given in the following form: 
\begin{align}\label{eq:npthidden2g}
    \mathcal{F}_{n-2}(1,\bar{\beta}_0,2)  \times &\mathcal{A}_{4}(\QQ_{n-2},n{-}1,n,-q)=\\
    &\sum_{\beta \in \bar{\beta}_0 \shuffle \{n-1,n\}} v[1,\beta,2]\mathcal{F}(1,\beta,2)+  \hskip -9pt \sum_{\beta' \in \bar{\beta}_0 \shuffle \{n,n-1\}} v[1,\beta',2]\mathcal{F}(1,\beta',2) \big|_{\QQ_{n-2}=q^2} \nonumber  \,,
\end{align}
where $\bar{\beta}_0=\{3,\ldots,(n{-}2)\}$ and the $\vec{v}$ is 
\begin{equation}
\begin{aligned}
    &v[1,\beta,2]=\frac{\tau_{n-1,\Xi_{L}(n-1;\alpha)}\tau_{n,\Xi_{R}(n;\alpha)}}{\tau_{n-1 \QQ_{n-2}}} \textbf{ if } \beta \in \bar{\beta}_0 \shuffle \{n{-}1,n\} \,, \\
    &v[1,\beta,2]=\frac{\tau_{n,\Xi_{L}(n-1;\alpha)}\tau_{n-1,\Xi_{R}(n;\alpha)}}{\tau_{n-1 \QQ_{n-2}}} \textbf{ if } \beta \in \bar{\beta}_0 \shuffle \{n,n{-}1\} \,,\\
    &v[1,\beta,2]=0 \textbf{ otherwise}. 
\end{aligned}
\end{equation}
The four- and five-point examples of \eqref{eq:npthidden2g} are already given above, and the reader can write down explicitly six-point relations based on the $\vec{v}$ vectors given in Appendix~\ref{app:vectors}. 
Below we prove \eqref{eq:npthidden2g} for the MHV form factors following the same approach as above for \eqref{eq:nptbcjN}. 

On the LHS of \eqref{eq:npthidden2g}, we still have the only pole $z_{P}=\langle 2 (n{-}2) \rangle/\langle 1 (n{-}2) \rangle$, and the MHV form factor factorizes as
\begin{equation}
    \text{Res}_{z=z_{P}}[\mathcal{F}_{n-2}(\hat{1},3,\ldots,n{-}2,2)]=-\mathcal{F}_{n-3}(\hat{1},3,\ldots,\hat{P}) \frac{1}{s_{2(n-2)}}\mathcal{A}_3(\hat{2},n{-}2,-\hat{P})\Big|_{z=z_{P}}\,,
\end{equation}
similar to that in \eqref{eq:FLzp}. There is no pole at infinity. 

As for the RHS, we first argue that there is only the $z_{P}$ pole for finite $z$. When inspecting the sum on the RHS of \eqref{eq:npthidden2g}, we see that: 
1) the vector element of $\vec{v}$ has no finite pole because its denominator does not have $z$ dependence after the shift,
and 2) in principle we can have $\mathcal{F}(\ldots,n,2)$ with the  $z{=}\langle 2n \rangle/\langle 1n \rangle$ pole, but it is cancelled by the $\tau_{\hat{2}n}$ factor in the vector element of $\vec{v}$. The argument is exactly the same as the one above in \eqref{eq:npt1ginfinitypole}. 

The more tricky part in this case is to argue the absence of the pole at infinity for the sum on the RHS of \eqref{eq:npthidden2g}. 
In other words, we need to show that the RHS of \eqref{eq:npthidden2g} goes to zero when $z\rightarrow \infty$, or more concretely,  we need to prove that everything scales as $z^{1}$ or $z^{0}$ at large $z$ has a zero coefficient. From the concrete expression of $\mathcal{F}$ and $v$, we know 
\begin{equation}
\begin{aligned}
    & v[\hat{1},\beta,\hat{2}]=z^2 v^{(2)}[1,\beta,2] + z^1 v^{(1)}[1,\beta,2] + \cdots ,\\
    & \mathcal{F}(\hat{1},\beta,\hat{2})=\frac{1}{z}\mathcal{F}^{(1)}(1,\beta,2)+\frac{1}{z^2}\mathcal{F}^{(2)}(1,\beta,2)+ \cdots \,,
\end{aligned}
\end{equation}
and our job is to show that
\begin{equation}\label{eq:nopoleinfty0}
    \sum_{\beta} v[\hat{1},\beta,\hat{2}] \mathcal{F}(\hat{1},\beta,\hat{2})\sim O(z^{-1}) \,.
\end{equation}
Notice that 
\begin{equation}
    {v}^{(2)}[1,\beta,2]=-\frac{\langle 1 |n |  2] \langle 1 |n{-}1|2] }{\tau_{n-1 \QQ_{n-2}}} \quad \text{ for } \forall \beta\,.
\end{equation}
This allows us to do the following operation
\begin{equation}
     \sum_{\beta} z^2 v^{(2)}[1,\beta,2]\mathcal{F}(\hat{1},\beta,\hat{2})=-\frac{\langle 1 |n |  2] \langle 1 |n{-}1|2] }{\tau_{n-1 \QQ_{n-2}}} \sum_{\beta} \mathcal{F}(\hat{1},\beta,\hat{2})\,.
\end{equation}
Interestingly, the sum over $\beta$ runs from $\beta \in \beta_0 \shuffle \{n{-}1,n\}$ or $\beta \in \beta_0 \shuffle \{n,n{-}1\}$, so that the sum is actually a KK relation. In the end, we have 
\begin{align}
     \sum_{\beta} z^2 v^{(2)}[1,\beta,2]&\mathcal{F}(\hat{1},\beta,\hat{2})=\\
     & -z^2\frac{\langle 1 |n |  2] \langle 1 |n{-}1|2] }{\tau_{n-1 \QQ_{n-2}}}  \left(\mathcal{F}(\hat{1},\beta_0,\hat{2},n{-}1,n) + \mathcal{F}(\hat{1},\beta_0,\hat{2},n,n{-}1)\right)\,. \nonumber
\end{align}
Plugging in the explicit formula for these two $\mathcal{F}$ and take $z \rightarrow  \infty$, we get
\begin{align}
    -z^2\frac{\langle 1 |n |  2] \langle 1 |n{-}1|2] }{\tau_{n-1 \QQ_{n-2}}} \bigg( & \frac{\langle 12 \rangle^2}{\langle 1 3 \rangle \cdots \langle (n{-}2)\hat{2} \rangle \langle  \hat{2} (n{-}1) \rangle  \langle (n{-}1)n \rangle \langle n 1 \rangle } \nonumber\\
    &+ \frac{\langle 12 \rangle^2}{\langle 1 3 \rangle \cdots \langle (n{-}2)\hat{2} \rangle \langle  \hat{2} n \rangle  \langle n(n{-}1) \rangle \langle (n{-}1) 1 \rangle } \bigg)\nonumber\\
    =-z^2\frac{\langle 1 |n |  2] \langle 1 |n{-}1|2] }{\tau_{n-1 \QQ_{n-2}}} 
 & \frac{\langle 12 \rangle^3}{\langle 1 3 \rangle \cdots \langle (n{-}2)\hat{2} \rangle  \langle  \hat{2} (n{-}1) \rangle   \langle  \hat{2} n \rangle \langle n 1 \rangle \langle (n{-}1) 1 \rangle } \,, \label{eq:nopoleinfty1}
\end{align}
in which we have used the Schouten rule to simplify the result. Clearly, we have $z^3$ on the denominator and $z^2$ on the numerator so \eqref{eq:nopoleinfty1} scales as $z^{-1}$ for large $z$. 

Then to verify \eqref{eq:nopoleinfty0} it suffices to discuss why the following identity holds 
\begin{equation}\label{eq:nopoleinfty2}
    \sum_{\beta } z^1 v^{(1)}[1,\beta,2] \frac{1}{z}\mathcal{F}^{(1)}({1},\beta,{2})=0\,,
\end{equation}
with 
\begin{equation}
\begin{aligned}
   &v^{(1)}[1,\beta,2]=\frac{\langle 1 | n{-}1|2] \tau_{n,\Xi_{R}(n;\alpha)}-\tau_{n-1,\Xi_{L}(n-1;\alpha)}\langle 1 | n|2]}{\tau_{n-1 \QQ_{n-2}}} \textbf{ if } \beta \in \bar{\beta}_0 \shuffle \{n{-}1,n\} \,, \\
   &v^{(1)}[1,\beta,2]=\frac{-\langle 1 | n|2] \tau_{n-1,\Xi_{R}(n-1;\alpha)}+\tau_{n,\Xi_{L}(n;\alpha)}\langle 1 | n{-}1|2]}{\tau_{n-1 \QQ_{n-2}}} \textbf{ if } \beta \in \bar{\beta}_0 \shuffle \{n,n{-}1\} \,.
\end{aligned}
\end{equation}
This can be directly verified by plugging in the explicit expression for both $v$ and $\cal{F}$. Similar to the discussion in \eqref{eq:npt1ginfinitypole}, the U(1) decoupling relation is required for  \eqref{eq:nopoleinfty2}. 

Thus, we have proved that there is no pole at infinity, and both sides of \eqref{eq:npthidden2g} have the same pole and residue for finite $z$. This concludes our proof.%
\footnote{Based on the similarity between the proofs for the two classes of MHV form factors, we see that it is possible to finish the proof for all MHV form factors. The concrete analysis, however, requires a closed formula for all the $\vec{v}$ vectors given in \cite{treepaper2}.} 

\vspace{5pt}

Before ending this subsection, we give some remarks on the hidden factorization relations. 
First, we emphasize again that \eqref{eq:nptgeneralizedBCJ}  both sides are finite and no residues are involved. Second, the interesting point of \eqref{eq:nptgeneralizedBCJ} is that even without taking residues, a special linear combination of ordered form factors can still factorize, with the price of considering special kinematics like $\QQ_m^2=q^2$. Third, to determine or check the $\vec{v}$ vectors, working with the MHV form factors is a convenient choice. 
Finally, we have also checked the relations using form factors in general Lorentz kinematics up to 8 points.
We will discuss more thoroughly the properties of the $\vec{v}$ vectors and also give the closed formulae for them in \cite{treepaper2}.

\subsection{Factorizations of double-copy form factors}\label{ssec:FFDCnptfac}

The discussions in the previous subsections reflect many interesting aspects of the form factor double copy. One reason for studying these properties is to understand the factorization of the form factor double copy. We have two types of poles: the ``physical"-type poles and the ``spurious"-type poles. 
In this subsection, we will prove that the form factor double copy has desired factorization properties on both types of poles, which confirms that the form factor double copy is indeed a physically meaningful quantity. 

\subsubsection*{Factorizations on ``spurious"-type poles}
Now we prove that, given the matrix decomposition in Section~\ref{ssec:bilinearDecomp} and the hidden factorization relation in Section~\ref{ssec:Hiddenfac}, the factorizations on spurious poles are transparent. 

\begin{figure}
    \centering
    \includegraphics[width=0.7\linewidth]{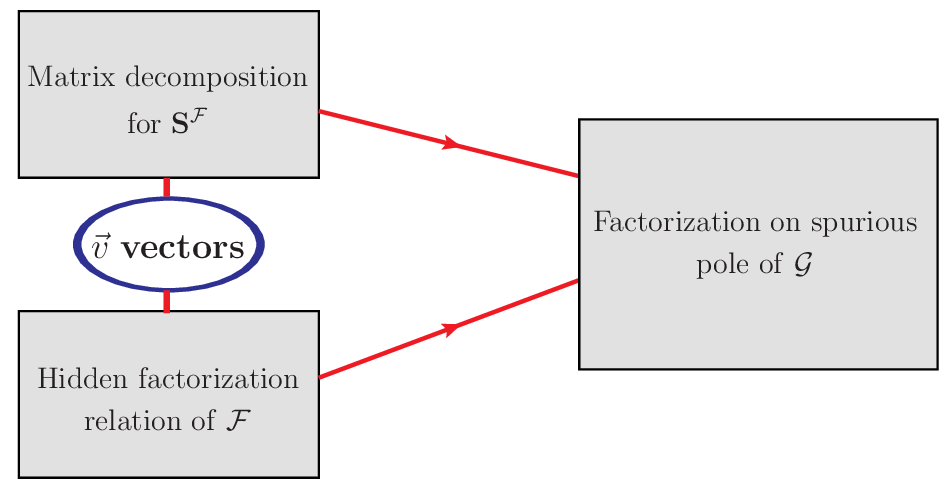}
    \caption{Relations between the factorizations on the ``spurious"-type poles involved in the form factor double copy. The matrix decomposition is the ``factorization" of the kernel, while the hidden factorization relation is the ``square root" of the factorization of the form factor double copy.  The $\vec{v}$ vectors play the role of a bridge. The factorization on the ``spurious"-type poles implies new Feynman diagrams with massive scalar propagators.}
    \label{fig:structure}
\end{figure}

The key observation is that the $\vec{v}$ vectors play the role of a bridge connecting the matrix decomposition and the hidden factorization relation: suppose the $\vec{v}$ vectors are known, then the same $\vec{v}$ can be used in both the matrix decomposition and the hidden factorization relation. 
Combining these two relations will give us the factorization of the form factor double copy on the ``spurious"-type pole, see Figure~\ref{fig:structure}.

Some examples have been given in Section~\ref{sec:34ptexample} such as \eqref{eq:4ptFactorization}. Here we provide two further explicit examples: 
\begin{enumerate}[topsep=3pt,itemsep=-1ex,partopsep=1ex,parsep=1ex]
    \item[(1)] The first is the four-point double copy on the $\QQ_2^2=q^2$ pole. 
    The ingredients that we need are already stated in the previous sections: the matrix decomposition hidden in \eqref{eq:4ptmatdecpA} and the factorization relation in \eqref{eq:4ptinitialcond}. 

    With the double copy representation in \eqref{eq:genericKLT}, we give the following detailed derivation
    \begin{align}
        \text{Res}_{\QQ_2^2=q^2}[{\cal G}_4]=&\sum_{\beta_{1,2}\in S_2}{\cal F}_4[\beta_1] \text{Res}_{\QQ_2^2=q^2}\left[\mathbf{S}_{4}^{\cal F}[\beta_1|\beta_2]\right] {\cal F}_4[\beta_2] |_{\QQ_2^2=q^2}\\
        =&\sum_{\beta_{1,2}\in S_2}{\cal F}_4[\beta_1] 
        \Big(v^{\scriptscriptstyle A}_{4,(\mathbf{1},\mathbf{1})}[\beta_1]\mathbf{S}_{2}^{\cal F}[\mathbf{1}|\mathbf{1}]\mathbf{S}_{4}^{\cal A}[\mathbf{1}|\mathbf{1}]  v^{\scriptscriptstyle A}_{4,(\mathbf{1},\mathbf{1})}[\beta_2] \Big) {\cal F}_4[\beta_2] |_{\QQ_2^2=q^2} \nonumber \\
        =& \Big(\sum_{\beta \in S_2} v^{\scriptscriptstyle A}_{4,(\mathbf{1},\mathbf{1})}[\beta] {\cal F}_4[\beta]|_{\QQ_2^2=q^2}\Big)^2 \mathbf{S}_{2}^{\cal F}[\mathbf{1}|\mathbf{1}]\mathbf{S}_{4}^{\cal A}[\mathbf{1}|\mathbf{1}] \nonumber \\
        =& \big[ \mathcal{F}_2(1,2) \mathcal{A}_4(\QQ_2,3,4,-q) \big]^2 \mathbf{S}_{2}^{\cal F}[\mathbf{1}|\mathbf{1}]\mathbf{S}_{4}^{\cal A}[\mathbf{1}|\mathbf{1}]  \nonumber\\
        =& \left[ \mathcal{F}_2(1,2)\right]^2 \mathbf{S}_{2}^{\cal F}(1,2|1,2)\times \left[ \mathcal{A}_4(\QQ_2,3,4,-q)\right]^2 \mathbf{S}_{4}^{\cal A}(\QQ_2,3,4,-q|\QQ_2,3,4,-q)\nonumber\\
        =& \mathcal{G}_2(1,2) \times \mathcal{M}_4(\QQ_2,3,4,-q)\,, \nonumber
    \end{align}    
which shows the factorization of $\mathcal{G}_4$ on the $\QQ_2^2= q^2$ pole. 
From this factorization, we can see that the hidden factorization relation can be understood as the ``square root" of the $\mathcal{G}$ factorization. 

\item[(2)] As another example, we consider the five-point case on the $\QQ^2_2=q^2$ pole. We need the matrix factorization \eqref{eq:5ptmatdecpA} and the hidden factorization relation \eqref{eq:5pthiddenA}. Inspecting $\text{Res}_{\QQ_2^2=q^2}[\mathcal{G}_5]$, we get
\begin{align}
    \text{Res}_{\QQ_2^2=q^2}[{\cal G}_5]&=\sum_{\beta_{1,2}\in S_3} {\cal F}_5[\beta_1] \text{Res}_{\QQ_2^2=q^2}\left[\mathbf{S}_{5}^{\cal F}[\beta_1|\beta_2]\right] {\cal F}_5[\beta_2] |_{\QQ_2^2=q^2} \\
    &=\sum_{\beta_{1,2}\in S_3} {\cal F}_5[\beta_1] \Big( \sum_{\bar{\rho}_{1,2}\in S_{2} } { v}^{A}_{5,(\mathbf{1},\bar\rho_1)}[\beta_1]\mathbf{S}^{\mathcal{F}}_{2}[\mathbf{1}|\mathbf{1}]  \mathbf{S}^{\mathcal{A}}_{5}[\bar{\rho}_1|\bar{\rho}_2] { v}^{A}_{5,(\mathbf{1},\bar\rho_2)}[\beta_2] \Big) {\cal F}_5[\beta_2] |_{\QQ_2^2=q^2} \nonumber \\
    &=(\mathcal{F}_{2}[\mathbf{1}])^2 \mathbf{S}^{\mathcal{F}}_{2}[\mathbf{1}|\mathbf{1}] \times \sum_{\bar{\rho}_{1,2}} \mathcal{A}_5[\bar{\rho}_{1}]\mathbf{S}^{\mathcal{A}}_{5}[\bar{\rho}_1|\bar{\rho}_2] \mathcal{A}_5[\bar{\rho}_{2}] \nonumber \\
    &=\mathcal{G}_2(1,2) \times \mathcal{M}_5(\QQ_2,3,4,5,-q) \,. \nonumber
\end{align}

\end{enumerate}

In general, to analyze the factorization on the ``spurious"-type poles, we start from the KLT double copy \eqref{eq:genericKLT}. In this equation, the ``spurious"-type poles only appear in the kernel $\mathbf{S}_{n}^{\cal F}$, of which the decomposition of the kernel has been discussed above. Next, we plug in the decomposition formula \eqref{eq:nptSFdecomposition} in the KLT matrix, giving
\begin{align}
\text{Res}_{\QQ_{m}^2=q^2}[\mathcal{G}_{n}]&=\sum_{\alpha_{1,2}\in S_{n-2}}\mathcal{F} _{n}[\alpha_1] \text{Res}_{\QQ_{m}^2=q^2}\big[\mathbf{S}_{n}^{\mathcal{F}}[\alpha_1|\alpha_2] \big] \mathcal{F} _{n}[\alpha_2] |_{\QQ_m^2=q^2} \\
	\nonumber & \hskip -55pt = 
 \sum_{\alpha_{1,2}\in S_{n-2}}    \mathcal{F} _{n}[\alpha_1]
 \Big( \sum_{\substack{\bar{\kappa}_{1,2}\in S_{m-2}  \\  \bar{\rho}_{1,2}\in S_{m^{\prime}-3} }}
  v_{n,(\bar{\kappa}_1,\bar{\rho}_1)}[\alpha_1]
	\mathbf{S}^{\mathcal{F}}_{m}[\bar{\kappa}_1|\bar{\kappa}_2] \mathbf{S}^{\mathcal{A}}_{m^{\prime}}[\bar{\rho}_1|\bar{\rho}_2] 
	 v_{n,(\bar{\kappa}_2,\bar{\rho}_2)}[\alpha_2] \Big) \mathcal{F} _{n}[\alpha_2] |_{\QQ_m^2=q^2} ,
\end{align}	
in which $m^{\prime}=n{-}m{+}2$.
Then we perform the sum over $\alpha_{1,2}$, where the hidden factorization relation \eqref{eq:npthidden} appears. In the end, we reach 
\begin{align}\label{eq:Gnspuriouspole}
\text{Res}_{\QQ_{m}^2=q^2}[\mathcal{G}_{n}] &=  
\sum_{\substack{\bar{\kappa}_{1,2}\in S_{m-2}  \\  \bar{\rho}_{1,2}\in S_{m^{\prime}-3} }} \bigg(
\prod_{i=1,2} \sum_{\alpha_{i}\in S_{n-2}}
    \mathcal{F} _{n}[\alpha_i] v_{n,(\bar{\kappa}_i,\bar{\rho}_i)}[\alpha_i]\bigg)
	\mathbf{S}^{\mathcal{F}}_{m}[\bar{\kappa}_1|\bar{\kappa}_2] \mathbf{S}^{\mathcal{A}}_{m^{\prime}}[\bar{\rho}_1|\bar{\rho}_2]  \nonumber \\ 
    &=\sum_{\bar{\kappa}_{1,2}\in S_{m-2}} \mathcal{F} _{m}[\bar{\kappa}_1] \mathbf{S}^{\mathcal{F}}_{{m}}[\bar{\kappa}_1|\bar{\kappa}_2] \mathcal{F} _{m}[\bar{\kappa}_2] 
    \sum_{\bar{\rho}_{1,2}\in S_{m'-3}} \mathcal{A} _{m'}[\bar{\rho}_1] \mathbf{S}^{\mathcal{A}}_{{m'}}[\bar{\rho}_1|\bar{\rho}_2] \mathcal{A} _{m'}[\bar{\rho}_2] \nonumber \\
	&= \mathcal{G}_{m} \times \mathcal{M}_{m'} \,.
\end{align}

We give two  concluding remarks here:
\begin{enumerate}[topsep=3pt,itemsep=-1ex,partopsep=1ex,parsep=1ex]
    \item[1)] As mentioned above in Figure~\ref{fig:structure}, the factorizations on the spurious poles in gravity require the hidden factorization relations in gauge theory (as the ``square root" of the gravity factorization) and the matrix decomposition (as the ``factorization" of the KLT kernel). We now see why the same $\vec{v}$ vectors make it possible to combine these two kinds of relations.

    \item[2)] As we showed in the three- and four-point examples in Section~\ref{sec:34ptexample} (see Figure~\ref{fig:F3treeGRA} and \ref{fig:F4treeGRA}), the existence of the ``spurious"-type poles, as well as the factorizations, means that there are new Feynman diagrams with massive propagators after double copy. 
    Physically, the appearance of these new diagrams is necessary, since gravitons couple to everything including the ``operator'' leg.
    One explanation of this picture is to interpret the form factors as amplitudes involving a color-single scalar particle, see Section~\ref{sec:generalize1}.
\end{enumerate}

\subsubsection*{Factorization on ``physical"-type poles}

Similar to the discussions above, we can examine the factorization properties on the ``physical"-type  poles. 
In this case, it is more convenient to focus on the BCJ form of the double copy in \eqref{eq:genericKLT},
and the factorization is similar to that  in the ordinary amplitudes double copy, 
which makes it relatively easy to understand the ``physical"-type pole factorizations for the form factor case. Also, we will use compact matrix notations in the following discussion in this subsection and leave the detailed definitions and explanations in Appendix~\ref{ap:physical}.  

When taking the residue on the pole $s_{13\cdots r}{=}0$, as mentioned in Section~\ref{ssec:bilinearDecomp}, the propagator matrix factorizes as \eqref{eq:nptmFDecomp2}, so that (here we have $r'=n{-}r{+}2$)
\begin{equation}
\begin{aligned}\label{eq:Gnphysicalpole}
	\text{Res}_{s_{13\cdots r}=0}[\mathcal{G}_{n}] =& \text{Res}_{s_{13\cdots r}=0}\Big[\big(\vec{N}^{\cal F}_n\big)^{\scriptscriptstyle \rm T} \cdot {\Theta}_{n}^{\cal F} \cdot \vec{N}^{\cal F}_n \Big]\\
  =& \big(\vec{N}^{\cal F}_n\big)^{\scriptscriptstyle \rm T} \cdot 
	 (\mathbb{I}_{r'} \otimes \mathbf{Y}_r^{\scriptscriptstyle \rm T}) \cdot 
    (\overline{\Theta}^{\cal A}_{r}\otimes \Theta^{\cal F}_{r'})\cdot (\mathbf{Y}_r\otimes \mathbb{I}_{r'}) \cdot \vec{N}^{\cal F}_n 
   \big|_{s_{13\cdots r}=0}
 \,.
\end{aligned}
\end{equation}
One has a similar form for amplitudes, using \eqref{eq:nptmADecomp}
\begin{equation}
\begin{aligned}\label{eq:Mnphysicalpole}
	\text{Res}_{s_{13\cdots r}=0}[\mathcal{M}_{n}] &=
 \text{Res}_{s_{13\cdots r}=0}\Big[\big(\vec{N}^{\cal A}_n\big)^{\scriptscriptstyle \rm T} \cdot \overline{\Theta}_{n}^{\cal A} \cdot \vec{N}^{\cal A}_n \Big]\\
	&= \big(\vec{N}^{\cal A}_n\big)^{\scriptscriptstyle \rm T} \cdot 
	 (\overline{\mathbb{I}}_{r'} \otimes \mathbf{Y}_r^{\scriptscriptstyle \rm T}) \cdot 
    (\overline{\Theta}^{\cal A}_{r}\otimes \overline{\Theta}^{\cal A}_{r'})\cdot (\mathbf{Y}_r\otimes \overline{\mathbb{I}}_{r'}) \cdot \vec{N}^{\cal A}_n
    \big|_{s_{13\cdots r}=0}
 \,.
\end{aligned}
\end{equation}
Here we use the superscript $\mathcal{A}$ or $\mathcal{F}$ to distinguish numerators for amplitudes or form factors.

Interestingly, the $\mathbf{Y}$ matrix, similar to the $\mathbf{V}$ matrix above, also leads to a factorization relation for numerators sketched as
\begin{equation}\label{eq:Numfact}
   (\mathbf{Y}_r\otimes \mathbb{I}_{r'}) \cdot \vec{N}^{\cal F}_n \big|_{s_{13\cdots r}=0} = \vec{N}^{\cal A}_r\otimes  \vec{N}^{\cal F}_{r'},\quad
   (\mathbf{Y}_r\otimes \mathbb{I}_{r'}) \cdot \vec{N}^{\cal A}_n \big|_{s_{13\cdots r}=0} = \vec{N}^{\cal A}_r\otimes \vec{N}^{\cal A}_{r'} \,.
\end{equation}

From \eqref{eq:Gnphysicalpole} and \eqref{eq:Numfact}, the ``physical"-type pole factorization is also transparent
\begin{equation}
\begin{aligned}\label{eq:Gnphysicalpole2}
	\text{Res}_{s_{13\cdots r}=0}[\mathcal{G}_{n}] &=
	(\vec{N}^{\cal A}_r\otimes \vec{N}^{\cal F}_{r'})^{\scriptscriptstyle \rm T}\cdot 
    (\overline{\Theta}^{\cal A}_{r}\otimes \Theta^{\cal F}_{r'})\cdot (\vec{N}^{\cal A}_r\otimes \vec{N}^{\cal F}_{r'})\\
    &= \left(\big(\vec{N}^{\cal A}_r\big)^{\scriptscriptstyle \rm T} \cdot \overline{\Theta}^{\cal A}_{r} \cdot \vec{N}^{\cal A}_r \right) \times \left(\big(\vec{N}^{\cal F}_r\big)^{\scriptscriptstyle \rm T} \cdot{\Theta}^{\cal F}_{r} \cdot \vec{N}^{\cal F}_r \right) \\
    &= \mathcal{M}_r\times \mathcal{G}_{r'}\,,
\end{aligned}    
\end{equation}
which is parallel to the amplitude case as 
\begin{equation}
\begin{aligned}\label{eq:Mnphysicalpole3}
\text{Res}_{s_{13\cdots r}=0}[\mathcal{M}_{n}] &=
	(\vec{N}^{\cal A}_r\otimes \vec{N}^{\cal A}_{r'})^{\scriptscriptstyle \rm T}\cdot 
    (\overline{\Theta}^{\cal A}_{r}\otimes \overline{\Theta}^{\cal A}_{r'})\cdot (\vec{N}^{\cal A}_r\otimes \vec{N}^{\cal A}_{r'})\\
    &= \left(\big(\vec{N}^{\cal A}_r\big)^{\scriptscriptstyle \rm T} \cdot \overline{\Theta}^{\cal A}_{r} \cdot \vec{N}^{\cal A}_r \right) \times \left(\big(\vec{N}^{\cal A}_r\big)^{\scriptscriptstyle \rm T} \cdot \overline{\Theta}^{\cal A}_{r} \cdot \vec{N}^{\cal A}_r \right) \\
    &= \mathcal{M}_r\times \mathcal{M}_{r'}\,.
\end{aligned}    
\end{equation}
See more detailed discussions in Appendix~\ref{ap:physical}.

\vspace{5pt}

In conclusion, we have verified that the form factor double copy has  the desired factorization properties, which serves as the final step of confirming that the form factor double copy is physically meaningful and new Feynman diagrams are necessary. 
We comment that the same procedure for verifying factorizations is universal and applicable to other form factors, as we will see in  the generalizations below.

\section{Generalization I: Higgs amplitudes in QCD}\label{sec:generalize1}

In this section, we will take a different point of view on the form factors, by interpreting them as certain amplitudes involving a color-singlet scalar particle. 
This will make it easy to generalize to the case of Higgs amplitudes in QCD. It will also provide a picture of the double-copy form factors as gravitational amplitudes.

\subsubsection*{Form factors as Higgs Amplitudes}
Let us first consider the scalar form factors considered in previous sections. Those form factors can be interpreted as amplitudes in the following YM-scalar-Higgs (YMSH) theory
\begin{equation}\label{eq:ymshl}
{\cal L}^{\rm YMSH} = - {1\over2} {\rm tr}(F_{\mu\nu}F^{\mu\nu}) + {1\over2}  {\rm tr}(D^\mu \phi D_\mu \phi) + \lambda_{\phi} H {\rm tr}(\phi^2) + {1\over2} \partial^\mu H \partial_\mu H - {1\over2} m_H^2 H^2 \,,
\end{equation}
where $\phi = \phi^a T^a$ is a gauged scalar while $H$ is a color-singlet scalar. The map between form factors and amplitudes is 
\begin{equation}
\itbf{F}_{n} \quad \Leftrightarrow \quad \itbf{A}^{\rm YMSH}_{n+1}(1^{\phi},2^{\phi},3^{g}, \ldots, n^g, q^H) =  \langle \phi(p_1) \, \phi(p_2) \, g(p_3) \ldots g(p_n) | H(q) \rangle  \,.
\end{equation} 

Via double-copy, one has the gravitational-scalar-Higgs (GSH) theory
\begin{equation}
{\cal L}^{\rm GSH} = \sqrt{-g} \Big({1\over16\pi G} R + {1\over2} g^{\mu\nu} \partial_\mu \phi \partial_\nu \phi + \lambda_{\phi}' S \phi^2 + {1\over2} g^{\mu\nu} \partial_\mu S \partial_\nu S - {1\over2} m_S^2 S^2 \Big) \,.
\end{equation}
This theory is related to the previous gauge theory \eqref{eq:ymshl} through the following double-copy map:
\begin{equation}
A^{a}_\mu \rightarrow g_{\mu\nu}\,, \quad \phi^a \rightarrow \phi \,, \quad H \rightarrow S \,,
\end{equation}
where both $\phi$ and $S$ are scalars in the gravitational theory.
The double copy of the form factor can be understood as the amplitude in the gravitational theory as
\begin{equation}
{\cal G}_n \quad \Leftrightarrow \quad {\cal M}^{\rm GSH}_{n+1}(1^{\phi},2^{\phi},3^{h}, \ldots, n^h, q^S) =  \langle \phi(p_1) \, \phi(p_2) \, h(p_3) \ldots h(p_n) | S(q) \rangle \,,
\end{equation} 
which contains $(n-2)$ external gravitons, 2 $\phi$ scalars, and one massive $S$ scalar.
We emphasize that all the amplitudes should be understood at the tree level in our discussion.

The above correspondence inspires us to consider the following generalization: 
one can apply the above double copy to the QCD amplitudes involving a Higgs particle. 
To be explicit, we consider the QCD Lagrangian involving gluons, fundamental quarks, and a color-single scalar $H$:
\begin{equation}
{\cal L}^{\rm HQCD} = -{1\over2} {\rm tr}(F_{\mu\nu}F^{\mu\nu}) + i \bar{\psi} \gamma^\mu D_\mu \psi + \lambda_{\psi} H \bar{\psi} \psi + {1\over2} \partial^\mu H \partial_\mu H - {1\over2} m_H^2 H^2  \,.
\end{equation}
The QCD amplitudes can be understood as a form factor with the operator ${\cal O}$ as the Yukawa interaction vertex $H{\cal O}= H \bar{\psi} \psi$,
namely,
\begin{equation}\label{eq:fermionFF}
\itbf{A}( 1^{\bar\psi}, 2^{\psi}, 3^g, \ldots, n^g,q^H) = \itbf{F}_{{\cal O}=\bar\psi\psi}(1^{\bar\psi}, 2^{\psi}, 3^g, \ldots, n^g) \,.
\end{equation}
We expect that the previous double copy procedure will apply to the Higgs-QCD amplitudes here.

The general double-copy principle tells us that the following double-copy map should hold \cite{Johansson:2014zca,Johansson:2015oia,Johansson:2019dnu}
\begin{equation}\label{eq:doublecopymap}
A^{a}_\mu \rightarrow g_{\mu\nu}\,, \quad \psi_i \rightarrow A^{\rm U(1)}_{\mu} \,, \quad H \rightarrow S \,,
\end{equation}
so that if we make a double copy of the Higgs amplitudes in \eqref{eq:fermionFF}, we should get amplitudes like $M( 1^{\gamma}, 2^{\gamma}, 3^h, \ldots, n^h,q^{S})$, where ${\gamma}$ represents the photon of the $U(1)$ gauge field in \eqref{eq:doublecopymap}, in the theory described by the Lagrangian 
\begin{equation}
{\cal L}^{\rm HG} = \sqrt{-g} \Big({1\over16\pi G} R - {1\over4} F_{\mu\nu}F^{\mu\nu} + \lambda_{A}S F_{\mu\nu}F^{\mu\nu} + {1\over2} g^{\mu\nu} \partial_\mu S \partial_\nu S - {1\over2} m_S^2 S^2 \Big) \,.
\end{equation}
We verify below that such anticipation is reasonable. 

\subsubsection*{The double copy of Higgs amplitudes in QCD}

To study the double copy, we notice that the two fermions $\{1^{\bar\psi},2^{\psi}\}$ in \eqref{eq:fermionFF} can play the role of the two scalars $\{1^{\phi},2^{\phi}\}$ in the form factors in Section~\ref{sec:nptscalar}. 
The color basis is now the DDM-like color basis with the positions of the two fundamental fermions fixed (this is actually a special case of the Melia basis \cite{Melia:2013bta,Melia:2013epa,Ochirov:2019mtf}). The basis numerators and color-ordered $n$-point form factors are $N_{n}[1^{\bar\psi},\alpha,2^{\psi}]$ and ${\cal F}_{n}[1^{\bar\psi},\alpha,2^{\psi}]$, respectively, where $\alpha \in S_{n-2}$.
We have the same relation as \eqref{eq:FthetaNn}: 
\begin{equation}
    \mathcal{F}_n(1^{\bar{\psi}},\alpha, 2^{\psi})=\sum_{\beta \in S_{n-2}}\Theta^{\cal F}_n[\alpha|\beta]N_n[1^{\bar{\psi}},\beta,2^{\psi}]\,.
\end{equation}
In particular, the propagator matrix $\Theta^{\cal F}_n[\alpha|\beta]$ in this case is exactly the one studied in previous sections, such as defined in \eqref{eq:deftheta}. Hence, the KLT kernel $\mathbf{S}^{\cal F}_n$, as well as its matrix decomposition, is also the same as what we have derived in Section~\ref{sec:nptscalar}. 

The double copy is straightforward to obtain as
\begin{align}
		\mathcal{\tilde{G}}_{n} &=\sum_{\alpha, \beta\in S_{n-2}}\mathcal{F} _{n}[1^{\bar{\psi}},\alpha,2^{\psi}] \mathbf{S}_{n}^{\mathcal{F}}[\alpha|\beta] \mathcal{F} _{n}[1^{\bar{\psi}},\beta,2^{\psi}] \label{eq:genericKLTferm}\\
      &=\sum_{\alpha, \beta\in S_{n-2}}N_n[1^{\bar{\psi}},\alpha,2^{\psi}] \Theta^{\cal F}_{n}[\alpha|\beta]N_n[1^{\bar{\psi}},\beta,2^{\psi}] \,. \label{eq:genericKLTferm2}
\end{align}
It is manifestly diffeomorphism invariant, and we can also check its factorizations. 
Two ingredients are essential for confirming the double copy factorizations: the matrix decomposition and the hidden factorization relation. The matrix decomposition is the same as in the previous discussion mentioned above. Here we only need to focus on the hidden factorization relations (in gauge theory), which take a similar form as \eqref{eq:nptgeneralizedBCJ}
\begin{equation}\label{eq:nptgeneralizedBCJferm}
    \sum_{\alpha \in S_{n-2}} {\vec v}_{(\bar\kappa,\bar\rho)}[\alpha] \mathcal{F} _{n}[1^{\bar{\psi}},\alpha,2^{\psi}] \big|_{\QQ_m^2=q^2} =\mathcal{F} _{m}[1^{\bar{\psi}},\bar{\kappa},2^{\psi}]\times  \mathcal{A} _{m^{\prime}}[\bar{\rho}]\,,
\end{equation}
where the ${\vec v}_{(\bar\kappa,\bar\rho)}$ vectors and the amplitude $\mathcal{A} _{m^{\prime}}$ are the same as in \eqref{eq:nptgeneralizedBCJ}. However, the ordered form factors $\mathcal{F}$ are different.  These $D$-dimensional fermion form factor expressions, involving $\gamma$-matrices and Dirac spinors, are more complicated than the scalar case, but still satisfy \eqref{eq:nptgeneralizedBCJferm}, with the help of $\gamma$-algebra. 

Below we provide a few examples with explicit expressions, which will also reveal an interesting connection between the fermion numerators and the scalar numerators, including the previously obtained ones in \eqref{eq:BCJnumF}. 

For the three-point case, the ordered form factor is 
\begin{equation}
    \mathcal{F}_{3}(1^{\bar{\psi}},3^{g},2^{\psi})=\frac{1}{2}\left(\frac{1}{s_{13}}+\frac{1}{s_{23}}\right){\bar{v}}_{1,\alpha} \left(\slashed{f_3}\right)^{\alpha}_{\beta} u_2^{\beta} + \frac{2 p_1\cdot f_3\cdot p_2}{s_{13}s_{23}} {\bar{v}}_{1,\alpha} u_2^{\alpha}\,,
\end{equation}
where ${\bar{v}}_{1,\alpha}(p_1)$ and $u_2^{\beta}(p_2)$ are Dirac spinor for the fermions ${\bar{\psi}}(p_1)$ and ${\psi}(p_2)$, respectively.
There are two cubic diagrams similar to that in Figure~\ref{fig:F3tree} (except that the scalar legs should be understood as fermion legs), and the color factors are 
\begin{equation}
C_1 = C_2 = (T^a)_i^{~j} \,.
\end{equation}
Similar to \eqref{eq:3ptnumsols}, we obtain the CK-dual numerator as
\begin{equation}
    N_{3}[1^{\bar{\psi}},3^{g},2^{\psi}]=\frac{1}{2}{\bar{v}}_{1,\alpha} \left(\slashed{f_3}\right)^{\alpha}_{\beta} u_2^{\beta} - \frac{2 p_1\cdot f_3\cdot p_2}{s_{12}-q^2} {\bar{v}}_{1,\alpha} u_2^{\alpha}\,.
\end{equation}
Note that the second term is nothing but ${\bar{v}}_{1,\alpha} u_2^{\alpha}$ times the scalar one in \eqref{eq:3ptnumsols}. 

This pattern continues to hold for higher-point form factors. In the four-point case, the form factor is a little bit lengthy but the master numerator remains strikingly simple 
\begin{equation}
\begin{aligned}
    N_{4}[1^{\bar{\psi}},3^{g},4^{g},2^{\psi}]=&
    \frac{1}{4}{\bar{v}}_{1,\alpha}(\slashed{f_3}\slashed{f_4})^{\alpha}_{\beta} u_{2}^{\beta}-\frac{1}{2}N_{4}[1^{\phi},3^{\phi},4^{g},2^{\phi}]
    {\bar{v}}_{1,\alpha}(\slashed{f_3})^{\alpha}_{\beta} u_{2}^{\beta}\\
    &-\frac{1}{2}N_{4}[1^{\phi},3^{g},4^{\phi},2^{\phi}] {\bar{v}}_{1,\alpha}(\slashed{f_4})^{\alpha}_{\beta} u_{2}^{\beta}+
    N_{4}[1^{\phi},3^{g},4^{g},2^{\phi}] {\bar{v}}_{1,\alpha} u_2^{\alpha}\,,
\end{aligned}
\end{equation}
in which the last term is proportional to the previous four-point scalar form factor; moreover,  we have the following two additional blocks that turn out to be master numerators for other scalar form factors (which will be discussed in Section~\ref{sec:generalize2})
\begin{equation}
    N_{4}[1^{\phi},3^{\phi},4^{g},2^{\phi}]= \frac{-2 (p_1+p_3)\cdot f_4 \cdot p_2}{s_{123}-q^2} , \quad
    N_{4}[1^{\phi},3^{g},4^{\phi},2^{\phi}]= \frac{-2 p_1\cdot f_3 \cdot (p_2+p_4)}{s_{124}-q^2}.
\end{equation}

Finally, given the pattern of the three- and four-point results, we make the following conjecture for general  $n$-point numerators, reading 
\begin{equation}
    N_{n}[1^{\bar{\psi}},3^{g},\ldots,n^{g},2^{\psi}]= \sum_{r=0}^{n-2} \sum_{i_1<i_2\cdots < i_r} \hskip -6pt \frac{{\bar{v}}_{1,\alpha} (\slashed{f_{i_1}}\slashed{f_{i_2}}\cdots \slashed{f_{i_r}} ) u_{2}^{\beta}}{(-2)^{r}} N_n[1^{\phi},3^{g},.. i_1^{\phi} ..i_2^{\phi} .. i_r^{\phi}.. , n^{g},2^{\phi}]\,,
\end{equation}
where the $N_n[1^{\phi},3^{g},.. i_1^{\phi} ..i_2^{\phi} .. i_r^{\phi}.. , n^{g},2^{\phi}]$ means that we have changed $r$ gluons $\{i_1,\ldots,i_r\}$ into scalars. 
Such a formula is originally motivated  in the study of kinematic algebra \cite{Chen:2022nei} and checked up to seven points, and can be understood in parallel to a relation satisfied between scalar-gluon numerators (which can be checked to be consistent with the results in Appendix~\ref{ap:nums})
\begin{equation}
    N_{n}[1^{\phi},3^{g},\ldots,n^{g},2^{\phi}]= \sum_{r=1}^{n-2} \sum_{i_1<i_2\cdots < i_r} \hskip -6pt \frac{p_1\cdot {f_{i_1}}\cdot{f_{i_2}}\cdots {f_{i_r}} \cdot p_2}{(-2)^{r} (s_{12}-q^2)} N_n[1^{\phi},3^{g},.. i_1^{\phi} ..i_2^{\phi} .. i_r^{\phi}.. , n^{g},2^{\phi}]\,.
\end{equation}
We will discuss how to construct these scalar-gluon numerators explicitly in the following sections.

\section{Generalization II: form factors with multiple external scalars}\label{sec:generalize2}

Now we are going to deal with a more non-trivial generalization, that is to introduce matter field self-interactions. 
For simplicity, we still consider scalars and a $\phi^3$ interaction only. The Lagrangian is given as
\begin{align}\label{eq:L}
    \mathcal{L}=&\mathcal{L}^{\rm YMS}  +\frac{g'}{3!}{\tilde f}^{IJK}f^{abc}\phi^{I,a}\phi^{J,b}\phi^{K,c}\,,
\end{align}
where we also generalize the scalars in \eqref{eq:LagrangianYMS} by including flavor indices denoted by $I,J,K$.
The operator in the form factors is the simplest $\operatorname{tr}(\phi^2)=\sum_{a,I}\phi^{a,I}\phi^{a,I}$. 

\subsubsection*{The three-point case with three external scalars}

\begin{figure}[t]
    \centering
 \includegraphics[width=0.7\linewidth]{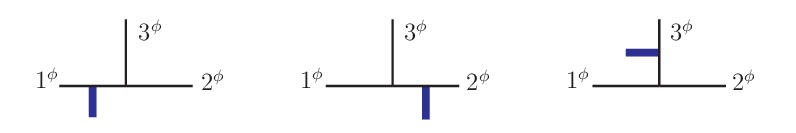}
    \caption{Cubic diagrams of the three-point form factor in $\itbf{F}_{{\rm tr}(\phi^2),3}(1^{\phi},2^{\phi},3^{\phi})$.}
    \label{fig:F3treephi2}
\end{figure}

We start with simple examples with three external scalars. The three-scalar three-point form factor is a little trivial and plays the role of a ``minimal" form factor in the double copy.
There are three cubic diagrams with the same color factors $C = f^{123}$, as shown in Figure~\ref{fig:F3treephi2}. By summing the three diagrams up, one has ($\tilde{f}^{123}$ is the flavor factor)
\begin{equation}\label{eq:3scalarFF}
    \itbf{F}_{{\rm tr}(\phi^2),3}(1^{\phi},2^{\phi},3^{\phi}) = {C} \mathcal{F}_3(1^{\phi},2^{\phi},3^{\phi})= f^{123} \tilde{f}^{123} \left({1\over s_{12}}{+}{1\over s_{13}}{+}{1\over s_{23}}\right) \,.
\end{equation}
Note that the kinematic numerator is $1$ for all three cubic diagrams, which automatically satisfy the CK duality. As a remark, since the flavor factor for the three-scalar form factor is always  $\tilde{f}^{123}$ and we can ignore it. 

Since there is a single color factor, the propagator matrix has only one element
\begin{equation}
\Theta'^{\cal F}_3 = \Big( {1\over s_{12}}{+}{1\over s_{13}}{+}{1\over s_{23}} \Big) =  {s_{12} s_{13} + s_{12} s_{23} + s_{13} s_{23} \over s_{12} s_{13} s_{23} }  \,,
\end{equation} 
and the KLT kernel is 
\begin{equation}\label{eq:phi3KLTS3}
\mathbf{S'}^{\mathcal{F}}_{3} = (\Theta'^{\mathcal{F}}_3)^{-1}  =  {s_{12} s_{13} s_{23} \over s_{12} s_{13} + s_{12} s_{23} + s_{13} s_{23} }  \,.
\end{equation}
Here we add a ``prime" in $\Theta'^{\cal F}$ and $\mathbf{S'}^{\mathcal{F}}$ to stress that they are different from those in the previous sections.
The double copy is straightforward to obtain: 
\begin{equation}\label{eq:phi3G3DC}
\mathcal{G}_3(1^{\phi},2^{\phi},3^{\phi}) = N_3 \Theta'^{\cal F}_3 N_3 = {\cal F}_3 \mathbf{S'}^{\mathcal{F}}_{3} {\cal F}_3 = {1\over s_{12}}{+}{1\over s_{13}}{+}{1\over s_{23}} \,. 
\end{equation} 
We comment that the KLT kernel \eqref{eq:phi3KLTS3} has a non-trivial pole $s_{12} s_{13} {+} s_{12} s_{23} {+} s_{13} s_{23}$, which is canceled in the form factor double copy.
A similar feature also appears in higher-point cases with more intriguing structures.

\subsubsection*{The four-point case with three external scalars}

The first ``non-trivial" example is the four-point form factor with three external scalars plus one gluon. The steps for this calculation are similar to the two-scalar-two-gluon one discussed in Section~\ref{ssec:4pteg}.  

\begin{figure}[t]
    \centering
 \includegraphics[width=1\linewidth]{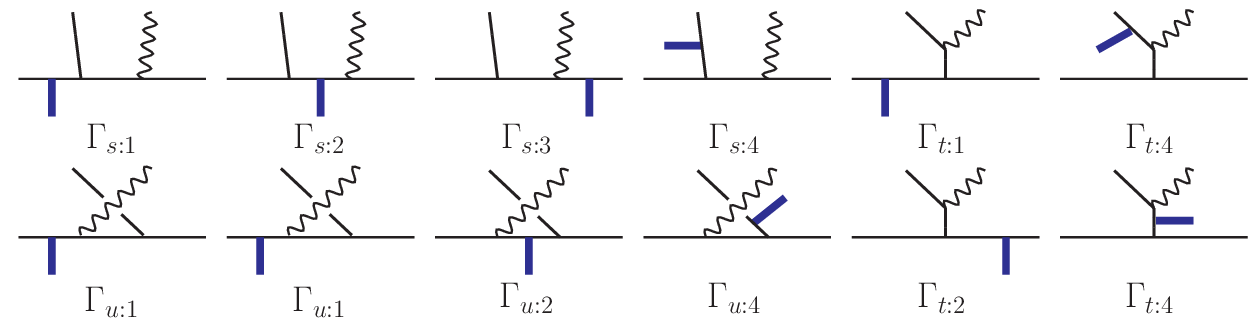}
    \caption{Cubic diagrams for the double copy of the four-point form factor in  $\itbf{F} _{4}$.}
    \label{fig:F4treephi2}
\end{figure}

The cubic diagrams for this form factor are given in Figure~\ref{fig:F4treephi2}, and the dual Jacobi relations relating to their numerators are
\begin{equation}
\begin{aligned}
    &N_{s:1}= N_{s:2}= N_{s:3}= N_{s:4}\equiv N'_{\{s\}}\,, \\
    &N_{t:1}= N_{t:2}= N_{t:3}= N_{t:4}\equiv N'_{\{t\}}\,, \\
    &N_{u:1}= N_{u:2}= N_{u:3}= N_{u:4}\equiv N'_{\{u\}}\,,\\
    &N'_{\{s\}}=N'_{\{t\}}+N'_{\{u\}} \,.
\end{aligned}
\end{equation}
Consequently, the full-color four-point form factor is 
\begin{equation}\label{eq:FCF4pt}
    \itbf{F}_{\operatorname{tr}(\phi^2),4}(1^{\phi},2^{\phi},3^{\phi},4^{g})=\frac{C_{\{s\}} N'_{\{s\}}}{P'_{\{s\}}}+\frac{C_{\{t\}} N'_{\{t\}}}{P'_{\{t\}}}+\frac{C_{\{u\}} N'_{\{u\}}}{P'_{\{u\}}} \,,
\end{equation}
where  $(P'_{\{s\}})^{-1}{=}\frac{1}{s_{24}s_{234}}{+}\frac{1}{s_{24}s_{13}}{+}\frac{1}{s_{13}s_{134}}{+}\frac{1}{s_{24}s_{124}}$ and $(P^{\prime}_{\{t\}})^{-1},(P'_{\{u\}})^{-1}$ can be defined similarly. 

Note that \eqref{eq:FCF4pt} takes the same form as \eqref{eq:4pttreefullcolor-CK}, so that we follow the same procedure dealing with it. 
Thus, here we also regard $C_{\{s\}},C_{\{u\}}$ as the color basis and $N_{\{s\}},N_{\{u\}}$ as the numerator basis. This will lead us to the propagator matrix form like $\vec{\cal F}=\Theta\cdot \vec{N}$, as in \eqref{eq:4ptFNeq0}; but the propagator matrix now is different from the one in \eqref{eq:4ptThetaMatrix}. The new propagator matrix is 
\begin{align}\label{eq:4ptThetaMatrixnew}
& \Theta'^{\mathcal{F}}_{4} = 
\begin{pmatrix}
{1\over P'_{\{s\}}} +  {1\over P'_{\{t\}}}& - {1\over P'_{\{t\}}} \\
-{1\over P'_{\{t\}}} &  {1\over P'_{\{t\}}} + {1\over P'_{\{u\}}}
\end{pmatrix}\,,
\end{align}
which is more complicated than the previous one \eqref{eq:4ptThetaMatrix}, {\emph e.g.} the upper-left element of the new matrix contains more terms 
\begin{equation}
    {1\over P'_{\{s\}}} {+}  {1\over P'_{\{t\}}}=\frac{1}{s_{24}s_{234}}{+}\frac{1}{s_{24}s_{13}}{+}\frac{1}{s_{13}s_{134}}{+}\frac{1}{s_{24}s_{124}}{+}\frac{1}{s_{34}s_{134}}{+}\frac{1}{s_{34}s_{234}} {+}\frac{1}{s_{34}s_{12}} {+}\frac{1}{s_{124}s_{12}}\,.
\end{equation}

\vspace{3pt}

Next, we need to figure out the new ``spurious"-type pole after double copy. Physically, it is easy to anticipate that we have the ``spurious"-type $s_{123}{-}q^2$ pole which will become a real physical pole after double copy. Practically, we can find these poles from the zeros of $\det\left(\Theta'^{\mathcal{F}}_{4}\right)$. However, the concrete evaluation gives the determinant as
\begin{equation}\label{eq:4ptphi3P0}
    \det(\Theta'^{\mathcal{F}}_4)= \frac{(s_{123}-q^2) \times \mathcal{P}}{s_{12}s_{23}s_{13}s_{14}s_{24}s_{34}s_{124}s_{134}s_{234}} \,,
\end{equation}
where the polynomial $\cal P$ is 
\begin{align}
    \mathcal{P}&=s_{34}s_{12}s_{124}(s_{12}-q^2)(s_{124}-q^2)+s_{14}s_{23}s_{234}(s_{23}-q^2)(s_{234}-q^2) \label{eq:4ptphi3P}  \\
    &+ s_{24}s_{13}s_{134}(s_{13}-q^2)(s_{134}-q^2)+(s_{23}s_{13}+s_{12}s_{13}+s_{12}s_{23})(s_{234}s_{134}+s_{124}s_{134}+s_{124}s_{234}) .\nonumber
\end{align}
We should be happy to see that the $s_{123}{-}q^2$ factor is there, which will become a new pole after double copy. However, there is also a nontrivial $\mathcal{P}$ factor in the numerator of \eqref{eq:4ptphi3P0}. 
Naively, since $\mathbf{S'}^{\mathcal{F}}_{4} = (\Theta'^{\mathcal{F}}_4)^{-1}$ is proportional to $\mathcal{P}^{-1}$, one might worry that $\mathcal{G}_4=\vec{\mathcal{F}}\cdot \mathbf{S'}^{\mathcal{F}}_{4} \cdot \vec{\mathcal{F}}$ has the undesired $\mathcal{P}^{-1}$ pole and the double copy prescription would hit a dead end.

Fortunately, computing the residue of $\mathcal{G}_4$ at ${\mathcal{P}=0}$, we get \textbf{zero}, namely,
\begin{equation}
\mathrm{Res}_{\mathcal{P}=0}[\mathcal{G}_4]=0 \,,
\end{equation}
which means the complicated $\mathcal{P}^{-1}$ pole in double copy is nothing but an artifact! 
This can be understood first from the master numerators. 
The master numerators, obtained via $\vec{N'}=\mathbf{S'}\cdot \vec{\cal F}$, take the following nice form
\begin{equation}\label{eq:4ptphi3numsolnew}
    N'_{\{s\}}=-\frac{2 f_{4}^{\mu\nu}(p_{1}+p_{3})_{\mu} p_{2,\nu}}{s_{123}-q^2} \,, \qquad 
    N'_{\{u\}}=-\frac{2 f_{4}^{\mu\nu}(p_{2}+p_{3})_{\mu} p_{1,\nu}}{s_{123}-q^2}\,,
\end{equation}
which has no $\mathcal{P}^{-1}$ pole.\footnote{The kernel $\mathbf{S}^{\cal F}_4$ in this case, of which the expression is pretty messy, indeed has the ``bad" ${\cal P}^{-1}$ pole. Magically, multiplying it with the special ordered form factors $\mathcal{F}_4$ will cancel the ${\cal P}^{-1}$ factor.}
Thus the double-copy formula
\begin{equation}
\mathcal{G}_4=\vec{N'}_4^{\scriptscriptstyle \rm T} \cdot \Theta'^{N}_{4} \cdot \vec{N'}_4= \vec{N'}_4^{\scriptscriptstyle \rm T} \cdot  \mathcal{F}_4\,, 
\qquad
 \vec N'_4 =
\begin{pmatrix}
N'_{\{s\}} \\
N'_{\{u\}}
\end{pmatrix} ,
\end{equation}
directly shows that $\mathcal{G}_4$ is free of the $\mathcal{P}^{-1}$ pole.
Another hint is from the matrix decomposition of $\mathbf{S'}^{\cal F}_4$. The $\mathcal{P}$ factor will completely disappear when calculating 
\begin{equation}
    \text{Res}[\mathbf{S'}^{\cal F}_4]_{s_{123}=q^2}=
     \begin{pmatrix}
        s_{42} \\ s_{42}+s_{43}
    \end{pmatrix}
\cdot \left( \mathbf{S'}^{\mathcal{F}}_{3} \times 1\right)\cdot     \left(s_{42} \ \ s_{42}+s_{43} \right)\,,
\end{equation}
where the KLT kernel $\mathbf{S'}^{\cal F}_3$ for the three-scalar form factor is given in \eqref{eq:phi3KLTS3}.

Finally, we examine some factorization properties. As argued above, it is okay to forget about the $\mathcal{P}$ factor and  only focus on the special kinematics $s_{123}=q^2$. As for the hidden factorization relation, we have 
\begin{equation}\label{eq:4ptbcjexpandnew}
\big[ s_{42} \mathcal{F}_{4}(1^{\phi},3^{\phi},4^{g},2^{\phi}) + (s_{42}+s_{43}) \mathcal{F}_{4}(1^{\phi},4^{g},3^{\phi},2^{\phi}) \big] \big|_{s_{123}=q^2}
 =\mathcal{F}_{3}(1^{\phi},2^{\phi},3^{\phi}) \ \mathcal{A}_{3}(\QQ_{3}^{S},4^{g},-q^{S}) \,,
\end{equation}
which has the same structure as \eqref{eq:4ptbcjexpand}. 

\begin{figure}[t]
    \centering
 \includegraphics[width=0.8\linewidth]{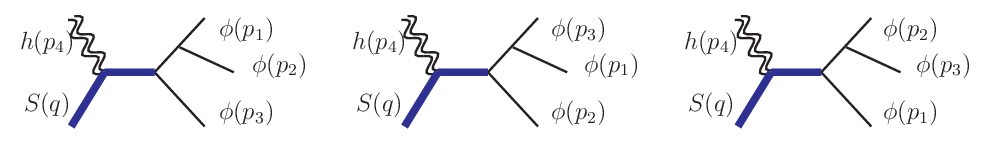}
    \caption{Cubic diagrams with the new $s_{123}{-}q^2$ pole for the form factor double copy  $G_{4}(1^{\phi},2^{\phi},3^{\phi},4^{h})$.}
    \label{fig:F4treeGRA2}
\end{figure}

The factorization of $\mathcal{G}_4$ on the $s_{123}{-}q^2$ pole is also easy to check:
\begin{equation}\label{eq:4ptfactorizationGRA3phi}
\begin{aligned}
    \text{Res}_{s_{123}=q^2}[\mathcal{G}_4]&=
    \text{Res}[ N'_{\{s\}} 
    \mathcal{F}_{4}(1^{\phi},3^{\phi},4^{g},2^{\phi}) + N'_{\{u\}} 
    \mathcal{F}_{4}(1^{\phi},4^{g},3^{\phi},2^{\phi})
    ]_{s_{123}=q^2}\\
    &=
    \mathcal{G}_3 (1^{\phi},2^{\phi},3^{\phi}) \times  \mathcal{M}_{3}(q_3^{S}, -q^{S},4^h)\,.
\end{aligned}
\end{equation}
This also means new Feynman diagrams must be included for $\mathcal{G}_4$, see Figure~\ref{fig:F4treeGRA2}. 

\vspace{3pt}

From these examples, we can learn that not all the zeros of $\det\left(\Theta_4'^{\cal F}\right)$ will appear as the new ``spurious"-type poles after double copy. 
Some of these zeros, such as  $\mathcal{P}=0$, are purely artifacts and do not enter the expressions of master numerators. 
The other zeros like $s_{123}=q^2$, on the other hand, indeed matter and contribute to the poles of the double copy.
On these poles, the matrix decomposition, hidden factorization relation, and the factorization of the double copy are all perfectly fine.

\subsubsection*{Generalizations to $n$-point cases}

Now we discuss the $n$-point generalizations.
We first clarify how to make a generalization to the $n$-point form factors with three external scalars and then for more than three external scalars.

We will follow the similar steps given in Section~\ref{ssec:FFDCnpt} and highlight the differences.

After writing down a cubic diagram expansion similar to \eqref{eq:FCFncubic}, we specify a set of color basis. The DDM color basis is also a suitable choice. In this case, we fix two of the scalars $\{ 1^\phi, 2^\phi\}$ and permute the third scalar and all the gluons. A typical basis element can be represented by 
\begin{equation}\label{eq:3scalarb}
\begin{aligned}
    \includegraphics[width=0.36\linewidth]{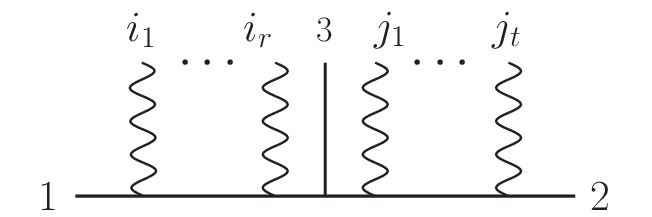}
\end{aligned} ,
\end{equation}
where $\{i_1,\ldots,i_r\}$ and  $\{j_1,\ldots,j_t\}$ are two subsets of the $(n{-}3)$ gluons $\mathbf{g}{\equiv} \{4^g,\ldots,n^g\}$. Note that the omitted $q$-leg can be also inserted on the scalar leg 3. 
The element of the DDM basis in \eqref{eq:3scalarb} corresponds to the ordering  $\{1,i_1,\ldots,i_r,3,j_1,\ldots,j_t,2\}$. 
One can define the kinematic numerator for a given ordering as
\begin{equation}\label{eq:ddmN2}
        N_{n}[1,i_1,\ldots,i_r,3,j_1,\ldots,j_t,2]\equiv N\bigg( \hskip -3pt \begin{aligned}
        \includegraphics[width=0.28\linewidth]{figure/3scalars.eps}
        \end{aligned}\hskip -6pt\bigg)\,.
\end{equation}

There are $(n-2)!$ DDM basis elements obtained by permutations $\alpha \in S_{n-2}$ acting on the set $\{3^{\phi},\mathbf{g}\}{=}\{3^{\phi},4^{g},\ldots,n^{g}\}$, denoted as $\alpha(3^{\phi},\mathbf{g})$. 
The color-ordered form factors can be related to the numerators in a similar manner to \eqref{eq:FthetaNn} as
\begin{equation}\label{eq:FthetaNnnew}
        {\cal F}_{n}[1^{\phi},\alpha(3^{\phi},\mathbf{g}),2^{\phi}]=\sum_{\beta \in S_{n-2}}\Theta'^{\cal F}_n[1^{\phi},\alpha(3^{\phi},\mathbf{g}),2^{\phi}|1^{\phi},\beta(3^{\phi},\mathbf{g}),2^{\phi}] N_{n}[1^{\phi},\beta(3^{\phi},\mathbf{g}),2^{\phi}]\,.
\end{equation}
As in Section~\ref{subsec:PmatrixFF}, the $\Theta$ matrix here can be also defined as bi-adjoint scalar form factors: 
\begin{align}\label{eq:defthetanew}
    \Theta'^{\cal F}_n[\alpha &|\beta]
    =\int d^{D}x\  e^{\mathrm{i}q\cdot x} \langle 1^\phi, 2^\phi, 3^\phi,4^{\Phi} \ldots n^\Phi | \mathcal{O}_{\phi}(x) | 0 \rangle \big|_{\operatorname{tr}_{\rm C}(\alpha) \operatorname{tr}_{\rm FL}(\beta)}\,.
\end{align}
The formulae \eqref{eq:defthetanew} and \eqref{eq:FthetaNnnew} are taking the same form as \eqref{eq:deftheta} and \eqref{eq:FthetaNn}, but the external states are different; especially we have the additional $3^{\phi}$ here. 

From \eqref{eq:FthetaNnnew} and \eqref{eq:defthetanew}, solving for the new master numerators $N_{n}$ can be done. Interestingly, these three-scalar numerators take a very similar form to the previous two-scalar ones in \eqref{eq:BCJnumF}. 
The result can be found in \cite{Chen:2022nei}
\begin{align}\label{eq:FFT2}
    N_n[1^{\phi},4^{g},\ldots,& m^{g},3^{\phi},(m{+}1)^{g},\ldots,n^{g},2^{\phi}]\nonumber \\
    =&\sum_{r=1}^{n-3}
	\sum_{\tau\in \mathbf{P}_{\mathbf{g}}^{(r)}}
	{(-2)^r\prod\limits_{i=1}^r \Big(p_{\Phi_{L}\Theta_L(i)}\cdot f_{(\tau_i)}\cdot p_{\Phi_{R}\Theta_R(i)}\Big)\over  (p_{123}^2{-}q^2) (p_{123\tau_1}^2{-}q^2)\cdots (p_{123\tau_1\cdots \tau_{r-1}}^2{-}q^2)} \,,
\end{align}
where $\mathbf{g}$ is the gluon set $\{4^g,\ldots,n^g\}$ and the meaning of other notations are the same as in \eqref{eq:BCJnumF}. 
The rules of assigning proper scalars to $\Phi_L$ and $\Phi_R$ are: if  max$(\tau_i)>m$, $\Phi_{L}=\{1^{\phi}\}$, otherwise $\Phi_{L}=\{1^{\phi},2^{\phi}\}$; if min$(\tau_i)\leq m$, $\Phi_{R}=\{3^{\phi}\}$, otherwise $\Phi_{R}=\{2^{\phi},3^{\phi}\}$.

Notably, only ``spurious"-type poles like $s_{123\cdots }{-}q^2$ appear in the numerator \eqref{eq:FFT2}. The determinant of the propagator matrix $\Theta'^{\cal F}_{n}$ of the form factor with three scalars is expected to be more complicated than the two-scalar one \eqref{eq:det}. However, among all the zeros of the determinant,  only the $s_{123\cdots }{-}q^2$ ones survive after double copy. We have the following representations for the form factor double copy which manifest its pole structures
\begin{equation}\label{eq:phi23scalarDC}
\begin{aligned}
	\mathcal{G}_{n}&=\sum_{\alpha,\beta \in S_{n-2}} N_n[1^{\phi},\alpha(3^{\phi},\mathbf{g}),2^{\phi}] \Theta'^{\cal F}_n[\alpha(3^{\phi},\mathbf{g})|\beta(3^{\phi},\mathbf{g})]N_n[1^{\phi},\alpha(3^{\phi},\mathbf{g}),2^{\phi}]\\
	&=\sum_{\alpha \in S_{n-2}} N_n[1^{\phi},\alpha(3^{\phi},\mathbf{g}),2^{\phi}] \mathcal{F}_n(1^{\phi},\alpha(3^{\phi},\mathbf{g}),2^{\phi})\,.
\end{aligned}
\end{equation}
One can check the factorization properties on the spurious poles and ensure that they are truly physical propagators after double copy. 

As discussed in the previous sections for  the two-scalar case, the gravitational factorization is a joint consequence of the matrix decomposition and the hidden factorization relation. The bridge connecting these two interesting aspects is still the  $\vec{v}$ vectors. 
Surprisingly, the same $\vec{v}$ vectors discussed in Section~\ref{sec:nptscalar} and Appendix~\ref{app:vectors} apply to the three-scalar case: that is they give rise to the matrix decomposition and induce the hidden factorization relation for the three-scalar form factors. We will perform a more complete investigation on this universality in \cite{treepaper2}. 

\vspace{3pt}

Apart from the three-scalar form factors, we mention that generalizations to form factors with four or more external scalars are also possible. 
Below we summarize a few important points in the $m$-scalar generalization and the general procedure is totally parallel to the aforementioned ones. 
\begin{enumerate}[topsep=3pt,itemsep=-1ex,partopsep=1ex,parsep=1ex]
    \item We choose the DDM color basis by considering half-ladder diagrams fixing two scalars $\{1^{\phi},2^{\phi}\}$ and permuting the particles $\{3^{\phi},\ldots,m^{\phi},(m{+}1)^{g},\ldots,n^{g}\}$ with $\beta\in S_{n-2}$. 
    \item The propagator matrix can also be understood as the bi-adjoint scalar form factors similar to \eqref{eq:deftheta} and \eqref{eq:defthetanew}. 
    \item The master numerators are chosen as $N_n[1^{\phi},\beta(3^{\phi},\ldots,m^{\phi},(m{+}1)^{g},\ldots,n^{g}),2^{\phi}]$ and can be obtained via a closed formula similar to \eqref{eq:BCJnumF} and \eqref{eq:FFT2}, see also \cite{Chen:2022nei}. 
    We also give a review of that closed formula in Appendix~\ref{ap:nums}, and just mention some useful features of those numerators   here.
    
    First, the poles of the numerators always take the form of $(\QQ_m{+}\ldots)^2{-}q^2$ where $\QQ_m$ is $\sum_{i=1}^{m}p_i$ as the sum of all scalar momenta and the ``$\ldots$" are some gluon momenta in $\{p_{m{+}1},\ldots,p_n\}$. 
They are exactly the anticipated ``spurious"-type poles as massive Feynman propagators. Second, the numerators are gauge invariant and linear in the field strengths. Third, those numerators in Appendix~\ref{ap:nums} are exactly the numerators of the half-ladder cubic diagram so that the numerators of other cubic diagrams can be obtained from simple commutators of these basis numerators. 
    \item The double copy can be obtained by 
    \begin{equation}\label{eq:multiscalarG}
        \hskip -25pt 
        \mathcal{G}_{n}{=}\hskip -3pt \sum_{\beta \in S_{n-2}} N_n[1^{\phi},\beta(3^{\phi},..,m^{\phi},(m{+}1)^{g},..,n^{g}),2^{\phi}] \mathcal{F}_n(1^{\phi},\beta(3^{\phi},..,m^{\phi},(m{+}1)^{g},..,n^{g}),2^{\phi})\,,
    \end{equation}
    which manifestly shows that $\mathcal{G}_n$ has only two types of simple poles: the ``physical"-type poles appearing in $\mathcal{F}$ and the ``spurious"-type poles appearing in $N$. 

    \item The factorizations of  $\mathcal{G}_n$ can be also checked. Again, the matrix decomposition and the hidden factorization relations are two central ingredients. Importantly, exactly the same $\vec{v}$ vectors can be used in the matrix decomposition and the hidden factorization relation, regardless of the number of external scalar particles. 
    We will discuss this point further \cite{treepaper2}. 
    \item As a minor final remark, we would like to emphasize that form factors with only single trace flavor factors are considered in this paper. Multi-trace form factors can be related to the single trace ones by transmutation operators \cite{Cheung:2017ems} in a straightforward yet complicated way, and we leave their study in the future. 
    
\end{enumerate}

\section{Discussion}\label{sec:discussion}

As mentioned in the introduction, this paper is the first of a series of two papers focusing on the double-copy construction of tree-level form factors. 
This paper discusses the basic foundations by focusing on the form factors of ${\rm tr}(\phi^2)$ in the Yang-Mills-scalar theory.
Such form factors will provide the prototype for various other generalizations. 
The most important new feature is the emergence of the ``spurious"-type poles via double copy. We explain various properties related to the factorizations on these poles in Section~\ref{sec:nptscalar},  which is the central part of this paper.
We also discuss two generalizations: first for the Higgs amplitudes in QCD, and then for the ${\rm tr}(\phi^2)$ form factors with multiple external scalar states.

In the second paper \cite{treepaper2}, we will discuss some more advanced topics, including (1) a detailed study of the $\vec{v}$ vectors, and (2)  other non-trivial generalizations for form factors with higher-length operators ${\rm tr}(\phi^L)$ and purely gluonic operators like ${\rm tr}(F^2)$. 
As outlined in Figure~\ref{fig:paperstructure}, the topics in these two papers are closely connected: the $\vec{v}$ vectors, which are crucially important in this paper, will have their closed formula given in the next paper. The universality of the $\vec{v}$ vectors will be covered there, of which the evidence already appears in the multi-scalar generalization in Section~\ref{sec:generalize2} in this paper. Besides, the generalizations to form factors of higher-length operators and $\operatorname{tr}(F^2)$ also require the knowledge of multi-scalar generalization.

Before ending this paper, we would like to discuss the practical and physical origin of the new features in the form factor double copy. 

Practically, as we have already seen from the discussion in Section~\ref{sec:34ptexample}, besides the color Jacobi relations, there are new types of operator-induced color relations for form factors, since the operator is a color singlet. 
The kinematic numerators must satisfy the corresponding dual relations, which are indispensable for a diffeomorphism invariant double copy. This is where the new ``spurious"-type poles (as well as the subsequent factorizations)  originate from.

Physically, we can see that instead of having a ``naked" color-singlet local operator in the gauge theory, one gets the operator ``graviton-dressed" after double copy, which is precisely what the new diagrams are describing. 
This is related to the puzzle of the form factor double copy: gravity theories cannot have local operators as physical observables, since a local operator ${\cal O}(x)$ is not diffeomorphism invariant, then what could the double copy of form factor correspond to?
The ``graviton-dressed" operator, like the thick blue lines in Figure~\ref{fig:F4treeGRA}, can have any allowed interaction with gravitons and may be considered as a non-local object, and the double copy of the gauge-theory form factor is just the form factor of such a ``graviton-dressed" operator.

An alternative physical  explanation is, of course, the perspective that we took in Section~\ref{sec:generalize1}, that is to view the form factors and their double copies as Higgs amplitudes. 
At the tree level, these two explanations are equivalent. 
The story at the loop level, however, is expected to be more involving, and the point of view of Higgs-like amplitudes in gravity may no longer apply. 
For example, the Higgs-scalar can be fully dynamical in the amplitudes and can appear in the internal loops. 
On the other hand, the double-copy of form factors are more constrained at the loop level in the sense that a construction allowing the massive propagators in the internal loops may not exist after double copy, which is supported by some primitive calculations.
This suggests that the double copy of the loop form factors involves a (semi-classical) non-local operator in gravity, which is like the ``graviton-dressed" operator at the tree level but receives no loop correction for the operator itself. 
Finally, we briefly comment that such a ``graviton-dressed" operator resembles the ``graviton-dressed" propagator in world line theories \cite{Bastianelli:2002fv,vanHolten:2001ea}.
We leave details of the loop generalization to further work.

\acknowledgments

We thank Gang Chen, Jin Dong, Song He, Henrik Johansson, Radu Roiban, Congkao Wen and  Mao Zeng for discussions. 
This work is supported in part by the National Natural Science Foundation of China (Grants No.~12175291, 11935013, 12047503, 12247103) and by the CAS under Grants No.~YSBR-101.
We also thank the support of the HPC Cluster of ITP-CAS.

\appendix

\section{Explicit expressions for the $\vec{v}$ vectors}\label{app:vectors}
In this appendix, we provide explicit expressions for the $\vec{v}$ vectors up to six-point form factors. As mentioned in the main text (see Section~\ref{ssec:Hiddenfac}), we can focus on $\vec{v}_{(\mathbf{1},\mathbf{1})}$ since $\vec{v}_{(\bar{\kappa},\bar{\rho})}$ can be obtained via $\vec{v}_{(\mathbf{1},\mathbf{1})}$ by permutations. The detailed and more systematic discussions will be given in \cite{treepaper2}. For simplicity, below we only give the non-zero components in $\vec{v}$.

\subsubsection*{Four-point $\QQ_2^2=q^2$}

\begin{align}
v[1,3,4,2]=\frac{\tau_{31}\tau_{42}}{s_{123}-q^2},\quad  v[1,4,3,2]= \frac{\tau_{32}\tau_{41}}{s_{123}-q^2} \,. \nonumber
\end{align}

\subsubsection*{Four-point $\QQ_3^2=q^2$}

\begin{align}
v[1,3,4,2]=-\tau_{42},\quad v[1,4,3,2]= -\tau_{4,(2+3)} \,. \nonumber
\end{align}
They appear in \eqref{eq:4ptinitialcond} and \eqref{eq:4ptBCJA}.\footnote{We have slightly modified the sign of the $\vec{v}$ vector to be consistent of the closed formula given in the next paper.}

\subsubsection*{Five-point $\QQ_2^2=q^2$}

\begin{align}
    &v[1,3,4,5,2]=-\frac{\tau_{31}\tau_{52}}{s_{1234}-q^2}\Big(\frac{\tau_{4,(1+3)}}{s_{123}-q^2}+1\Big), \quad v[1,3,5,4,2]= -\frac{\tau_{31}\tau_{42}\tau_{5,(1+3)}}{(s_{123}-q^2)(s_{1234}-q^2)},\nonumber\\
    &v[1,5,3,4,2]=-\frac{\tau_{31}\tau_{42}\tau_{51}}{(s_{123}-q^2)(s_{1234}-q^2)}, \qquad\ \   v[1,4,3,5,2]=\frac{\tau_{32}\tau_{41}\tau_{52}}{(s_{123}-q^2)(s_{1234}-q^2)}, \nonumber \\
    & v[1,5,4,3,2]=\frac{\tau_{32}\tau_{51}}{s_{1234}-q^2}\Big(\frac{\tau_{4,(2+3)}}{s_{123}-q^2}+1\Big),\qquad v[1,4,5,3,2]= \frac{\tau_{32}\tau_{41}\tau_{5,(2+3)}}{(s_{123}-q^2)(s_{1234}-q^2)} \,. \nonumber 
\end{align}

\subsubsection*{Five-point $\QQ_3^2=q^2$}

\begin{align}
    &v[1,3,4,5,2]=\frac{\tau_{4,(1+3)}\tau_{52}}{s_{1234}-q^2},\quad v[1,3,5,4,2]= \frac{\tau_{42}\tau_{5,(1+3)}}{s_{1234}-q^2},\quad v[1,5,3,4,2]=\frac{\tau_{42}\tau_{51}}{s_{1234}-q^2}, \nonumber \\ 
    &v[1,4,3,5,2]=\frac{\tau_{41}\tau_{52}}{s_{1234}-q^2}, \quad v[1,5,4,3,2]=\frac{\tau_{4,(2+3)}\tau_{51}}{s_{1234}-q^2},\quad v[1,4,5,3,2]= \frac{\tau_{41}\tau_{5,(2+3)}}{s_{1234}-q^2} \,. \nonumber 
\end{align}

\subsubsection*{Five-point $\QQ_4^2=q^2$}

\begin{align}
    &v[1,3,4,5,2]=-{\tau_{52}},\quad v[1,3,5,4,2]= -{\tau_{5,(2+4)}},\quad v[1,5,3,4,2]=-{\tau_{5,(2+3+4)}} \,. \nonumber 
\end{align}
They appear in \eqref{eq:5pthiddenA}, \eqref{eq:5pthiddenB} and \eqref{eq:5pthiddenC}.

\subsubsection*{Six-point $\QQ_2^2=q^2$}

\begin{align}
    &v[1,3,4,5,6,2]= \frac{\tau_{31}\tau_{62}}{s_{12345}-q^2}\Big(\frac{\tau_{4,(1+3)}}{s_{123}-q^2}+1\Big) \Big(\frac{\tau_{5,(1+3+4)}}{s_{1234}-q^2}+1\Big), \nonumber\\
    &v[1,3,4,6,5,2]= \frac{\tau_{31}\tau_{52}\tau_{6,(1+3+4)}}{(s_{1234}-q^2)(s_{12345}-q^2)}\Big(\frac{\tau_{4,(1+3)}}{s_{123}-q^2}+1\Big), \nonumber\\
    &v[1,3,5,4,6,2]= \frac{-\tau_{31}\tau_{42}\tau_{5,(1+3)}\tau_{62}}{(s_{123}-q^2)(s_{1234}-q^2)(s_{12345}-q^2)},
    \nonumber \\
     &v[1,3,5,6,4,2]= \frac{-\tau_{31}\tau_{42}\tau_{5,(1+3)}\tau_{6,(2+4)}}{(s_{123}-q^2)(s_{1234}-q^2)(s_{12345}-q^2)},
    \nonumber \\
    &v[1,3,6,4,5,2]= \frac{\tau_{31}\tau_{52}\tau_{6,(1+3)}}{(s_{1234}-q^2)(s_{12345}-q^2)}\Big(\frac{\tau_{4,(1+3)}}{s_{123}-q^2}+1\Big),
    \nonumber \\
    &v[1,3,6,5,4,2]= \frac{-\tau_{31}\tau_{42}\tau_{6,(1+3)}}{(s_{123}-q^2)(s_{12345}-q^2)}\Big(\frac{\tau_{5,(2+4)}}{s_{1234}-q^2}+1\Big),
    \nonumber \\
    &v[1,4,3,5,6,2]= \frac{-\tau_{23}\tau_{41}\tau_{26}}{(s_{123}-q^2)(s_{12345}-q^2)}\Big(\frac{\tau_{5,(1+3+4)}}{s_{1234}-q^2}+1\Big), \nonumber \\
    &v[1,4,3,6,5,2]= \frac{-\tau_{32}\tau_{41}\tau_{52}\tau_{6,(1+3+4)}}{(s_{123}-q^2)(s_{1234}-q^2)(s_{12345}-q^2)}, \nonumber \\
     &v[1,4,5,3,6,2]= \frac{-\tau_{32}\tau_{41}\tau_{62}}{(s_{123}-q^2)(s_{12345}-q^2)}\Big(\frac{\tau_{5,(1+4)}}{s_{1234}-q^2}+1\Big), \nonumber \\
     &v[1,4,5,6,3,2]= \frac{-\tau_{32}\tau_{41}\tau_{6,(2+3)}}{(s_{123}-q^2)(s_{12345}-q^2)}\Big(\frac{\tau_{5,(1+4)}}{s_{1234}-q^2}+1\Big), \nonumber \\
     &v[1,4,6,3,5,2]= \frac{-\tau_{32}\tau_{41}\tau_{52}\tau_{6,(1+4)}}{(s_{123}-q^2)(s_{1234}-q^2)(s_{12345}-q^2)}, \nonumber \\
     &v[1,4,6,5,3,2]= \frac{-\tau_{32}\tau_{41}\tau_{5,(2+3)}\tau_{6,(1+4)}}{(s_{123}-q^2)(s_{1234}-q^2)(s_{12345}-q^2)}, \nonumber \\
    &v[1, 5, 3, 4, 6, 2]= \frac{-\tau_{31}\tau_{42}\tau_{51}\tau_{62}}{(s_{123}-q^2)(s_{1234}-q^2)(s_{12345}-q^2)},
    \nonumber \\
    &v[1, 5, 3, 6, 4, 2]= \frac{-\tau_{31}\tau_{42}\tau_{51}\tau_{6,(2+4)}}{(s_{123}-q^2)(s_{1234}-q^2)(s_{12345}-q^2)},
    \nonumber \\
    &v[1, 5, 4,3,6, 2]= \frac{\tau_{32}\tau_{51}\tau_{62}}{(s_{1234}-q^2)(s_{12345}-q^2)}\Big(\frac{\tau_{4,(2+3)}}{s_{123}-q^2}+1\Big),
    \nonumber \\
    &v[1, 5, 4,6,3, 2]= \frac{\tau_{32}\tau_{51}\tau_{6,(2+3)}}{(s_{1234}-q^2)(s_{12345}-q^2)}\Big(\frac{\tau_{4,(2+3)}}{s_{123}-q^2}+1\Big),
    \nonumber \\
     &v[1, 5, 6, 3, 4, 2]= \frac{-\tau_{31}\tau_{42}\tau_{51}\tau_{6,(2+3+4)}}{(s_{123}-q^2)(s_{1234}-q^2)(s_{12345}-q^2)},
    \nonumber \\
     &v[1, 5, 6, 4, 3, 2]= \frac{\tau_{32}\tau_{51}\tau_{6,(2+3+4)}}{(s_{1234}-q^2)(s_{12345}-q^2)}\Big(\frac{\tau_{4,(2+3)}}{s_{123}-q^2}+1\Big),
    \nonumber \\
    &v[1, 6, 3, 4, 5, 2]= \frac{\tau_{31}\tau_{52}\tau_{61}}{(s_{1234}-q^2)(s_{12345}-q^2)}\Big(\frac{\tau_{4,(2+3)}}{s_{123}-q^2}+1\Big),
    \nonumber \\
    &v[1, 6, 3, 5 ,4 , 2]= \frac{-\tau_{31}\tau_{42}\tau_{61}}{(s_{123}-q^2)(s_{12345}-q^2)}\Big(\frac{\tau_{5,(2+4)}}{s_{1234}-q^2}+1\Big),
    \nonumber \\
    &v[1, 6, 4, 3,5 , 2]= \frac{-\tau_{32}\tau_{41}\tau_{52}\tau_{61}}{(s_{123}-q^2)(s_{1234}-q^2)(s_{12345}-q^2)},
    \nonumber \\
    &v[1, 6, 4, 5,3, 2]= \frac{-\tau_{32}\tau_{41}\tau_{5,(2+3)}\tau_{61}}{(s_{123}-q^2)(s_{1234}-q^2)(s_{12345}-q^2)},
    \nonumber 
\end{align}
\begin{align}
    &v[1, 6, 5,3,4 , 2]= \frac{-\tau_{31}\tau_{42}\tau_{61}}{(s_{123}-q^2)(s_{12345}-q^2)}\Big(\frac{\tau_{5,(2+4)}}{s_{1234}-q^2}+1\Big),
    \nonumber \\
    &v[1,6,5,4,3,2]= \frac{\tau_{32}\tau_{61}}{s_{12345}-q^2}\Big(\frac{\tau_{4,(2+3)}}{s_{123}-q^2}+1\Big) \Big(\frac{\tau_{5,(2+3+4)}}{s_{1234}-q^2}+1\Big) \,. \nonumber
\end{align}

\subsubsection*{Six-point $\QQ_3^2=q^2$}

\begin{align}
    &v[1,3,4,5,6,2]=  - \frac{\tau_{4,(1+3)}\tau_{62}}{s_{12345}-q^2} \Big(\frac{\tau_{5,(1+3+4)}}{s_{1234}-q^2}+1\Big), \ v[1,3,4,6,5,2]= -\frac{\tau_{4,(1+3)}\tau_{52}\tau_{6,(1+3+4)}}{(s_{1234}-q^2)(s_{12345}-q^2)}, \nonumber\\
    &v[1,3,5,4,6,2]= \frac{\tau_{42}\tau_{5,(1+3)}\tau_{62}}{(s_{1234}-q^2)(s_{12345}-q^2)},
    \ v[1,3,5,6,4,2]= \frac{\tau_{42}\tau_{5,(1+3)}\tau_{6,(2+4)}}{(s_{1234}-q^2)(s_{12345}-q^2)},
    \nonumber \\
    &v[1,3,6,4,5,2]= -\frac{\tau_{4,(1+3)}\tau_{52}\tau_{6,(1+3)}}{(s_{1234}-q^2)(s_{12345}-q^2)},
    \ v[1,3,6,5,4,2]= \frac{\tau_{42}\tau_{6,(1+3)}}{(s_{12345}-q^2)}\Big(\frac{\tau_{5,(2+4)}}{s_{1234}-q^2}+1\Big),
    \nonumber \\
    &v[1,4,3,5,6,2]= -\frac{\tau_{41}\tau_{26}}{s_{12345}-q^2}\Big(\frac{\tau_{5,(1+3+4)}}{s_{1234}-q^2}+1\Big), \ v[1,4,3,6,5,2]= -\frac{\tau_{41}\tau_{52}\tau_{6,(1+3+4)}}{(s_{1234}-q^2)(s_{12345}-q^2)}, \nonumber \\
     &v[1,4,5,3,6,2]= - \frac{\tau_{41}\tau_{62}}{s_{12345}-q^2}\Big(\frac{\tau_{5,(1+4)}}{s_{1234}-q^2}+1\Big), \ v[1,4,5,6,3,2]= -\frac{\tau_{41}\tau_{6,(2+3)}}{s_{12345}-q^2}\Big(\frac{\tau_{5,(1+4)}}{s_{1234}-q^2}+1\Big), \nonumber \\
     &v[1,4,6,3,5,2]= -\frac{\tau_{41}\tau_{52}\tau_{6,(1+4)}}{(s_{1234}-q^2)(s_{12345}-q^2)}, \ v[1,4,6,5,3,2]= -\frac{\tau_{41}\tau_{5,(2+3)}\tau_{6,(1+4)}}{(s_{1234}-q^2)(s_{12345}-q^2)},  \nonumber \\
    &v[1, 5, 3, 4, 6, 2]= \frac{\tau_{42}\tau_{51}\tau_{62}}{(s_{1234}-q^2)(s_{12345}-q^2)},
    \ v[1, 5, 3, 6, 4, 2]= \frac{\tau_{42}\tau_{51}\tau_{6,(2+4)}}{(s_{1234}-q^2)(s_{12345}-q^2)},
    \nonumber \\
    &v[1, 5, 4,3,6, 2]= \frac{\tau_{4,(2+3)}\tau_{51}\tau_{62}}{(s_{1234}-q^2)(s_{12345}-q^2)},
    \ v[1, 5, 4,6,3, 2]= \frac{\tau_{4,(2+3)}\tau_{51}\tau_{6,(2+3)}}{(s_{1234}-q^2)(s_{12345}-q^2)},
    \nonumber \\
     &v[1, 5, 6, 3, 4, 2]= \frac{\tau_{42}\tau_{51}\tau_{6,(2+3+4)}}{(s_{1234}-q^2)(s_{12345}-q^2)},
    \ v[1, 5, 6, 4, 3, 2]= \frac{\tau_{4,(2+3)}\tau_{51}\tau_{6,(2+3+4)}}{(s_{1234}-q^2)(s_{12345}-q^2)},
    \nonumber \\
    &v[1, 6, 3, 4, 5, 2]= \frac{\tau_{4,(2+3)}\tau_{52}\tau_{61}}{(s_{1234}-q^2)(s_{12345}-q^2)},
    \ v[1, 6, 3, 5 ,4 , 2]= \frac{\tau_{42}\tau_{61}}{s_{12345}-q^2}\Big(\frac{\tau_{5,(2+4)}}{s_{1234}-q^2}+1\Big),
    \nonumber \\
    &v[1, 6, 4, 3,5 , 2]= -\frac{\tau_{41}\tau_{52}\tau_{61}}{(s_{1234}-q^2)(s_{12345}-q^2)},
    \ v[1, 6, 4, 5,3, 2]= - \frac{\tau_{41}\tau_{5,(2+3)}\tau_{61}}{(s_{1234}-q^2)(s_{12345}-q^2)},
    \nonumber \\
    &v[1, 6, 5,3,4 , 2]= \frac{\tau_{42}\tau_{61}}{s_{12345}-q^2}\Big(\frac{\tau_{5,(2+4)}}{s_{1234}-q^2}+1\Big),
    \ v[1,6,5,4,3,2]= \frac{\tau_{4,(2+3)}\tau_{61}}{s_{12345}-q^2}\Big(\frac{\tau_{5,(2+3+4)}}{s_{1234}-q^2}+1\Big). \nonumber
\end{align}

\subsubsection*{Six-point $\QQ_4^2=q^2$}

\begin{align}
    &v[1,3,4,5,6,2]= \frac{\tau_{5,(1+3+4)}\tau_{62}}{s_{12345}-q^2}, \ v[1,3,4,6,5,2]= \frac{\tau_{52}\tau_{6,(1+3+4)}}{s_{12345}-q^2}, \ v[1,3,5,4,6,2]= \frac{\tau_{5,(1+3)}\tau_{62}}{s_{12345}-q^2}, \nonumber \\
    & v[1,3,5,6,4,2]= \frac{\tau_{5,(1+3)}\tau_{6,(2+4)}}{s_{12345}-q^2}, \  v[1,3,6,4,5,2]= \frac{\tau_{52}\tau_{6,(1+3)}}{s_{12345}-q^2}, \
    v[1,3,6,5,4,2]= \frac{\tau_{5,(2+4)}\tau_{6,(1+3)}}{(s_{12345}-q^2)}, \nonumber \\
    &v[1, 5, 3, 4, 6, 2]= \frac{\tau_{51}\tau_{62}}{s_{12345}-q^2},
    \ v[1, 5, 3, 6, 4, 2]= \frac{\tau_{51}\tau_{6,(2+4)}}{s_{12345}-q^2}, \ v[1, 5, 6, 3, 4, 2]= \frac{\tau_{51}\tau_{6,(2+3+4)}}{s_{12345}-q^2},
    \nonumber \\
    &v[1, 6, 3, 4, 5, 2]= \frac{\tau_{52}\tau_{61}}{s_{12345}-q^2},
    \ v[1, 6, 3, 5 ,4 , 2]= \frac{\tau_{5,(2+4)}\tau_{61}}{s_{12345}-q^2}, 
    \ v[1, 6, 5,3,4 , 2]= \frac{\tau_{5,(2+4)}\tau_{61}}{s_{12345}-q^2} \,.
     \nonumber
\end{align}

\subsubsection*{Six-point $\QQ_5^2=q^2$}
\begin{align}
    &v[1,3,4,5,6,2]= -\tau_{62}, \qquad \qquad v[1,3,4,6,5,2]= -\tau_{6,(2+5)}, \nonumber \\
    & v[1,3,6,4,5,2]= -\tau_{6,(2+4+5)}, \quad v[1, 6, 3, 4, 5, 2]= -\tau_{6,(2+3+4+5)} \,. \nonumber 
\end{align}

\subsubsection*{Comment on obtaining the $\vec{v}$ vectors}
Based on the discussion in Section~\ref{ssec:Hiddenfac}, the denominators, or say the pole structures, of $\vec{v}$s are transparent: the $\vec{v}$ vectors contain the spurious pole $\QQ_m^2{-}q^2$ is\footnote{Even without the explicit expression, this is also understandable due to the Feynman propagator ${s_{12\cdots m \cdots (m+i) }-q^2}$ ($0{<}i{<}n{-}m$) in the amplitudes on the RHS of \eqref{eq:nptgeneralizedBCJ}.} 
    \begin{equation}\label{eq:generalvdeno}
	\vec{v}\ \propto \  \prod_{k=m+1}^{n-1} \frac{1}{s_{12\cdots k}-q^2} \,,
    \end{equation}
    which is a degree $(n{-}m{-}1)$ polynomial. 
    By dimension analysis, the $\vec{v}$ vector must have a numerator factor of a degree-$(n{-}m)$ polynomial,  and the major task is to study this degree $(n{-}m)$ polynomial.

The conditions imposed on the degree-$(n{-}m)$ polynomial are the null vector condition (see Appendix~A in the forthcoming paaper) and the hidden factorization relation \eqref{eq:nptgeneralizedBCJ}, of which the MHV version is particularly useful.\footnote{The MHV form factors are given simply by the Parke-Taylor form $$\mathcal{F}(1^{\phi},3^{+},..,n^{+},2^{\phi})=\frac{\langle 12 \rangle^2 }{\langle 13 \rangle \cdots \langle n2 \rangle \langle 21 \rangle}$$ and the MHV amplitude (with two massive scalars and many positive helicity gluons) can be found in \cite{Badger:2005zh}. } 
One can make an ansatz for the polynomial and fix it with the aforementioned conditions to get the data above.
However, we should remind that the naive polynomial ansatz for high-point $\vec{v}$s become cumbersome so that it is convenient to directly check the closed formula in Section~3.2 in the forthcoming paper. 

\section{On the factorizations on physical poles}\label{ap:physical}

In this appendix, we provide more details about the factorization of the double copy on the ``physical"-type poles,
which has been discussed in Section~\ref{sssec:thetadecomp} and Section~\ref{ssec:FFDCnptfac}.
Our discussion will be parallel to the discussion for the ``spurious"-type poles:
\begin{itemize}
\item 
When discussing the  ``spurious"-type poles, we focus on  the KLT kernel $\mathbf{S}^{\cal F}$. We know from Section~\ref{sssec:KLTkerneldecomp} that the decomposition of the KLT kernel gives the factorized sub-kernels and the $\mathbf{V}$ matrix (\emph{i.e.}~$\vec{v}$ vectors). Furthermore, $\mathbf{V}$  leads to the hidden factorization relations for form factors as discussed in Section~\ref{ssec:Hiddenfac}. 
\item
Here, to consider the  ``physical"-type poles, the focus is on the propagator matrix $\Theta^{\cal F}$. The decomposition of the propagator matrix will give the factorized sub-propagator-matrices and the $\mathbf{Y}$ matrix. Moreover, $\mathbf{Y}$ will lead to factorization relations for the CK-dual numerators $N$. 
\end{itemize}

Let us start with the propagator matrix decomposition in Section~\ref{ssec:bilinearDecomp}. In \eqref{eq:nptmFDecomp1}, we proposed a decomposition for the propagator matrix $\Theta^{\cal F}$. Specified with the $s_{13\cdots r}$ pole, the decomposition is as follows
\begin{equation}
    \text{Res}_{s_{13\cdots r}=0}[\Theta^{\cal F}_n]\sim \Theta^{\cal A}_{r}\otimes \Theta^{\cal F}_{r'}\,, 
\end{equation}
where $r'=n{-}r{-}2$ and the two smaller propagator matrix are 
\begin{equation}
\begin{aligned}
    \Theta^{\cal A}_{r}&=\Theta^{\cal A}_{r}[1,\bar{\alpha}_1(3,\ldots,r),I|1,\bar{\alpha}_2(3,\ldots,r),I], \text{ with } \bar{\alpha}_{1,2} \in S_{r-2}\,,\\
    \Theta^{\cal F}_{r'}&=\Theta^{\cal F}_{r'}[{-}I,\bar{\beta}_1(r{+}1,\ldots,n),2|{-}I,\bar{\beta}_2(r{+}1,\ldots,n),2], \text{ with } \bar{\beta}_{1,2} \in S_{n-r} \,.
\end{aligned}
\end{equation}
The first matrix has the rank $(r{-}1)!$ permitting the following decomposition
\begin{equation}\label{eq:propdecappB}
    \Theta^{\cal A}_r=\mathbf{Y}^{\scriptscriptstyle \rm T} \cdot \overline{\Theta}^{\cal A}_r \cdot \mathbf{Y}
\end{equation}
where we will explain $\mathbf{Y}$ later and  $\overline{\Theta}^{\cal A}_r$ is a $(r{-}1)!$ by $(r{-}1)!$ minor of $\Theta^{\cal A}_r$ as 
\begin{equation}
    \overline{\Theta}^{\cal A}_r =\Theta^{\cal A}_{r}[1,3,\bar{\gamma}_1(4,\ldots,r),I|1,3,\bar{\gamma}_2(4,\ldots,r),I], \text{ with } \bar{\gamma}_{1,2} \in S_{r-3}\,.
\end{equation}
We abbreviate the matrix elements of the above matrices as $\Theta^{\cal A}_r[\bar{\alpha}_1|\bar{\alpha}_2]$, $\Theta^{\cal F}_{r'}[\bar{\beta}_1|\bar{\beta}_2]$ and $\overline{\Theta}^{\cal A}_{r}[\bar{\gamma}_1|\bar{\delta}_1]$, with the permutations defined above. Using these notations, we spell out the decomposition \eqref{eq:propdecappB} more explicitly as
\begin{align}
	 \Theta^{\mathcal{A}}_{r}[\bar{\alpha}_1|\bar{\alpha}_2] &=\sum_{\bar{\gamma}_{1,2}\in S_{r-3}} \mathbf{Y}[\bar{\gamma}_1|\bar{\alpha}_1]\ \overline{\Theta}^{\mathcal{A}}_{r}[\bar{\gamma}_1|\bar{\gamma}_2] \  \mathbf{Y}[\bar{\gamma}_2|\bar{\alpha}_2] \,,
\end{align}
in which the matrix element for the matrix $\mathbf{Y}$ is labeled as \emph{e.g.}~$\mathbf{Y}[\bar{\delta}_1|\bar{\beta}_1]$ with $\bar{\delta}_1\in S_{r-3}$ permuting $\{4,\ldots,r\}$ and $\bar{\beta}_1\in S_{r-2}$ permuting $\{3,\ldots,r\}$. 

Synthesizing everything above, we have 
\begin{align}\label{eq:nptmFDecomp4}
	\text{Res}_{s_{13\cdots r}=0}[\Theta_n^{\mathcal{F}}[1,\alpha,2|1,\beta,2]]=&\ \Theta_{r}^{\mathcal{A}}[1,\bar{\alpha}_1,I|1,\bar{\alpha}_2,I] \    \Theta_{r'}^{\mathcal{F}} [{-}I,\bar{\beta}_1,2|{-}I,\bar{\beta}_2,2] \\
 =& \hskip -5pt \sum_{\bar{\gamma}_{1,2}\in S_{r-3}} \hskip -5pt  \mathbf{Y}[\bar{\gamma}_1|\bar{\alpha}_1] 
 \ \overline{\Theta}^{\mathcal{A}}_{r}[\bar{\gamma}_1|\bar{\gamma}_2] \Theta_{r'}^{\mathcal{F}} [\bar{\beta}_1|\bar{\beta}_2]
 \ \mathbf{Y}[\bar{\gamma}_2|\bar{\alpha}_2]\nonumber\,,
\end{align}
which is the component form of \eqref{eq:nptmFDecomp2}, as a refinement of \eqref{eq:nptmFDecomp}. 

We also comment that the amplitudes propagator matrix has a similar decomposition as given in \eqref{eq:nptmADecomp}, which takes the following more detailed form 
\begin{align}\label{eq:nptmADecomp2}
	&\text{Res}_{s_{13\cdots r}=0}[\overline{\Theta}_n^{\mathcal{A}}[1,\sigma,n,2|1,\theta,n,2]]\nonumber \\
 =&\ \Theta_{r}^{\mathcal{A}}[1,\bar{\sigma}_1,I|1,\bar{\sigma}_2,I]  \  \overline{\Theta}_{r'}^{\mathcal{A}} [{-}I,\bar{\theta}_1,n,2|{-}I,\bar{\theta}_2,n,2] \\
 =& \hskip -5pt \sum_{\bar{\gamma}_{1,2}\in S_{r-3}} \hskip -5pt  \mathbf{Y}[\bar{\gamma}_1|\bar{\sigma}_1] 
 \ \overline{\Theta}^{\mathcal{A}}_{r}[1,3,\bar{\gamma}_1,I|1,3,\bar{\gamma}_2,I] 
 \ \overline{\Theta}_{r'}^{\mathcal{A}} [{-}I,\bar{\theta}_1,n,2|{-}I,\bar{\theta}_2,n,2] 
 \mathbf{Y}[\bar{\gamma}_2|\bar{\sigma}_2]\nonumber\,,
\end{align}
where the permutations involved here are 
\begin{equation}
\begin{aligned}
    &\sigma=\{\sigma_1,\theta_1\},\theta=\{\sigma_2,\theta_2\}\in S_{n-3} \text{ permuting } \{3,\ldots,(n{-}1)\}\,, \\
    &\bar{\sigma}_{1,2}\in S_{r-2} \text{ permuting } \{3,\ldots,r\}\,, \\
    &\bar{\theta}_{1,2}\in S_{n-r-1}=S_{r'-3} \text{ permuting } \{(r{+1}),\ldots,(n{-}1)\}\,, \\
    &\bar{\gamma}_{1,2}\in S_{r-3} \text{ permuting } \{4,\ldots,r\}\,.
\end{aligned}
\end{equation}
For the corresponding compact matrix form, see \eqref{eq:nptmFDecomp2} and \eqref{eq:nptmADecomp}.\footnote{Note that to express the tensor product correctly, we introduced $\mathbb{I}$ and $\bar{\mathbb{I}}$, where $\mathbb{I}_a$ is a $(a{-}2)!$ by $(a{-}2)!$ identity matrix and $\bar{\mathbb{I}}_a$ is a $(a{-}3)!$ by $(a{-}3)!$ identity matrix, 
in \eqref{eq:nptmFDecomp2} and \eqref{eq:nptmADecomp}. But these matrices have matrix elements like $\mathbb{I}[\bar{\beta}_1|\bar{\beta}_2]=\delta_{\bar{\beta}_1\bar{\beta}_2}$ so we omit these matrix elements.}

After introducing all the orderings and components for the propagator matrix decomposition, we use these notations to express the physical pole factorization.
Starting from the propagator matrix form of double copy  \eqref{eq:genericBCJKLT}, we take the $s_{13\cdots r}$ residue, getting 
\begin{align}\label{eq:Gnphysicalpole1}
	&\text{Res}_{s_{13\cdots r}=0}[\mathcal{G}_{n}] \nonumber \\
 &=\sum_{\alpha,\beta\in S_{n-2}} N_{n}^{\cal F}[1,\alpha,2] \ 
 \text{Res}_{s_{13\cdots r}=0} \big[\Theta_{n}^{\cal F}[1,\alpha,2|1,\beta,2]\big] \ 
 N_{n}^{\cal F}[1,\beta,2]\big|_{s_{13\cdots r}=0}\\
 &=
\sum_{\substack{\bar{\alpha}_{1,2}\in S_{r-2} \\ \bar{\beta}_{1,2}\in S_{n-r} }}
	N_{n}^{\cal F}[1,\bar{\alpha}_1,\bar{\beta}_1,2] 
	\Theta^{\mathcal{A}}_{r}[1,\bar{\alpha}_1,I|1,\bar{\alpha}_2,I]  
	\Theta^{\mathcal{F}}_{r'}[{-}I,\bar{\beta}_1,2|{-}I,\bar{\beta}_2,2]  
	N_{n}^{\cal F}[1,\bar{\alpha}_2,\bar{\beta}_2,2] \nonumber\\
 &=\sum_{\substack{\bar{\gamma}_{1,2}\in S_{r-3}\\ \bar{\beta}_{1,2}\in S_{n-r} }} \overline{\Theta}^{\mathcal{A}}_{r}[1,3,\bar{\gamma}_1,I|1,3,\bar{\gamma}_2,I] 
 \ \overline{\Theta}_{r'}^{\mathcal{F}} [{-}I,\bar{\beta}_1,2|{-}I,\bar{\beta}_2,2] \nonumber \\
	&\qquad \qquad   \times \Big(\sum_{\bar{\alpha}_{1}\in S_{r-2}}  \mathbf{Y}[\bar{\gamma}_1|\bar{\alpha}_1] N_{n}^{\cal F}[1,\bar{\alpha}_1,\bar{\beta}_1,2]\Big) 
 \Big(\sum_{\bar{\alpha}_{2}\in S_{r-2}}  \mathbf{Y}[\bar{\gamma}_2|\bar{\alpha}_2] N_{n}^{\cal F}[1,\bar{\alpha}_2,\bar{\beta}_2,2]\Big)\nonumber \,.
\end{align}
This looks like the factorization of amplitudes double copy
\begin{align}\label{eq:Ampphysicalpole}
    &\text{Res}_{s_{13\cdots r}=0}[\mathcal{M}_{n}] \nonumber \\
    &= \sum_{\sigma,\theta\in S_{n-3}} 
    N_{n}^{\cal A}[1,\sigma,n,2] \ 
    \text{Res}_{s_{13\cdots r}=0}[\overline{\Theta}_n^{\mathcal{A}}[1,\sigma,n,2|1,\theta,n,2]] \ 
    N_{n}^{\cal A}[1,\theta,n,2]\big|_{s_{13\cdots r}=0}\\
    &=\sum_{\substack{\bar{\sigma}_{1,2}\in S_{r-2} \\ \bar{\theta}_{1,2}\in S_{r'-3} }} N_{n}^{\cal A}[1,\bar{\sigma}_1,\bar{\theta}_1,n,2]  
    \Theta_{r}^{\mathcal{A}}[1,\bar{\sigma}_1,I|1,\bar{\theta}_1,I]  \overline{\Theta}_{r'}^{\mathcal{A}} [{-}I,\bar{\theta}_1,n,2|{-}I,\bar{\theta}_2,n,2] 
	N_{n}^{\cal A}[1,\bar{\sigma}_2,\bar{\theta}_2,n,2] \nonumber \\
 &=\sum_{\substack{\bar{\gamma}_{1,2}\in S_{r-3}\\ \bar{\theta}_{1,2}\in S_{r'-3} }} \overline{\Theta}^{\mathcal{A}}_{r}[1,3,\bar{\gamma}_1,I|1,3,\bar{\gamma}_2,I] 
 \ \overline{\Theta}_{r'}^{\mathcal{A}} [{-}I,\bar{\theta}_1,n,2|{-}I,\bar{\theta}_2,n,2] \nonumber \\
	&\qquad \qquad   \times \Big(\sum_{\bar{\sigma}_{1}\in S_{r-2}}  \mathbf{Y}[\bar{\gamma}_1|\bar{\sigma}_1] N_{n}^{\cal A}[1,\bar{\sigma}_1,\bar{\theta}_1,n,2]\Big) 
 \Big(\sum_{\bar{\sigma}_{2}\in S_{r-2}}  \mathbf{Y}[\bar{\gamma}_2|\bar{\sigma}_2] N_{n}^{\cal A}[1,\bar{\sigma}_2,\bar{\theta}_2,n,2]\Big)\nonumber \,.
\end{align}

To further write \eqref{eq:Gnphysicalpole1} as factorized lower-point form factor and amplitude, we need a factorization formula for the master numerators. We can expect such a factorization as 
\begin{align}
\sum_{\bar{\alpha}_{1}\in S_{r-2}}  \mathbf{Y}[\bar{\gamma}_1|\bar{\alpha}_1] N^{\cal F}_{n}[1,\bar{\alpha}_1,\bar{\beta}_1,2] \big|_{s_{13\cdots r}=0} &=N_r^{\mathcal{A}}[1,{3},\bar{\gamma}_1,I] N_{r'}^{\mathcal{F}}[{-}I,\bar{\beta}_1,2]\,,\label{eq:NfactFF}\\
\sum_{\bar{\sigma}_{1}\in S_{r-2}}  \mathbf{Y}[\bar{\gamma}_1|\bar{\sigma}_1] N_{n}^{\cal A}[1,\bar{\sigma}_1,\bar{\theta}_1,n,2] \big|_{s_{13\cdots r}=0} &=N_r^{\mathcal{A}}[1,{3},\bar{\gamma}_1,I] N_{r'}^{\mathcal{A}}[{-}I,\bar{\theta}_1,n,2]\,. \label{eq:NfactAmp}
\end{align} 
These relations correspond to the component form of the compact formulae \eqref{eq:Numfact}, where we have the tensor product of two (column) vectors $\vec{N}_r\otimes \vec{N}_{r'}$. The elements of $\vec{N}^{\cal A}_{r}$, $\vec{N}^{\cal F}_{r'}$ and $\vec{N}^{\cal A}_{r'}$ are respectively $N_r^{\mathcal{A}}[1,{3},\bar{\gamma}_1,I]$, $N_{r'}^{\mathcal{F}}[{-}I,\bar{\beta}_1,2]$ and  $N_{r'}^{\mathcal{A}}[{-}I,\bar{\theta}_1,n,2]$, respectively. 

Below we justify these relations via the following argument:
for amplitudes, we know that they have the factorization property as
\begin{equation}\label{eq:Mnphysicalres}
\begin{aligned}
	&\text{Res}_{s_{13 .. r}=0}[\mathcal{M}_{n}] = \mathcal{M}_{r} \times \mathcal{M}_{r'} \\
	& = \sum_{\substack{\bar{\gamma}_{1,2}\in S_{r-3}\\ \bar{\theta}_{1,2}\in S_{r'-3} }} 
	N_r^{\mathcal{A}}[1^{\phi},{3},\bar{\gamma}_1,I]
	 \Theta^{\mathcal{A}}_{r}[1,3,\bar{\gamma}_1,I|1,3,\bar{\gamma}_2,I] 	 
	 N_r^{\mathcal{A}}[1^{\phi},{3},\bar{\gamma}_2,I]\\
	&  \qquad \qquad \times  N_{r'}^{\mathcal{A}}[{-}I,\bar{\theta}_1,n,2]
	 \Theta^{\mathcal{A}}_{r'}[{-}I,\bar{\theta}_1,n,2|{-}I,\bar{\theta}_2,n,2]
	 N_{r'}^{\mathcal{A}}[{-}I,\bar{\theta}_2,n,2]\,.
\end{aligned}
\end{equation}
By inspecting the equivalence between \eqref{eq:Mnphysicalres} and \eqref{eq:Ampphysicalpole}, we know that \eqref{eq:NfactAmp} must be true. 
Then we can convince ourselves that \eqref{eq:NfactFF} is also true, since \eqref{eq:NfactFF} has a parallel structure as \eqref{eq:NfactAmp}: 
(i) they have exactly the same sum over permutations $S_{n-2}$; (ii) the $\mathbf{Y}$ matrices involved are exactly the same; (iii) the master numerator expressions are similar\footnote{For the amplitude numerators, one need to use the momentum conservation to remove one particle's momentum. In the expression for $N^{\mathcal{A}}$, we remove the momentum $p_{1}^{\phi}$ and the next momentum $p_{3}^{g}$ plays the similar role of $p_{1}^{\phi}$ in $N^{\mathcal{F}}$. }
\begin{equation}
    {N}^{\cal F}_n[1^{\phi},3^{g},\ldots, n^{g},2^{\phi}]=\sum_{k=1}^{n-2}\sum_{\tau\in \mathbf{P}_{\mathbf{g}}^{(k)}}(-2)^k{\prod\limits_{i=1}^k \Big(p_{1\Xi_L(i)}\cdot f_{(\tau_i)}\cdot p_{2 {\Xi}_R(i)}\Big)\over  (p_{12}^2{-}q^2) (p_{12\tau_1}^2{-}q^2)\cdots (p_{12\tau_1\cdots \tau_{k-1}}^2{-}q^2)},
\end{equation}
\begin{equation}\label{eq:BCJnumA}
 N^{\mathcal{A}}_n[1^{\phi},3^{g},\ldots, n^{g}, 2^{\phi}]=\sum_{k=1}^{n{-}3}\sum_{\tau\in \mathbf{P}_{\mathbf{g}\backslash  \{3\}}^{(k)}} (-2)^k {p_{2}\cdot f_{(3\tau_1)}\cdot p_{2}\prod\limits_{i=2}^k \Big(p_{\Xi_L(i)}\cdot f_{(\tau_i)}\cdot p_{2\Xi_R(i)}\Big)\over  (p_{23}^2{-}m^2) (p_{23\tau_1}^2{-}m^2)\cdots (p_{23\tau_1\cdots \tau_{k-1}}^2{-}m^2)}\,,
\end{equation}
of which the notations are given after \eqref{eq:BCJnumF}. 
One can also check that \eqref{eq:NfactFF} is indeed true via explicit calculations. 

In the end, after giving an argument on the numerator factorization relation \eqref{eq:NfactFF}, we plug \eqref{eq:NfactFF} into  \eqref{eq:Gnphysicalpole1} so that 
\begin{equation}
\begin{aligned}\label{eq:Gnphysicalpolefinal}
&\text{Res}_{s_{13\cdots r}=0}[\mathcal{G}_{n}] \nonumber \\
 &=\sum_{\substack{\bar{\gamma}_{1,2}\in S_{r-3}\\ \bar{\theta}_{1,2}\in S_{r'-2} }} \overline{\Theta}^{\mathcal{A}}_{r}[1,3,\bar{\gamma}_1,I|1,3,\bar{\gamma}_2,I] 
 \ \Theta^{\mathcal{F}}_{r'}[{-}I,\bar{\beta}_1,2|{-}I,\bar{\beta}_2,2]  \nonumber \\
	&\qquad \qquad   \times \Big(N_r^{\mathcal{A}}[1,{3},\bar{\gamma}_1,I] N_{r'}^{\mathcal{F}}[{-}I,\bar{\beta}_1,2]\Big) 
 \Big(N_r^{\mathcal{A}}[1,{3},\bar{\gamma}_2,I] N_{r'}^{\mathcal{F}}[{-}I,\bar{\beta}_2,2]\Big)\nonumber \\
 &= \bigg(\sum_{\bar{\gamma}_{1,2}\in S_{r-3} } N_r^{\mathcal{A}}[1,{3},\bar{\gamma}_1,I] 
 \overline{\Theta}^{\mathcal{A}}_{r}[1,3,\bar{\gamma}_1,I|1,3,\bar{\gamma}_2,I]  
 N_r^{\mathcal{A}}[1,{3},\bar{\gamma}_2,I]\bigg) \nonumber \\
 &\qquad \times \bigg(\sum_{ \bar{\theta}_{1,2}\in S_{r'-2}} N_{r'}^{\mathcal{F}}[{-}I,\bar{\beta}_1,2] 
 \Theta^{\mathcal{F}}_{r'}[{-}I,\bar{\beta}_1,2|{-}I,\bar{\beta}_2,2] 
 N_{r'}^{\mathcal{F}}[{-}I,\bar{\beta}_2,2]\bigg) \nonumber \\
 &= \mathcal{M}_r\times \mathcal{G}_{r'}\,.
\end{aligned}
\end{equation}
This provides also an explicit form for \eqref{eq:Gnphysicalpole}.

\section{On the master numerators of general $\operatorname{tr}(\phi^2)$ form factors}\label{ap:nums}

In this appendix, we give the results and properties of the master numerators\footnote{Technically, here we mean the kinematic part of the  numerators. When considering multi-scalar  (scalar number $\geq 4$) numerators, the flavor factors are also important ingredients. But the flavor factors of the half-ladder diagrams are easy to spell out, see \cite{Chen:2022nei} for details.} for generic $\operatorname{tr}(\phi^2)$ form factors, \emph{i.e.} form factors with $m$-scalar external lines.
The construction was first finished in \cite{Chen:2022nei,Brandhuber:2022enp} by one of the authors and other collaborators using kinematic Hopf algebra, and here we just quote the result and highlight some important properties for the form factor double copy. We also comment that these numerators will be important in the forthcoming paper \cite{treepaper2} when discussing high-length operators. 

We consider the numerator $N[1^{\phi},\alpha(3^{\phi},..,m^{\phi},(m{+}1)^{g},..,n^{g}),2^{\phi}]$, which is also the numerator of the half-ladder diagram in the DDM basis fixing $1^{\phi},2^{\phi}$. The permutation $\beta$ mixes the positions of gluons and scalars, so that the ordering now looks like \emph{e.g.} $\{1^{\phi},5^{g},4^{\phi},6^{g},3^{\phi},8^{g},$ $4^{\phi},9^{g},7^{g},2^{\phi}\}$. 
Focusing only on the external scalars or gluons, one gets two ordered sets $\mathbf{s}=\{1^{\phi},i_1,i_2,\ldots,i_{m-2},2^{\phi}\}$ and $\mathbf{g}=\{j_1,j_2,\ldots,j_{n-m}\}$, for the ordering above, $\mathbf{s}=\{1^{\phi},4^{\phi},3^{\phi},2^{\phi}\}$ and $\mathbf{g}=\{5^{g},6^{g},8^{g},9^{g},7^{g}\}$. 

Then we can define the ordered partition for the gluon set as $\mathbf{P}_{\mathbf{g}}^{(r)}$, which is the collection of all possible ways dividing the gluon set into $r$ (ordered) subsets. 
Then the numerator in general takes the following form 
\begin{equation}\label{eq:BCJnumFgen}
	N[1^{\phi},\alpha(3^{\phi},..,m^{\phi},(m{+}1)^{g},..,n^{g}),2^{\phi}]={\rm tr}_{\rm FL}(t^{I_1}t^{I_{i_1}}\cdots t^{I_{i_{m-2}}}t^{I_{2}})\sum_{r=1}^{|\mathbf{g}|}\sum_{\tau\in \mathbf{P}_{\mathbf{g}}^{(r)} } n^{(1\alpha 2)}_{(\tau_1),\ldots,(\tau_r)}\,,
\end{equation}
where $|\mathrm{g}|$ is the length of the gluon set, $\tau$ is one particular way dividing $\mathbf{g}$ into $r$ subsets, and these $r$ subsets are $(\tau_1),\ldots,(\tau_r)$ and the flavor trace ${\rm tr}_{\rm FL}$ is defined according to the scalar ordering $\mathbf{s}=\{1^{\phi},i_1,i_2,\ldots,i_{m-2},2^{\phi}\}$ . We are going to write down the formulae for $n^{(1\alpha 2)}_{(\tau_1),\ldots,(\tau_r)}$. 

To do this, we need to introduce the ``musical'' diagram first. Given the scalar set $\mathbf{s}$ and the particular ordered partition $\tau$ of $\mathbf{g}$, we can draw the ``musical'' diagram as follows. First, we embed the scalars as well as the partitions of gluons $\tau_1$ to $\tau_{r}$ into different levels: $\mathbf{s}$ lives on the bottom line, $\tau_1$ is on a line above it, then $\tau_2$, until $\tau_r$. Next, we require that when projecting the elements in all the levels onto the bottom line, the ordering should be exactly ${\alpha}$ -- the color ordering of all the external particles. These requirements uniquely fix the relative positions in both the vertical and the horizontal directions on the ``musical diagram''. 
For example, for $\mathbf{s}=\{1^{\phi},4^{\phi},3^{\phi},2^{\phi}\}$, $\tau_1=\{6^{g},7^{g},9^{g}\}$ and $\tau_2=\{5^{g},8^{g}\}$, we have the diagram 
\begin{equation}\label{eq:musical}
\begin{aligned}
    \includegraphics[width=0.55\linewidth]{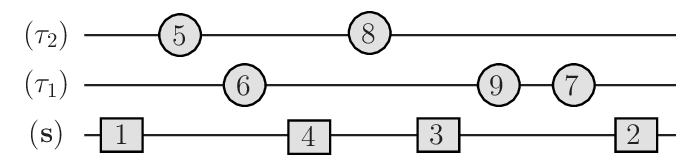}
\end{aligned}\,.
\end{equation}

Based on the musical diagram, we can write down the following expression for $n^{(1\alpha 2)}_{(\tau_1),\ldots,(\tau_r)}$ as 
 \begin{align}
 n^{(1\alpha 2)}_{(\tau_1),\ldots,(\tau_r)}={(-2)^r\prod\limits_{i=1}^r \Big(p_{\widetilde{\Xi}_L(\tau_i)}\cdot f_{(\tau_i)}\cdot p_{\widetilde{\Xi}_R(\tau_i)}\Big) \over  (p_{\mathbf{s}}^2{-}q^2) (p_{\mathbf{s}\tau_1}^2{-}q^2)\cdots (p_{\mathbf{s}\tau_1\cdots \tau_{r-1}}^2{-}q^2)} \,,
\end{align}
where $\widetilde{\Xi}_{L}(\tau_i)$ is the collection of  lower-left indices of the first element of $\tau_{i}$ and $\widetilde{\Xi}_{R}(\tau_i)$  is the collection of  lower-right indices of the last element of $\tau_{i}$, including the scalars. One can show that this definition is equivalent to the one that we wrote down before in \eqref{eq:BCJnumF} and \eqref{eq:BCJnumA}. Using the same example in \eqref{eq:musical}, we have 
\begin{equation}
	p_{\widetilde{\Xi}_{L}(\tau_1)}=p_1, \quad p_{\widetilde{\Xi}_{R}(\tau_1)}=p_2, \quad p_{\widetilde{\Xi}_{L}(\tau_2)}=p_1,\quad p_{\widetilde{\Xi}_{R}(\tau_2)}=p_2{+}p_{3}{+}p_{7}{+}p_9=p_{2379},
\end{equation}
so that
 \begin{align}
 n^{(156483972)}_{(679),(58)}={(-2)^2 \Big(p_{1}\cdot f_{6}\cdot f_{9}\cdot f_{7}\cdot p_{2}\Big) \Big(p_{1}\cdot f_{5}\cdot f_{8}\cdot p_{2379}\Big)  \over  (p_{1234}^2{-}q^2) (p_{1234679}^2{-}q^2)}\,.
\end{align}
In this way, we can spell out all the $n$s involved in \eqref{eq:BCJnumFgen} and get a complete formula for $N$. 

Below we analyze the properties of the numerators \eqref{eq:BCJnumFgen}. 
\begin{enumerate}[topsep=3pt,itemsep=-1ex,partopsep=1ex,parsep=1ex]
\item The denominator of the $n^{(1\alpha 2)}_{(\tau_1),\ldots,(\tau_r)}$ poles as massive Feynman diagram propagators. In particular we have the following diagram 
\begin{equation}
    \begin{aligned}
        \includegraphics[width=0.45\linewidth]{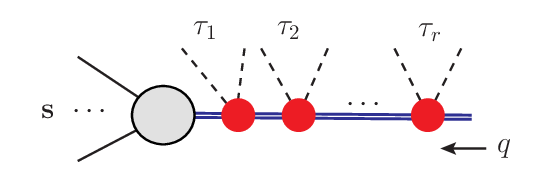}
    \end{aligned}\,
\end{equation}
where the explicitly shown double-line propagators are exactly the denominator of  $n^{(1\alpha 2)}_{(\tau_1),\ldots,(\tau_r)}$. This contributes to the 4-th point (around \eqref{eq:multiscalarG}) in the generalizations discussed at the end of  Section~\ref{sec:generalize2}.

\item The $N[1,\alpha,2]$ in \eqref{eq:BCJnumFgen} is also defined as ``prenumerators" in \cite{Chen:2022nei}. Here we regard it as the numerator of an ordering. To relate $N[1,\alpha,2]$ to the numerator of a specific cubic diagram, \emph{i.e.} $N(\Gamma_i)$, we need to take proper commutators. The details of taking commutators are included in  \cite{Chen:2022nei}, and here we just quote one special conclusion that if we take $\Gamma_i$ to be the basis DDM diagram, then we have 
\begin{equation}
	 N\bigg( \hskip -3pt \begin{aligned}
        \includegraphics[width=0.28\linewidth]{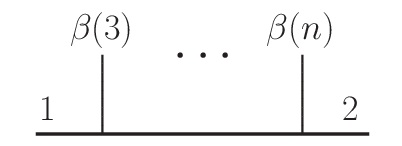}
        \end{aligned}\hskip -6pt\bigg)=\tilde{f}^{1i_1\mathrm{x}_1}\tilde{f}^{\mathrm{x}_1i_2\mathrm{x}_2}\cdots \tilde{f}^{\mathrm{x}_{m-3}i_{m-2}2}\sum_{r=1}^{|\mathbf{g}|}\sum_{\tau\in \mathbf{P}_{\mathbf{g}}^{(r)} } n^{(1\alpha 2)}_{(\tau_1),\ldots,(\tau_r)} \,,
\end{equation}
where $\beta(i)$ can be either scalars or gluons. The important point is that the kinematic part of $N(\Gamma_i)$ and $N[1,\alpha,2]$ are the same
\begin{equation}
    N[1,\alpha,2]|_{\text{flavor factor}\rightarrow 1}=N\bigg( \hskip -3pt \begin{aligned}
        \includegraphics[width=0.28\linewidth]{figure/mscalars.eps}
        \end{aligned}\hskip -6pt\bigg)_{\text{flavor factor}\rightarrow 1}\,.
\end{equation}

With these $N(\Gamma_i)$ at hand, it is easy to obtain numerators for other cubic diagrams. 

\item Given the CK-numerators for all the cubic diagrams, we can check the BCJ relations. The easiest example is the dual Jacobi relations for three scalar numerators. with the $n$-point numerator expressed in \eqref{eq:FFT2}. 

To be more precise, we consider $s$ and $t$ channel diagrams as two diagrams in \eqref{eq:3scalarb} with $\{i_1,\ldots,j_1\}=\emptyset$ and $\{i_1,\ldots,j_1\}=\{4^{g}\}$ respectively.
The dual Jacobi relation will naturally involve a $u$ channel diagram, which is not in our basis \eqref{eq:3scalarb} but can be described by \eqref{eq:FFT2} plus a trivial relabeling $1^{\phi}$ and $3^{\phi}$ from the $t$ channel diagram. 
The dual Jacobi relation holds at the $n^{(1\alpha 2)}_{(\tau_1),\ldots,(\tau_r)}$ level. 
We define the following short notations
\begin{equation}
\begin{aligned}
    n^{(134\cdots 2)}_{(\tau_1),\ldots,(\tau_r)}=\frac{(-2)^r n_s}{(p_{123}^2{-}q^2) (p_{123\tau_1}^2{-}q^2)\cdots (p_{123\tau_1\cdots \tau_{r-1}}^2{-}q^2)}, \\
    n^{(143\cdots 2)}_{(\tau_1),\ldots,(\tau_r)}=\frac{(-2)^r n_t}{(p_{123}^2{-}q^2) (p_{123\tau_1}^2{-}q^2)\cdots (p_{123\tau_1\cdots \tau_{r-1}}^2{-}q^2)}, \\
    n^{(341\cdots 2)}_{(\tau_1),\ldots,(\tau_r)}=\frac{(-2)^r n_u}{(p_{123}^2{-}q^2) (p_{123\tau_1}^2{-}q^2)\cdots (p_{123\tau_1\cdots \tau_{r-1}}^2{-}q^2)},
\end{aligned}
\end{equation}
with
\begin{equation}\label{eq:3scalarjacobi}
\begin{aligned}
    n^{s}=&{\prod\limits_{i=1}^r \Big(p_{\Phi^{s}_{L}\Xi_L(i)}\cdot f_{(\tau_i)}\cdot p_{\Phi^{s}_{R}\Xi_R(i)}\Big)}\text{ in which } \Phi^{s}_{L}=\{1,3\}\\
    n^{t}=&{\prod\limits_{i=1}^r \Big(p_{\Phi^{t}_{L}\Xi_L(i)}\cdot f_{(\tau_i)}\cdot p_{\Phi^{t}_{R}\Xi_R(i)}\Big)} \text{ in which } \Phi^{t}_{L}=\{1\} \textbf{ if }4\in \tau_i; \textbf{ else } \Phi^{t}_{L}=\{1,3\} \\
    n^{u}=&{\prod\limits_{i=1}^r \Big(p_{\Phi^{u}_{L}\Xi_L(i)}\cdot f_{(\tau_i)}\cdot p_{\Phi^{u}_{R}\Xi_R(i)}\Big)}\text{ in which } \Phi^{u}_{L}=\{3\}\textbf{ if }4\in \tau_i; \textbf{ else } \Phi^{u}_{L}=\{1,3\}\,. 
\end{aligned}
\end{equation}
Here we emphasize again that the definition of $\Xi_{L,R}$ is similar to $\widetilde{\Xi}_{L,R}$ but without including scalars.  
Now one can directly check the dual Jacobi relation 
\begin{equation}
    n^{s}-n^{t}-n^{u}=0
\end{equation}
using the fact that \textbf{ if } $\tau_i$ contains $4^{g}$ and other gluons,  then $ p_{\Phi_{L}^{t}}+p_{\Phi_{L}^{u}}=p_{\Phi_{L}^{s}}$, and $\Phi_{R}^{s,t,u}=\{C\}$; \textbf{ else}, $\tau_i=\{3\}$, then 
\begin{equation}
\begin{aligned}
    &\Xi_{L}=\emptyset \\
    &\Phi_{L}^{s}=\{1,3\},\ \Phi_{R}^{s}=\{2\};\quad \Phi_{L}^{t}=\{1\},\ \Phi_{R}^{t}=\{3,2\}; \quad \Phi_{L}^{u}=\{3\},\ \Phi_{R}^{u}=\{1,2\}\\
    &p_{13}\cdot f_{4}\cdot p_{2\Xi_R(i)}=p_{1}\cdot f_{4}\cdot p_{23\Xi_R(i)}+p_{2}\cdot f_{4}\cdot p_{12\Xi_R(i)}\,.
\end{aligned}
\end{equation}

\end{enumerate}

For more complicated cases, one can follow a similar argument and check the dual Jacobi relation by direct calculation.

Surprisingly, these CK-numerators can be mapped to CK-numerators for high-length operators; the dual Jacobi relations here can be mapped to more complicated dual relations for higher-length operators. 
We will give more concrete examples about the ${\rm tr}(\phi^2)$ numerators and explain how these maps work in \cite{treepaper2}.

\providecommand{\href}[2]{#2}\begingroup\raggedright\endgroup

\end{document}